%% file: ms9.tex
\documentclass[iop,numberedappendix]{emulateapj}

\pdfoutput=1

\usepackage[usenames,dvipsnames,svgnames]{xcolor}
\usepackage{natbib}
\usepackage{graphicx}

\usepackage{rotating}
\usepackage{apjfonts}
\usepackage{amsmath}

\usepackage[normalem]{ulem}
\usepackage{soul}

\usepackage{pdfpages}

\newcommand{\kms}       {km~s$^{-1}$}

\newcommand{\etal}      {{et~al.}}

\newcommand{\pcm}       {cm$^{-2}$}
\newcommand{\lya}       {Ly$\alpha$}
\newcommand{\lyb}       {Ly$\beta$}

\newcommand{\flux}       {ergs cm$^{-2}$ s$^{-1}$ \AA$^{-1}$}

\newcommand{\CH} {{COS-Halos}}
\newcommand{\vrot} {$v_{\rm{rot}}$}

\newcommand{\bturb} {$b_{\rm{turb}}$}
\newcommand{\btherm} {$b_{\rm{therm}}$}

\newcommand{\vtwo}{{\it{Q0244}}}
\newcommand{\vfour}{{\it{S393Z082}}}
\newcommand{\vsix}{{\it{Q0246}}}
\newcommand{\vseven}{{\it{HE0241}}}
\newcommand{\vthree}{{\it{S394Z150}}}

\usepackage[pdfnewwindow=true,colorlinks=false]{hyperref}

\shorttitle{ }
\shortauthors{D. V. Bowen et al}

\received{10-Feb-2016}
\accepted{13-May-2016}

\begin{document}

\title{The Structure of the Circumgalactic Medium of Galaxies:  Cool
  Accretion Inflow Around NGC 1097$^1$}

\author{David V.~Bowen\altaffilmark{2}, 
Doron Chelouche\altaffilmark{3}, 
Edward B.~Jenkins\altaffilmark{2}, 
Todd M. Tripp\altaffilmark{4},
Max Pettini\altaffilmark{5},
Donald G.~York\altaffilmark{6},\\
Brenda L. Frye\altaffilmark{7}
}

\altaffiltext{1}{Based on observations with the NASA/ESA {\it Hubble
    Space Telescope} (HST) obtained at the Space Telescope Science
  Institute, which is operated by the Association of Universities for Research in Astronomy, Inc., under NASA contract NAS 5-26555.}
\altaffiltext{2}{Princeton University Observatory, Ivy Lane, Princeton, NJ 08544.} 
\altaffiltext{3}{Dept.~of Physics, University of Haifa,
Mount Carmel, Haifa 31905, Israel.}
\altaffiltext{4}{Dept.~of Astronomy, University of Massachusetts,
  710 North Pleasant Street, Amherst, MA~01003.}
\altaffiltext{5}{Institute of Astronomy, University of Cambridge,
  Madingley Road, Cambridge, CB3 0EZ, UK.}
\altaffiltext{6}{Dept.\ of Astronomy and Astrophysics, University of
 Chicago,  Enrico Fermi Institute, 5640 South Ellis Avenue, Chicago,
 IL~60637.} 
\altaffiltext{7}{Dept.~of Astronomy/Steward Observatory, University of Arizona, 933 N.~Cherry Ave,
Tucson, AZ 85721}

\begin{abstract}

  We present {\it Hubble Space Telescope} far-UV spectra of 4 QSOs whose
  sightlines pass through the halo of NGC~1097 at impact parameters of
  $\rho = 48-165$~kpc.  NGC~1097 is a nearby spiral galaxy that has
  undergone at least two minor merger events, but no apparent major
  mergers, and is relatively isolated with respect to other nearby bright
  galaxies. This makes NGC~1097 a good case study for exploring baryons in
  a paradigmatic bright-galaxy halo.  \lya\ absorption is detected along
  all sightlines and Si~III~$\lambda 1206$ is found along the 3 smallest
  $\rho$ sightlines; metal lines of C~II, Si~II and Si~IV are only found
  with certainty towards the inner-most sightline.  The kinematics of the
  absorption lines are best replicated by a model with a disk-like
  distribution of gas approximately planar to the observed 21~cm H~I disk,
  that is rotating more slowly than the inner disk, and into which gas is
  infalling from the intergalactic medium.  Some part of the absorption
  towards the inner-most sightline may arise either from a small-scale
  outflow, or from tidal debris associated with the minor merger that gives
  rise to the well known `dog-leg' stellar stream that projects from
  NGC~1097.  When compared to other studies, NGC~1097 appears to be a
  `typical' absorber, although the large dispersion in absorption line
  column density and equivalent width in a single halo goes perhaps some
  way in explaining the wide range of these values seen in higher-$z$
  studies.
\end{abstract}


\section{INTRODUCTION}
\label{sect_intro}

It is now thought that the circumgalactic medium (CGM) of a galaxy
represents a genuine transition zone for the recycling of baryons.  The
term CGM refers to the gas that exists in some sense `outside' of the
interstellar medium (ISM) of a galaxy, yet `inside' its virial radius,
distinct from the surrounding intergalactic medium (IGM) \citep[see,
e.g.][]{shull14}.  The idea of the CGM has replaced that of the classic
``gaseous halo'' or ``galactic corona'', in large part because of a
recognition of the many physical processes that probably compete to sustain
it: baryons may be flowing out of galaxies as part of stellar or AGN winds
\citep[][and refs.~therein]{rubin14, bordoloi14b}, or from inflows from the
IGM \citep{ford14,martin12,rubin12}, or from the interplay between merging
galaxies, stripped via tidal interactions
\citep[e.g.][]{rupke13,bessiere12}; or they may simply be part of in-situ
cloud formations that arise from thermal instabilities in a halo
\citep[][and refs.~therein]{mo96,hobbs13}. These processes drive galactic
evolution, so characterizing the CGM around galaxies is essential for
understanding how galaxies evolve.

One way to study galaxy CGMs is through the absorption lines they cause in
the spectra of background objects. The method of detecting QSO absorption
line systems (QSOALSs) in background QSO spectra, and then searching for
galaxies at the same redshift as the absorption, has been in practice for
three decades.  Since the launch of the {\it Hubble Space Telescope} (HST),
QSOALSs have been found in the ultraviolet (UV) at very low redshifts; in
particular, the installation of the {\it Cosmic Origins Spectrograph}
\citep[COS,][]{green12} aboard HST has enabled the production of several
extensive surveys that have targeted the CGMs of specific samples of
relatively low-$z$ galaxies, including, for example, ``COS-Halos''
\citep{tumlinson13,werk13a}, ``COS-Dwarfs'' \citep{bordoloi14},
``COS-GASS'' \citep{borthakur15} as well as several unnamed surveys
\citep[e.g.][]{stocke13,borthakur13,burchett15,burchett16}.  If the goal of
a study is to understand the nature of the CGM around galaxies as a
function of their properties, then surveys that look for absorption lines
from selected galaxies are perhaps the most valuable, because selecting
galaxies prior to searching for any absorption reduces the bias in only
finding CGMs around particular types of galaxies or in particular
environments.

All these surveys study the statistical properties of absorbing clouds
that arise from single sightlines through single galaxies.
Conclusions are drawn about the CGMs of galaxies assuming that
galaxies are sufficiently similar that probing many galaxies with
single sightlines is the same as probing a single galaxy with many
sightlines.  Or at least --- as larger samples of galaxy-absorber
pairs have become available, and a more nuanced approach has been
possible --- that galaxies with similar characteristics (the same
luminosity, halo mass, type, star-formation rates, etc.) will have
similar enough CGMs that the properties of the absorbing clouds --- if
really linked to these properties --- will become apparent.

Is it valid to assume that similar galaxies can have the same CGMs?  Given
two identical galaxies, can their properties dominate the state of the
baryons within $R_v$ so directly that they end up having highly similar
CGMs? Or are the processes that manufacture the CGM inherently too chaotic
to lead to reproducible structures?  Indeed, can QSOALSs provide any unique
signature of a CGM's origin at all?

One way to test these assumptions is to {\it probe the CGM of individual
  galaxies along multiple sightlines}.  Absorption by the CGM of one galaxy
should depend on far fewer variables in such experiments because,
obviously, galaxy properties are fixed for a single galaxy.  Such studies
ought to be able to more easily expose which property of a galaxy a CGM
most depends on.

There are several ways to probe galaxies along more than a single
sightline.  Multiple sightlines have been observed from either closely
separated QSO sightlines \citep{bechtold94,fang96,dodorico98, dinshaw98,
  dinshaw97} or lines of sight towards the multiple images of lensed QSOs
\citep{petry98, smette92, monier98, rauch01b}. In many cases, such
investigations have focussed on determining the sizes of absorbing clouds
when lines were detected (or not) in common along the sightlines, although
the use of these background sources also provided examples of rarer Damped
Ly$\alpha$ (DLA) systems \citep{cooke10, ellison04, churchill03b,
  kobayashi02, lopez99,zuo97,smette95}, the detection of which might be
expected for lensed QSOs when the lensing object is predicted to be
relatively close to the sightlines. For these latter absorbers, multiple
sightlines have now been used to provide evidence for changes in gas-phase
metallicities on scales of a few kpc \citep{lopez05}.

For lensed QSOs, lensing galaxies have been at $z \ga 1$, and the absorbing
galaxy has been undetected (even if a lens redshift is inferred from the
configuration of the lensed images), and the galaxy's properties remain
unknown. There are only a small number of examples where {\it both}
multiple background sightlines have been observed at a high enough
resolution to show absorption systems, {\it and} a galaxy has been
confirmed spectroscopically to be at the same redshift as an absorber
\citep{rauch02,rogerson12,chen14,muzahid14,zahedy16}. Explanations for the
observed absorption have covered the entire gamut of CGM origins, from
ancient, as well as recent, outflows, to tidal debris, inflowing streams
and stripped gas. Even with these studies then, there is no clear agreement
on a common origin for absorption from lensing galaxies.

An alternative method for studying multiple sightlines through a galaxy
(and the one adopted in this paper) is to select {\it nearby} galaxies for
study. Galaxies at very low-redshifts have large angular footprints on the
sky, which make it easier to identify multiple QSO sightlines that pass
through the galaxy at interesting impact parameters.  The main disadvantage
of working at low-redshifts is that most of the absorption lines of
interest all lie in the UV, and can only be reached using HST. Not only is
access to the satellite limited due to high demand, but the telescope's
modest mirror size restricts spectroscopic observations to only the
brightest objects. Nevertheless, it was always expected prior to its launch
that HST would probe galaxy halos along more than single sightlines
\citep[e.g.][]{monk86} and indeed, some attempts were made with the first
generation of the HST spectrographs \citep[e.g.][]{norman96,bowen97_leo1}.
Significant progress has been made in the last few years, however, as COS
has enabled observations of the fainter QSOs that constitute a higher
surface density on the sky.

\citet{keeney13} first made use of the enhanced throughput of COS by
observing 3 QSO sightlines 74$-$172~kpc from the edge-on spiral
ESO~157$-$49 at $cz=1673$~\kms ; they detected absorption from \lya , and
both low- and high-ion species, and suggested that some of the absorbing
clouds are the remnants of recycled galactic fountain gas. Perhaps a more
obvious set of galaxies to probe with multiple QSOs, however, might be M31
and M33. Being the closest bright galaxies to the Milky Way, they have very
large angular diameters on the sky, and finding a high number of background
probes at interesting impact parameters is straightforward. The problem
with both these galaxies is that their velocities are low. Maps of 21~cm
emission show that the high H~I column density gas in the disk of M33
extends from $\sim -60$ to $-300$~\kms\ \citep[e.g.][]{putman09}, and from
$\sim -50$ to $-600$~\kms\ for M31 \citep[e.g.][]{corbelli10} (the latter
having more than twice the angular diameter on the sky than M33). The
complexity of the H~I on the largest scales between M31, M33 and the
Magellanic Stream\footnote{The Magellanic Clouds (MCs) also have very large
  angular extents of course; background QSOs have been cataloged
  \citep[e.g.][and refs. therein]{cioni13}, but the MCs are at even lower
  velocities than M31 and M33, and separating absorption from the MW disk
  \& HVCs, and from the MS stream, is complicated. The situation is just as
  difficult when stars in the MCs themselves are observed, and these stars
  probe only the very centers of the galaxies.  Many HST spectra exist of
  stars in the MCs, but given that they are not ideal galactic halo probes,
  we do not include results from these studies in this paper.}  and the
difficulties in separating out Galactic, Local Group, and extragalactic H~I
can be seen in the maps of \citet{braun04}. These low velocities make it
difficult to separate any absorption from the galaxies with absorption from
HVCs in the Milky Way, particularly Complexes H \& G \citep[e.g. Fig.~1
of][]{tripp12}; they also make it impossible to observe any extragalactic
\lya\ absorption, which is lost in the saturated \lya\ profile of the
Galaxy. Nevertheless, attempts to utilize the many QSOs behind both
galaxies have been made recently by \citet{rao13} and \citet{lehner14}. We
return to these studies later (see \S\ref{sect_others}).

In this paper, we investigate absorption --- including that of \lya\ --- by
multiple QSO sightlines that pass through the halo of NGC~1097, a galaxy
that lies well away from the Local Group. 
Our paper is divided into several sections.
In \S\ref{sect_galaxy} we summarise the properties of
NGC~1097, its local environment, and the large scale structure (LSS) in which it is found.
In \S\ref{sect_targets} we briefly discuss some of the
work that led to the selection of NGC~1097 and the
background QSOs for study. 
In \S\ref{sect_cos} we describe the HST observations and the reduction of
the data, and in \S\ref{sect_results} we analyse in detail the
characteristics of the absorption lines detected.
 \S\ref{sect_photou_point} outlines the results of photoionization modelling of
 the results, although full details are presented in the Appendix.
In the remaining
sections we discuss the interpretation of our results.  
We consider possible configurations of the CGM of NGC~1097 based
on the kinematics of the absorbing gas in \S\ref{sect_kinematics},
examine the
statistical properties of the detected absorption in \S\ref{sect_global},
and
compare our results to other studies of absorbing galaxies
in order to determine whether NGC~1097 is a
`typical' absorbing galaxy in comparison to the growing set of
higher-$z$ absorbing galaxies (\S\ref{sect_others}). 
We discuss our results in \S\ref{sect_discussion}, where we 
suggest the most likely
origin for the absorption  (\S\ref{sect_best}), and briefly
compare our data with the results from numerical simulations
(\S\ref{sect_sims}). We summarise our results in \S\ref{sect_summary}.

Throughout this paper we adopt the
cosmology $H_0 = 73$~\kms~Mpc$^{-1}$, $\Omega_m = 0.27$, and 
$\Omega_\Lambda = 0.73$, when required.

\begin{figure*}
\vspace*{0cm}\hspace*{0cm}\includegraphics[width=18.1cm]{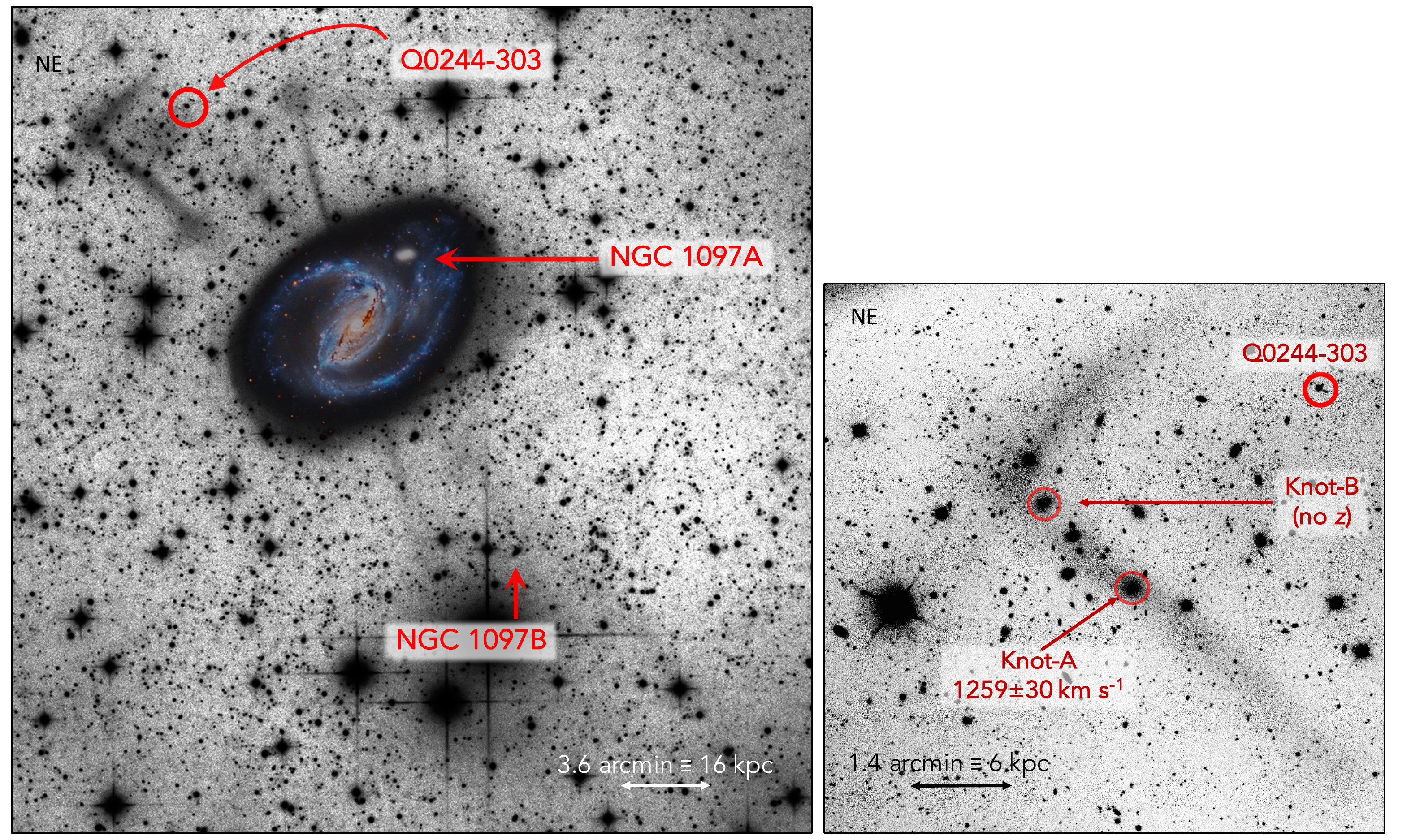}
\caption{The field of NGC~1097  ($cz=1270$~\kms), showing its stellar tidal streams.
  {\bf Left:} Part of a multicolor (22 hr total) composite
  image of the galaxy taken with a
Ceravolo 30~cm telescope and
Apogee U16M camera by Rick Stevenson in Queensland, Australia in 2013.
The two
  most prominent streams, towards the north-east, are easily seen:
  the well-known `dog-leg' stream is to the left, with its right-angle
  kink, while the 
  stream to the right appears to end in a bulbous cloud of stars. To
  the south-west, the left-most stream is fainter and shorter than the
  northern streams, but is clearly visible. The southern
  stream to the right is very faint ($\sim 27.6$~mag arcsec$^{-2}$)
  but may be the longest of the four, and perhaps wider and more amorphous
  than the other three streams. The streams themselves are likely the stellar remains of
  a low-mass galaxy that has passed through the disk of NGC~1097 \citep{higdon03,amorisco15}. 
NGC~1097's companion galaxies are also
  marked --- NGC~1097A
  ($cz=1368$~\kms ) and NGC~1097B (see Appendix \ref{sect_n1097b}). The
  position of one of the QSOs discussed in this paper, Q0244$-$303, is
  marked, but the remaining sightlines are outside the field of view
  (see instead Fig.~\ref{fig_map}). 
{\bf Right:} Portion of a 400 sec $V$-band FORS2 image extracted from the ESO Archive
  [originally published by \citet{galianni10}]. The dog-leg stellar
  trail is shown more clearly in relation to the position of Q0244$-$303,
  and the only part of the stream with a spectroscopic redshift,
  Knot-A, is indicated. 
\label{fig_pretty}\\}
\end{figure*}

\section{NGC~1097 and its Environment}
\label{sect_galaxy}

\subsection{Properties of NGC~1097}
\label{sect_galaxy_indi}


\begin{deluxetable}{llc}
\tablecolumns{3} 
\tablecaption{Parameters for NGC~1097 \label{tab_n1097}}
\tablehead{
\colhead{} & \colhead{} & \colhead{Ref.}
}
\startdata
RA, Dec:                                     & 02:46:18.95,  $-$30:16:28.8           & 1 \\
Heliocentric velocity $v_\odot$:          & 1271$\pm10$~\kms             & 2 \\
Hubble flow velocity:                    & 1105$\pm12$~\kms                    & 3 \\
Inclination, PA of major-axis:        & $40\pm 5$\degr , 130\degr         & 4 \\
Mag $b_J$, $M(b_J)$, $L$:           & 10.1, $-20.8$, $1.4\,L_*$             & 5 \\
Diameter $D(0)_{25}(B)$:               & 9.55$'$ $\equiv 42$ kpc             & 6 \\ 
H~I mass$^a$:                             & $7.7\times10^9 M_\odot$             & 2 \\
Star Formation Rate$^a$:                & $5 \pm 1 \:M_\odot$ yr$^{-1}$      & 7 \\
Stellar mass [$\log(M_\odot)$]$^a$: &  $10.5\pm 0.1$                         & 8 \\
Total mass [$\log(M_\odot)$]:           & $\simeq 12.0\pm 0.2$              & 9 \\
Adopted Distance $D$:                & $15.1\pm1.1$ Mpc                     & 10 \\
Virial Radius, $R_{\rm{vir}}$:            & 280 kpc                                        & 10 
\enddata
\tablenotetext{a}{Published values corrected for $D=15.1$~Mpc}
\tablecomments{References:
(1) Position of nucleus in HST WFPC2 F218W image  
from data in the {\it Hubble Legacy Archive}; 
(2) From 21~cm emission line data of \citet{koribalski04};
(3) Heliocentric velocity converted to CMB velocity using the
conversion given by the NED;
(4) \citet{higdon03};
(5) \citet{doyle05};
(6) \citet{RC3} (RC3);
(7) \citet{calzetti10};
(8) \citet{skibba11};
(9) Converted from stellar mass --- see \S\ref{sect_galaxy_indi};
(10) See \S\ref{sect_galaxy_indi}.
 }
\end{deluxetable}

NGC~1097 is an SB(s)b galaxy (Figs.~\ref{fig_pretty})
which hosts a Type 1 AGN that is
surrounded by a $\sim 700$~pc ring of on-going star-formation
\citep[e.g.][]{barth95,prieto05}. 
A global systemic
velocity for the galaxy can be taken from 21~cm emission line
measurements, the two most pertinent H~I maps having been made 
by \citet{ondrechen89} and \citet{higdon03}. 
 Ondrechen~et~al. cited a
heliocentric systemic velocity of $v_\odot = 1280\pm10$~\kms , but Higdon
\& Wallin did not derive $v_\odot$ from their data.  Since these maps
were produced, the {\it H~I Parkes All-Sky Survey} (HIPASS)
single-dish survey Bright Galaxy Catalog has listed
$v_\odot = 1271\pm3$~\kms\ \citep{koribalski04}.

Converting the heliocentric systemic
velocity to a Hubble flow distance is not necessarily correct, since
corrections for local group motions, and motions of NGC~1097 towards
the Fornax cluster (see \S\ref{sect_lss}) are not accounted for. The difference
between $v_\odot$ and the CMB background is 166~\kms, giving a corrected
velocity of $1105\pm12$~\kms\ if the HIPASS velocity is used, and
a corresponding distance of $15.1\pm1.1$~Mpc, which we
adopt in this paper. This gives a scale of 4.4~kpc~arcmin$^{-1}$ on
the sky. This distance agrees well with a distance of
$14.2\pm2.4$~Mpc given by \citet{tully08} based on a distance modulus
derived from the 21~cm line width-galaxy luminosity relationship of
\citet{tully-fisher77}. The apparent magnitude of NGC~1097 is
$b_J = 10.1$ \citep{doyle05}, which would translate to an absolute
magnitude of $M_*(b_J) = -20.8$ for $D=15.1$~Mpc. If an $L^*$ galaxy
has an absolute magnitude of $M_*(b_J) = -20.4$ \citep{norberg02b},
then $L/L_* = 1.4$ for NGC~1097.

We calculate the virial radius $R_v$ for NGC~1097 as follows.
With the stellar mass of the galaxy known (Table~\ref{tab_n1097}), the DM
mass can be calculated using the stellar-mass to halo-mass correlations
given by \citet{behroozi13}: for $\log M^* = 10.5\pm0.1$,
$\log M_h (M_\odot) = 12.1$. The error in converting between stellar
mass and halo mass is of the same order as the error in the stellar
mass itself, $\sim 0.2$ dex.  This halo mass is similar to
$\log M_h (M_\odot) = 11.8$ found by \citet{higdon03}
when fitting a simple disk+halo model to the observed 21~cm rotation curve.
Using the approximations given by \citet{maller04}, the
virial halo radius is $R_v \simeq 280$~kpc. Deviations of order 0.2~dex in
stellar mass would change this value by $\pm 40$~kpc, so the value of $R_v$
is not precisely known.  The value of $R_v = 280$~kpc  agrees with the
conversion between galaxy luminosity and $R_v$, $M_h$ given by
\citet[][their Fig.~1]{stocke13}, providing their halo abundance
matching curves are used; their ``adopted minimum $M/L$ ratio'' 
gives an $R_v$ that is 30\% smaller. For a halo with this mass, the
corresponding virial velocity is 140~\kms, and the virial temperature
is $6.9\times 10^5$~K.

These, and other relevant physical parameters of NGC~1097 (some of which are discussed later in this paper) 
are listed in Table~\ref{tab_n1097}.

\subsection{Local Environment}
\label{sect_local}

\begin{figure*}[t]
\centering
\includegraphics[width=17cm]{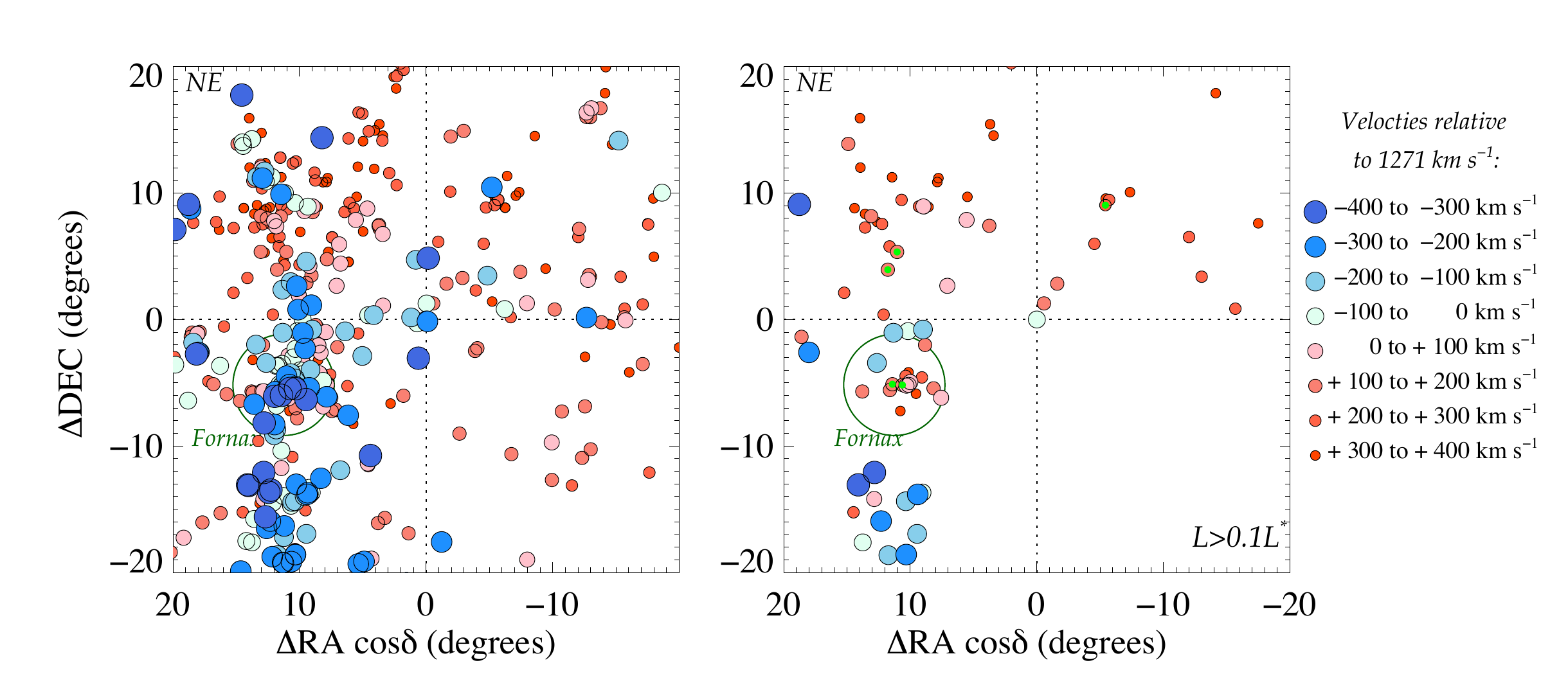}
\caption{Distribution of galaxies within $\pm 400$~\kms\ of NGC~1097. The position of
  NGC~1097 is at the center of each plot (dotted lines).  {\it Left:} All galaxies
  with redshifts cataloged by the NED are plotted. Galaxies which have
  redshifts greater than NGC~1097 are shown as small, red circles;
  those with redshifts less that NGC~1097 have larger, bluer
  circles. The magnitudes of the velocity differences, as a function of
  the color and size of the circles, are shown in the legend on the
  extreme right of the figure. The largest, bluest circles are for
  galaxies with the
  highest blueshifts relative to NGC~1097, while the smallest, reddest
  circles have the highest redshift relative to NGC~1097. 
  {\it Right:} Same as left panel, but only galaxies with
  luminosities $\ga 0.1L^*$ are plotted. The 5 red circles filled with green
  are the 5 closest galaxies with luminosities greater than 0.8 times
  that of NGC~1097, assuming that their radial redshifts really
  represent Hubble flow distances. The labelled (unfilled) green circle in each
  panel shows a 4 degree radius from the center of the Fornax cluster,
  which corresponds to $\approx 1.4$~Mpc at the distance of Fornax.\\
\label{fig_lss}}
\end{figure*}

At the outskirts of the galaxy, NGC~1097 shows clear signs of minor
merger events, one on-going, the other occurring in the more
distant past.  The most obvious feature is the presence of NGC~1097A
(see Fig.~\ref{fig_pretty}), an elliptical galaxy which has a velocity of
$1365\pm45$~\kms , and which appears to be interacting with (and
disrupting) the outer north-west spiral arm of the parent galaxy. The
interaction produces two additional H~I arms to the south of the
galaxy, well beyond the optical disk (see Fig.~4b of Higdon \& Wallin
2003, and the outer contour of the 21~cm emission
outlined in Fig.~\ref{fig_map}). The galaxy has a magnitude of
$m_{B(0)} = 13.6$ or $M_B = -17.3$ at a distance of 15.1~Mpc, largely
equivalent to the Large Magellanic Cloud (LMC) near our own Galaxy.

More uniquely, NGC~1097 also shows four
optical filaments  or streams \citep{wolstencroft75, arp76,
  lorre78}.  The structures show no optical, X-ray, or radio
emission \citep[][and refs. therein]{wehrle97}. All four streams can be seen in 
the left panel of 
Figure~\ref{fig_pretty}. 
The filament oriented to the
north-east is remarkable for showing an acute right-angle bend,
producing an 'L-shaped' structure which has become known as
the `dog-leg'.

In discussing the possible origin of the streams, \citet{higdon03}
suggested that they were caused by the minor merger of a low-mass dwarf
disk galaxy, with the lack of any detectable H~I and H~II emission due to
ram-pressure stripping of the dwarf's gas by the disk of NGC~1097.  
Higdon \& Wallin were able to show
that X-shaped tidal streams --- including features resembling the dog-leg seen
towards NGC~1097 --- could be reproduced in $N$-body simulations of a dwarf
disk galaxy with a mass of only 0.1~\% of the host galaxy mass (so in this case,
$M_*\:\sim\: 6\times 10^8 \:M_\odot$) passing through a disk-dominated
potential. The streams are, therefore, most likely the stellar remains
of a disrupted galaxy, and are not caused by outflowing gas from the
center of NGC~1097. 
More recently, \citet{amorisco15} modelled the streams using the infall of
a disky dwarf galaxy that has passed its pericenter three times, and which
has lost two dex in mass.  They too were able to reproduce the right-angle
kink seen in the dog-leg stream.

The interpretation that the streams are composed mainly of stars is
consistent with spectra of two regions of the dog-leg obtained by
\citet{galianni10}.  Within the dog-leg stream itself, there are several
`knots' of stars. The only feature with a confirmed redshift is Knot-A,
which Galianni~\etal\ measured to be $cz = 1259\pm30$~\kms , and which
likely gives a redshift for the stream at that position
(Fig.~\ref{fig_pretty}, right panel). Galianni~et~al also observed a second
source near the right angle of the trail (their 'Knot-B', see Fig.~\ref{fig_pretty}), but were unable
to derive a redshift; for both sources, no emission lines were detected, as
might be expected if the streams were actually hot outflows from the center
of the galaxy. Moreover, the SEDs of the knots are inconsistent with
thermal or synchrotron emission \citep{higdon03}.

\subsection{Global Environment \label{sect_lss}}

In order to examine how the CGM of NGC~1097 might be related to (and
influenced by) the wider IGM, we would like to know the galactic structures
in which the galaxy is found.
To examine the relationship of NGC~1097 to its environment, we
extracted all galaxies with known redshifts within $\pm 20$~degrees of
the galaxy from the {\it NASA/IPAC Extragalactic Database} (NED). 
If we naively assume that the difference in velocity between the
surrounding galaxies and NGC~1097 in the line-of-sight direction gives a
distance of $\Delta v /H_0$, then the nearest galaxies that have
luminosities $\geq 0.8$ times that of NGC~1097 lie $\sim 4-5$ Mpc away
(NGC~908, 1371, 1399 and 1398, and 6dF~J0342193$-$352334). 

NED's listing is not a magnitude-limited sample (there
is currently no magnitude-limited survey covering this wide a field around NGC~1097)
but the collation is useful in showing
the proximity of NGC~1097 to the Fornax Cluster, which lies 12 degrees
away.  The
galaxy surface density of Fornax reaches the background at only $\sim 6$~degrees
\citep{ferguson89} from its center, which almost certainly places NGC~1097 outside of the
cluster. The left-hand panel of Figure~\ref{fig_lss} shows a roughly linear
distribution of galaxies running North-South on the sky, defining the filamentary
structure containing Fornax cluster galaxies, and the color coding in the
figure highlights the filament's velocity gradient 
\citep{waugh02}. NGC~1097 appears to lie close to, but not obviously within, the
filament; not only does its distance from the central concentration of galaxies
suggest that it lies away from the filament, but its velocity is several hundred
\kms\ lower than the majority of the background galaxies in that direction. If the
distance from us to the center of Fornax is $19-20$~Mpc \citep[e.g.][]{madore99,dunn06}, then
NGC~1097 is $\sim 5-6$~Mpc from its center.  

The right-hand panel of Figure~\ref{fig_lss} shows the same distribution of galaxies,
but only after selecting galaxies with $L > 0.1 L^*$. The magnitudes of galaxies listed by the
NED are inhomogeneous in their filter selection, and the conversion of those
magnitudes to a luminosity is approximate. 
The five closest galaxies mentioned above that have comparable luminosities to that of
NGC~1097, are highlighted as circles with green
centers, and are  all over $4-5$~Mpc away.
The panel demonstrates that
there are far fewer {\it bright} galaxies near NGC~1097 than the left hand panel
suggests, and that there are no companion galaxies to NGC~1097 with
comparable luminosity.
Figure~\ref{fig_lss} therefore suggests that NGC~1097 is a relatively isolated
galaxy, probably lying at the edge of a large scale structure filament.

While it is possible that NGC~1097 may 
be moving towards the cluster, there is no evidence that it has experienced any recent
tidal stripping from any intracluster medium. In this sense, the CGM
of the galaxy has most recently been defined primarily by the local  
interactions discussed in the previous section.

\section{QSO Target Selection}
\label{sect_targets}

\begin{figure*}
\centering
\includegraphics[width=\textwidth]{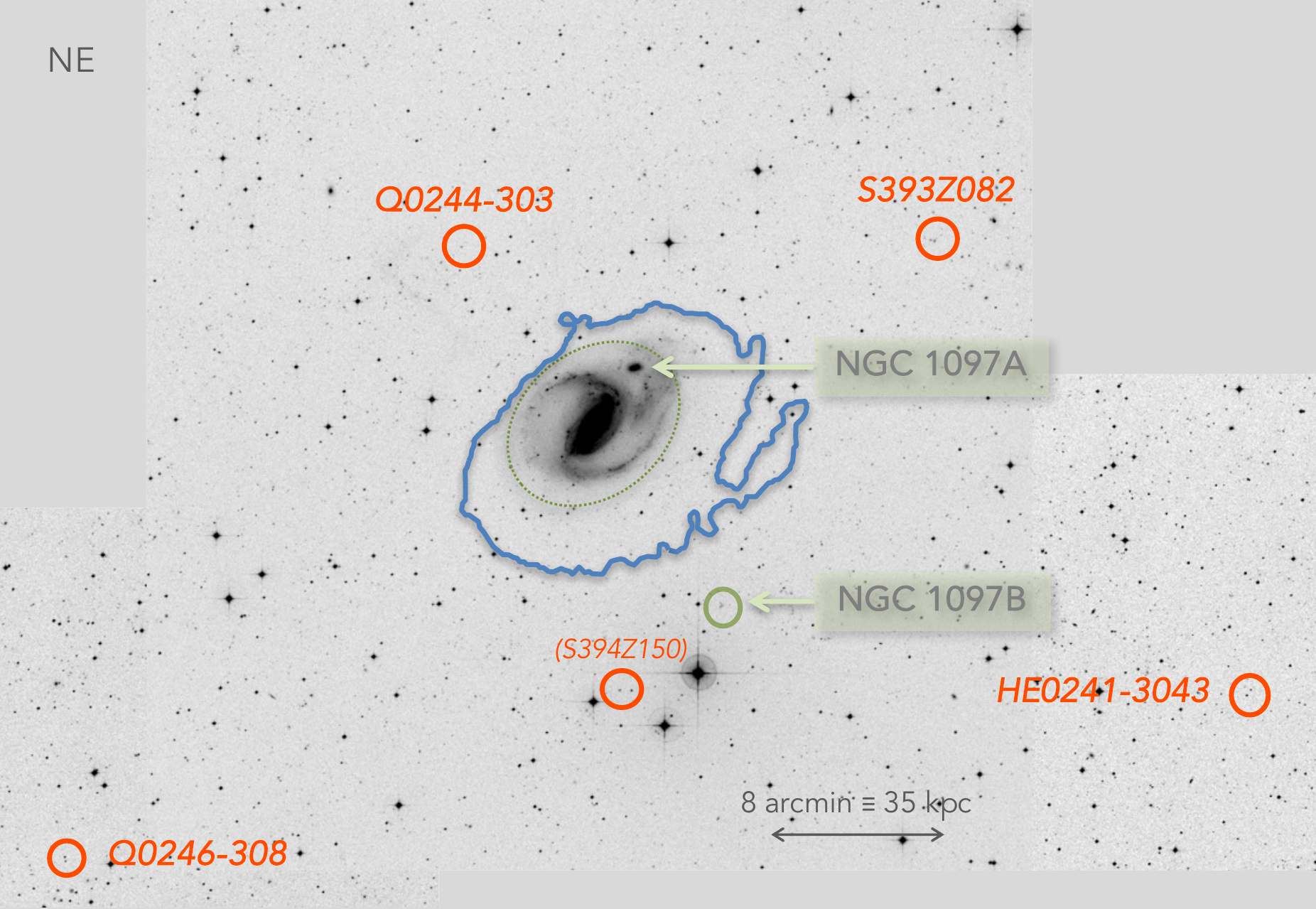}
\caption{Positions of QSO sightlines observed in this paper relative
  to NGC~1097. The image is constructed from portions
  of the STScI {\it Digitized Sky Survey} data, aligned with each
  other to reproduce the correct angular distances on the sky.
  QSOs observed in our
  COS program are labelled with circles; 
  we also show the position of S394Z150 even though our COS data
  found it to be a white dwarf. The green dotted line around NGC~1097 marks its
  optical diameter $D(0)_{25}$, while the thick blue line shows the
  maximum extent of the galaxy detected by \citet{higdon03} at 21~cm, at a
  level of $\simeq 1.3\times10^{19}$~\pcm .  If the virial radius
  $R_v$ of
  NGC~1097 is 280~kpc then the furthest
  sightline from the galaxy, towards HE~0241$-$3043, is at an impact parameter of $\simeq
  0.6 \:R_v$ in this figure.  
\label{fig_map} \\}
\end{figure*}

Over the last few years we have been searching the fields of nearby galaxies
to find those with multiple QSOs behind them.  
We began by compiling a list of the
top 250 low-redshift galaxies with the largest angular diameters on
the sky, and with velocities  $\geq 700$~\kms\ to
ensure that \lya\ absorption would be well away from the geocoronal \lya\
emission line (see \S\ref{sect_airglow}) in COS data. For each galaxy, we used
available {\it Galaxy Evolution Explorer} (GALEX) satellite catalogs
to collate all sources within 200~kpc, that had a measured Far-UV
(FUV) flux greater than 40~$\mu$Jy [$F_\lambda =
5\times10^{-16}$~\flux, or $m$(FUV)=19.9].
An additional cut was made
to remove objects with GALEX FUV widths greater than 15 arcsec FWHM, in
order to avoid extended sources that would pass only a small fraction of
their cataloged flux through the small COS aperture.

To corroborate which of the UV sources were extragalactic, and not Milky Way stars,
the GALEX-generated lists of FUV-bright objects were cross-correlated with the
latest release of the SDSS redshift
catalogs, the NED, and the SIMBAD
database. The galaxies were then ranked
simply by the surface density of the background probes.
NGC~1097 was one of a small number of
galaxies that had more than $5$ sources satisfying the criteria
described above 
and became one of the potential targets for HST follow-up.

\begin{deluxetable*}{llcccccrc crc}
\tablecolumns{12} 
\tablewidth{0pc} 
\tablecaption{Journal of Observations\tablenotemark{a} \label{tab_qsos}}
\tablehead{
\colhead{}             
& \colhead{}     
& \colhead{}         
& \colhead{}           
& \colhead{}              
& \colhead{ }                    
& \colhead{}   
& \colhead{}              
& \colhead{}      
& \multicolumn{3}{c}{Impact Parameters}\\
\cline{10-12}
\colhead{QSO}             
& \colhead{RA \& DEC}     
& \colhead{Galactic}         
& \colhead{$B(J)$\tablenotemark{b}}           
& \colhead{FUV\tablenotemark{c}}              
& \colhead{ }                    
& \colhead{Observation}   
& \colhead{Time}              
& \colhead{Observed}      
& \colhead{$\rho$}
& \colhead{$\rho$}
&\colhead{$\rho/$}\\
\colhead{name}            
& \colhead{(J2000.0)}       
& \colhead{$l$ \& $b$}     
& \colhead{(mag)}            
& \colhead{($\mu$Jy)}      
& \colhead{$z_{\rm{em}}$}   
& \colhead{Date}              
& \colhead{(min)}             
&  \colhead{Flux\tablenotemark{d}}  
& \colhead{$(')$}
& \colhead{(kpc)\tablenotemark{e}}
&\colhead{$R_v$\tablenotemark{f}}
}
\startdata
Q0244$-$303    & \phs02:46:49.87,        & 226.573,        & 18.4  & 126 & 0.53  &  2013$-$08$-$09 
& 204      & 1.14        &   11.0  & 48.2  & 0.17 \\
                                       & $-$30:07:41.3          &  $-$64.569     &          &         &&&& &&& \\ 
2dFGRS             & \phs02:45:00.77,       & 226.556,         & 19.1  & 57 & 0.34 & 2013$-$05$-$27
& 294  & 0.70       & 19.2  & 84.2  & 0.30  \\
             S393Z082          & $-$30:07:22.4           & $-$64.962      &          &         &&&&  &&& \\
Q0246$-$308    & \phs02:48:22.00,       & 227.747,         & 17.3  & 97 & 1.09 & 2013$-$08$-$06 
& 245  & 1.28  & 34.2 & 149.9  & 0.54 \\
                                       & $-$30:38:06.8        & $-$64.238              &         &         &&&& &&& \\
HE0241$-$3043 & \phs02:43:37.66,     & 227.488,        & 17.2  & 155 & 0.67 & 2013$-$06$-$21 
& 116  & 3.90   & 37.6  & 164.8  & 0.59 \\
                                        & $-$30:30:48.1        & $-$65.259            &          &         &&&& &&& \\
\hline
2dFGRS           &   \phs02:46:13.34      &   227.431     & 19.2         &   63    &   0.0\tablenotemark{g}    & 2012$-$10$-$17/20
& 308  &  1.0\tablenotemark{h}      & 13.3 & ... & ... \\
            S394Z150            &    $-$30:29:42.2    &   $-$64.700                      &          &         &&&& &&&
\enddata 
\tablenotetext{a}{All data were taken using COS and the G130M
grating centered at 1291~\AA\ \& 1327~\AA , and the PSA aperture.}
\tablenotetext{b}{Optical mags are $B(J)$ taken from the superCOSMOS Sky Survey web server or 2dFGRS database.}
\tablenotetext{c}{GALEX FUV flux.}
\tablenotetext{d}{Flux at 1240~\AA\ in units of $10^{-15}$~\flux .}
\tablenotetext{e}{Assuming a distance of 15.1~Mpc to NGC~1097.}
\tablenotetext{f}{Assuming a virial radius $R_v$ of 280~kpc.}
\tablenotetext{g}{The object was expected to have $z_{\rm{em}} = 0.131$,
  but turned out to be a white dwarf star --- see Appendix~\ref{sect_galex}}
\tablenotetext{h}{Flux at 1240~\AA\ is zero due to strong \lya\ absorption; the stated value is the flux in the continuum at 1350~\AA .\\}
\end{deluxetable*}

In addition, NGC~1097 was observed as part of a GALEX Cycle 1 program that we
designed to obtain FUV grism spectra of UV-bright objects in nearby galaxy fields.
Unfortunately, 
we did not discover any new
QSOs near NGC~1097, but did recover many of the objects already known.
Further details on the original discoveries of the QSOs selected
for our HST study
are given in Appendix~\ref{sect_galex} along with notes on our GALEX grism spectra.

Table~\ref{tab_qsos} lists the QSOs observed in this paper; in order
to provide a more succinct nomenclature for the objects,
we abbreviate the QSO names to the simpler forms of \vtwo , \vfour,
\vsix\ and \vseven\ throughout the text. 
A fifth object, 2dFGRS S394Z150, believed to be an AGN at $z=0.131$, 
was observed as part of our program. Its identification 
proved to be incorrect, however --- the object is likely a
white dwarf star instead. Additional details are given in 
Appendix \ref{sect_galex}, but this object is listed at the end of
Table~\ref{tab_qsos} as it was used in the analysis of the other spectra,
discussed below.

\section{COS Observations and Data Reduction }
\label{sect_cos}

\subsection{Co-addition of Sub-Exposures}

Our GO program (12988) observations were made with
COS using
the G130M grating and the Primary Science Aperture (PSA). A journal of the observations
is given in Table~\ref{tab_qsos}. Data were taken in {\tt TIME-TAG}
mode. Two different grating positions were used, centered at 1291 and
1327~\AA , to provide (after coadding all the exposures) some data in
the gap between the two segments of the photon-counting micro-channel
plate detector \citep{holland12}. Each sightline was observed over
$3-5$ orbits per visit, with all of the four available FP-POS offsets used to
reduce fixed-pattern noise.  

While the sub-exposures were wavelength calibrated by the pipeline, their
wavelengths disagreed with each other in pixel-space, because of
shifts in the dispersed light across the detector, caused by
the difficulty of returning the grating to exactly the same position
after any movement. In order to coadd spectra for a given sightline, we
interpolated the flux arrays onto a common wavelength grid 
at a dispersion of 0.01~\AA~pix$^{-1}$, close to the initial value from the
pipeline. To account for wavelength shifts in individual exposures that were not corrected for by the pipeline
calibration, we compared positions of the strongest lines visible in
each exposure, and measured the fractional pixel shifts needed 
to align the features.  
Any offsets found were added as
zero-point shifts to the wavelength arrays.

To co-add data at wavelengths obtained by only some of the exposures,
and to account for exposures with unequal exposure times, we
calculated a weighted
average from the {\tt COUNT-RATE} arrays in the {\tt
  x1d} files, with weights
formed from the inverse of a spectrum's variance.
 We propagated an error array $\sigma_{i}$ at each pixel $i$ from the
gross counts $n$ of a spectrum using the lower Poisson confidence
levels given by \citet{gehrels86}.
Each error array was smoothed by
a 10 pixel boxcar average to eliminate effects from fluctuations in
individual pixels.  The average count-rate of all the exposures was
formed by weighting each sub-exposure by the inverse of the variance array.

The long and short segments were 
 co-added using the same weighting scheme described
above. The final spectrum for each sightline was rebinned by a factor
of 3 to 0.03~\AA~pix$^{-1}$ ($\equiv 7$~\kms~pix$^{-1}$ at 1220~\AA )
to better sample the data given the resolution of COS. 
The final spectra had S/N ratios of $6-8$ per rebinned
pixel at 1220~\AA , and $8-10$ pix$^{-1}$ at 1340~\AA .

\subsubsection{Background Contribution from \lya\ Airglow \label{sect_airglow}}

Searching for \lya\ absorption lines at the redshifts of nearby galaxies is
hampered by two problems. First, at redshifts close to zero, \lya\ lines
are unobservable because the QSO flux is completely extinguished by the
damped \lya\ absorption from our own Milky Way. Second, geocoronal \lya\
emission filling the 2.5 arcsec diameter circular COS PSA will contribute
additional flux at the wavelengths of the expected extragalactic
absorption. This would make lines seem weaker than they really are, and
lead to an under-estimate of H~I column densities.  The \lya\ airglow is
distributed over a wide spectral range, and when observing objects with low
count rates, the emission can be a significant additional background source
near the rest wavelength of \lya .

\begin{figure}
\centering
\includegraphics[width=9cm]{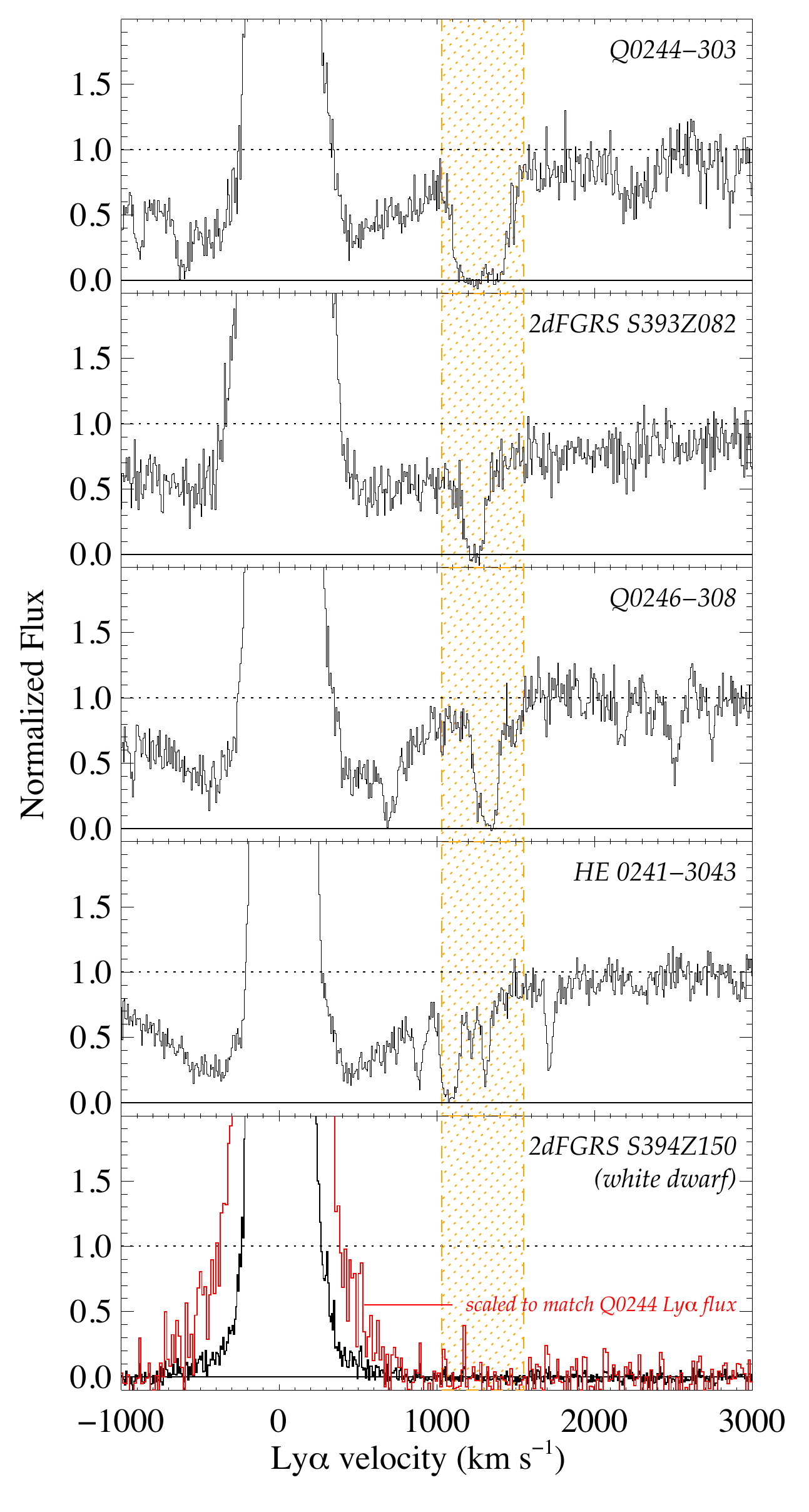}
\caption{The top 4 panels show COS spectra for the QSOs whose sightlines
  pass through the halo of NGC~1097. The orange dashed regions shows the
  approximate extent of \lya\ absorption from NGC~1097.
The bottom panel shows the spectrum of the WD
star \vthree\ that was also observed in our program. The black line shows the measured normalized
flux, while the red line shows the same flux (binned by a factor of 2 for
clarity) scaled (by a factor of 5.6) such that the geocoronal \lya\
emission line would have the same flux as
the line seen in the spectrum of \vtwo , which 
had the highest \lya\ emission line flux of the 4 QSOs. This panel 
shows that at velocities similar to that of NGC~1097, emission from the wings of \lya\
geocoronal emission add no additional background counts to the
observed \lya\ absorption lines seen from NGC~1097. \label{fig_spectra} }
\end{figure}

Portions of the co-added spectra near rest-frame \lya\ are shown in
Figure~\ref{fig_spectra}.
At the lowest velocities, geocoronal \lya\ emission dominates.
The spectrum of \vtwo\ shows the widest \lya\ absorption from
NGC~1097 (the extent of which is indicated by orange lines), but there is no flux at
the center of the absorption line profile, suggesting that at least by
a redshift of $cz=1270$~\kms, \lya\ airglow does not contribute to the
spectrum's flux.  Additional information comes from the observation of the WD
star, \vthree , where \lya\ absorption in the star's atmosphere is so
strong that there is no flux over a wavelength range much wider than
that which is usually extinguished from interstellar \lya\
in the Milky Way.  The spectrum of the WD is shown in the bottom panel
of Figure~\ref{fig_spectra}, where only geocoronal \lya\ emission is
left at the bottom of the damped \lya\ trough (black line). In our sample, the
strongest geocoronal \lya\ was observed towards \vtwo , and
Figure~\ref{fig_spectra} shows the extent of the emission towards the WD
after the peak of the \lya\ emission has been scaled to that of \vtwo\ (red line). The figure shows that by
$\simeq 800$~\kms , any contribution from airglow is
negligible and any \lya\ absorption lines at those velocities would
be unaffected.

\subsubsection{Final Wavelength Accuracy \label{sect_wavezero}}

To provide a precise 
zero-point to the wavelength calibration, we tied the wavelengths of
features in the COS data to the velocity
of 21~cm emission measured towards the QSOs from 
H~I profiles extracted from the {\it Parkes Galactic All-Sky Survey} (GASS) survey
\citep{mcclure-griffiths09} .
The median heliocentric velocities of the
21~cm H~I emission along all four QSO sightlines were nearly identical,
$+9$~\kms . We matched this velocity to those of the strongest
low-ionization ISM
lines seen in the COS spectra, which are expected to arise in the same
gas detected at 21~cm, namely Si~II~$\lambda 1190, 1193$, the
S~II~$\lambda\lambda 1250, 1253, 1259$ triplet, Si~II~$\lambda 1260$
and C~II~$\lambda 1334$. In fact, the velocities of these lines did
not all agree; in collating line centroids for all 4 sightlines,
differences of $0-12$~\kms\ ($\sim 0-2$ rebinned pixels) were found
between the different ions.  
There are several possible reasons for this: line centers may be
measured incorrectly because of low S/N in the line profile; weaker
components of the strongest lines may have different profiles compared
to weaker lines where such components are not detected; or errors may
arise in the wavelength solution applied to the detectors, or to {\it x}-walk along
the detector in regions where the gain-sag is high.  
Without knowing which of these effects dominate, 
we decided to match the 21cm emission velocity of +9~\kms\ to the {\it
  average} velocity of all seven MW lines; as a consequence, between
$1190-1334$~\AA , the error in measuring the wavelength of any narrow
feature ought to be $\sim \pm6$~\kms\ or about 1 rebinned
pixel. This wavelength range covers all the lines we expected to
see from NGC~1097, except the Si~IV~$\lambda\lambda 1393,
1402$ doublet.

\section{Results}
 \label{sect_results}

\subsection{Absorption Line Measurements \label{sect_measure}}

In the final co-added COS spectra we searched for absorption from
NGC~1097 from a variety of species, which are listed in
Tables~\ref{tab_fits_v20}$-$\ref{tab_fits_v70}.
In this section, we describe the analysis of detected absorption
lines, while in \S\ref{sect_ews} we discuss how limits to column
densities were
derived when no absorption was detected.
Much of this analysis 
follows the methods described by \citet{bowen08}.
In particular, the way in which the data were normalized, the procedures
used to measure velocities $v$, column densities $N$, and Doppler
parameters $b$, of the absorbing gas by generating theoretical Voigt line
profiles and fitting these to the data, and the use of Monte Carlo (MC)
simulations to account for errors in the fits arising from Poisson
statistics, were identical to those described by \citet{bowen08}, and are
not repeated here.

The results of the line profile fits are given in
Tables~\ref{tab_fits_v20}$-$\ref{tab_fits_v70}.  For all the fits
performed, $b$ and $N$ were allowed to vary as free parameters. In all
cases, however, $v$ was fixed to the values shown in column 2. The
alignment of components between metal lines and \lya\ is discussed in
detail below for each sightline.  
Values of
$\sigma (b)_T$ and $\sigma (N)_T$ are given in these tables (columns 7
\& 9): these represent the errors in $b$ and $N$ resulting from a
combination of continuum fitting errors and theoretical line profile
fitting errors \citep[again, see][]{bowen08}. The errors in velocity
(column 4) are simply those found from
the MC simulations, and do not include possible systematic errors
discussed in \S\ref{sect_wavezero}. 

For multicomponent fits, the total (summed) column density over all
components is given at the bottom of column~8 for each species. The
error on this total, given at the bottom of column 9 for each species,
was calculated from the distribution of summed column densities
derived for each MC run (and not from any combination of the errors
from individual components, because these are not independent for
multicomponent fits).

Tables~\ref{tab_fits_v20}$-$\ref{tab_fits_v70} also include the
equivalent widths (EWs) $W$ of detected lines (or upper limits when
no lines were detected---see below). Several of the absorption
lines from NGC~1097 were blended with other higher redshift lines,
in which case their EWs were derived directly from their theoretical line profiles
and not measured from the data. The EWs of the individual components
are given in column 5 of the tables, but when more than 1 component
exists, the {\it total} EW in column 5 (referred to in column 1 as
``total [p]'') is given for the EW of the blended components,
which need not equal the sum of the individual component
EWs.

Unfortunately, the resulting errors in $W$ for EWs
that rely on the profile fits cannot be
made using the errors in $N$ and $b$ because they are correlated in
such a way that $N$ and $b$ can vary while $W$ stays the
same. Instead, total EWs were recorded for each of the 300 MC
profile fits, and used to form a distribution of EWs.  
The distributions of $W$ were
always normally distributed,
yielding an error $\sigma(W)$.
Again, $W$ was measured for 
three possible continuum fits, yielding errors
that were added in quadrature with $\sigma (W)$ to give a total
error. These are the errors that accompany the EW totals in column 5.

Tables~\ref{tab_fits_v20}$-$\ref{tab_fits_v70} also list total EWs
measured in the traditional way, directly from the spectra, calculated over a specific number
of pixels (referred to as ``total [$N=$]'' in column 1).  The values
of $N$ here are chosen simply to be large enough to cover all the
detected components.  These EWs are 
independent of the assumed LSF or any
other assumptions about how the absorption line profile should be
modelled. The errors listed include errors from Poisson noise and
deviations from the continuum fitting, analogous to the column
density errors described above.

\subsection{Equivalent Width Limits and Column Density Limits for
  Undetected Metal Lines \label{sect_ews}}

In many cases, no metal lines were detected towards the QSO sightlines,
and column density limits had to be derived from EW limits. Traditionally,
the EW limit at some wavelength is taken to be $\sigma^2(W) =
\sum^{N}_i\: (\sigma_i\:\delta \lambda )^2$, where $\delta\lambda$ is
the wavelength dispersion in \AA~pixel$^{-1}$ and $N$ is the number of
pixels that the sum is made over, centered at the wavelength of
interest. Or more simply, if $\sigma_i$ is close to being the same
value $\sigma$ over $N$ pixels (because the errors change slowly with
wavelength) then $\sigma(W) = \sigma \delta\lambda \sqrt{N}$.
The COS LSF 
consists of a Gaussian
core superimposed on much broader wings, as a result of mid-frequency
wavelength errors in the HST mirrors \citep{kriss11,holland12}. If the LSFs generated
for the profile fitting described above are integrated, then 95.5\% of the area of a line
is enclosed in 14 of our re-binned pixels.
We adopted $N= 14$ as the value over which to
measure EW limits, knowing that the most a line EW might be under-estimated
would be by $\simeq 5$~\%.

For undetected lines in the QSO spectra, we list upper limits of
$2\sigma (W)$ in Tables~\ref{tab_fits_v20}$-$\ref{tab_fits_v70}. To
derive column density limits
we used $b$ values from other lines where
possible, and the values adopted are listed in parentheses in column
6 of Tables~\ref{tab_fits_v20}$-$\ref{tab_fits_v70}.
 
\medskip
In the following sections, we look in detail at the absorption lines
detected along each sightline, and discuss the measurements made from the
line profiles.

\subsection{Absorption Lines Towards Q0244$-$303 \label{sect_v20fits}}

\subsubsection{Metal Lines \label{sect_v20fits_metals}}

\begin{figure}
\vspace*{0cm}\hspace*{0cm}\includegraphics[width=8.5cm]{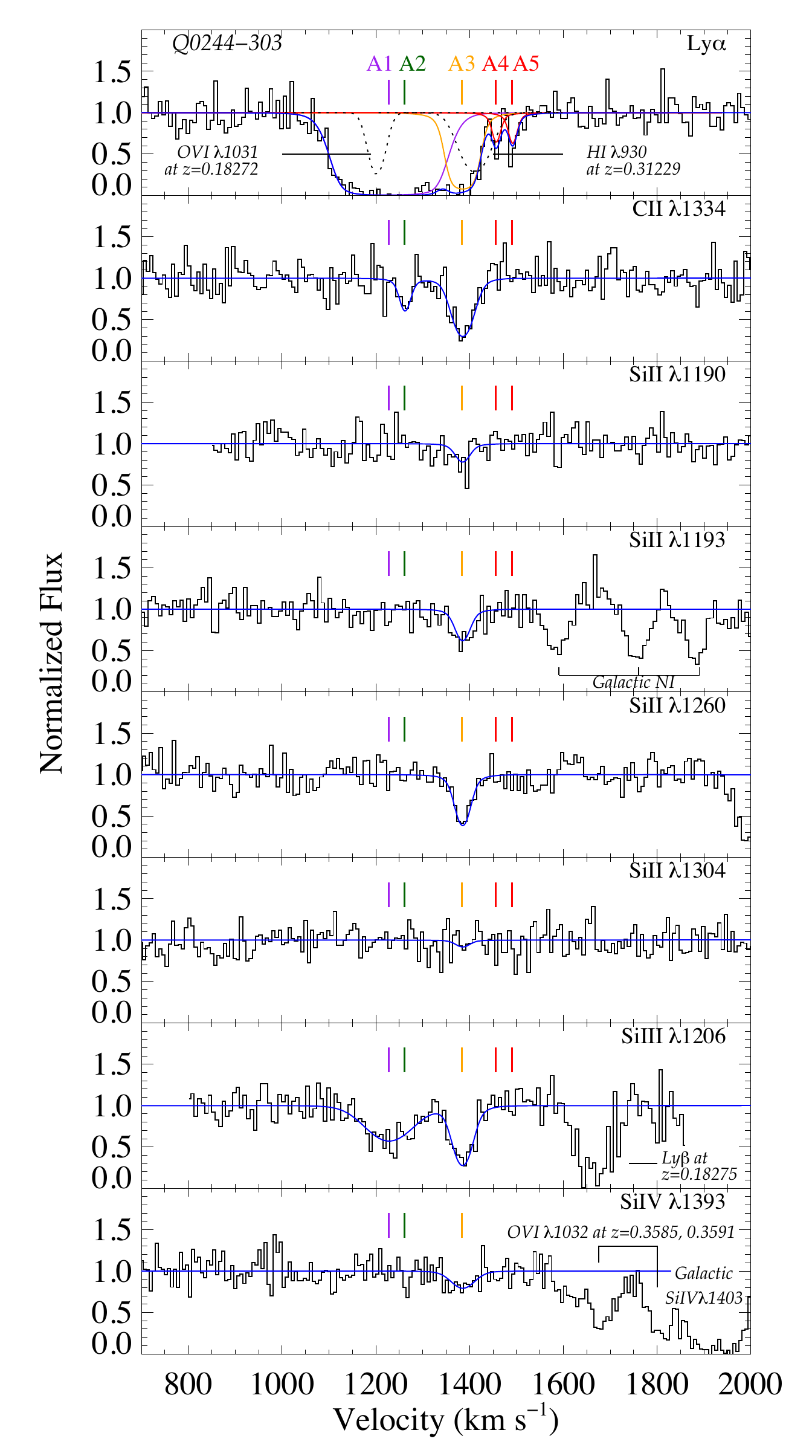}
\caption{Absorption lines found towards \vtwo. 
The \lya\ line is contaminated by two other lines, 
O~VI~$\lambda 1031$ at $z=0.18272$ and H~I~$\lambda 930$ 
at $z=0.31229$; the expected profiles of these two lines are shown as dotted lines, although the 
data shown are already corrected for them (see text).
As discussed in the text, only two components (A1 and A3) are fitted
to the body of the \lya\ line, but the position of a possible
component seen in C~II~$\lambda 1334$ is labelled (A2). 
\label{fig_stack_v20} }
\end{figure}

The absorption lines seen towards \vtwo\ are shown 
in Figure~\ref{fig_stack_v20}.  Metal lines are detected towards
\vtwo\ at at least two velocities; in the first case, C~II~$\lambda
1334$, Si~II~$\lambda 1193$, Si~II~$\lambda 1260$ Si~III~$\lambda
1206$ and Si~IV~$\lambda 1393$ are found at a velocity of 1384~\kms
. This component is labelled A3 in Table~\ref{tab_fits_v20}, and
categorizes the strongest metal-line absorbers along the
sightline. There is also a marginal detection of Si~II~$\lambda 1190$
at this velocity, although its reality is based largely on the
presence of the other Si~II lines since its strength is below our
detection threshold. As expected therefore, the weakest line of the
four Si~II lines, Si~II~$\lambda 1304$, is absent. 
Si~IV~$\lambda 1403$ at 1384~\kms\ is contaminated by an
O~VI~$\lambda 1032$ line at $z=0.35818$, from a system which also
shows corresponding Ly$\beta$, Ly$\gamma$, Ly$\delta$ and
O~VI~$\lambda 1037$. Hence the $\lambda 1403$ line is not used in the
derivation of the Si~IV column densities.

A second component is detected from Si~III~$\lambda 1206$ at
1227~\kms ; although there are no other metal lines to confirm the
reality of this component, there is clear evidence of \lya\ at a
similar velocity (see below), and so we consider this component to be
real. This component is labelled A1.

Finally, a weak feature is detected near the C~II~$\lambda 1334$ A3
component, which, if a lower velocity C~II absorption feature, would be at
$v=1261$~\kms.  This velocity is clearly different
from that of A1 and A3. We label this possible system A2 in
Table~\ref{tab_fits_v20} and discuss its reality below.

In order to establish the velocity of components A1 \& A3, all the
metal lines were fit simultaneously. Fits were first made without
including the \lya\ line. (The \lya\ components are clearly
blended, so there is little information in the line profile to help
establish the components' velocities). Values of $b$ and $N$ were allowed to
vary independently for each component of each species, and while $v$
was also allowed to vary, it was constrained to be the same for each component
of each ion.

Although A2 is defined only by the possible existence of a weak C~II
line, Table~\ref{tab_fits_v20} lists EW limits and column density
limits for the other species at the same velocity. For Si~III, a limit
is hard to define, since any absorption would be blended with the A1
component. The Si~III A1 component is broad, and the low S/N of the
data makes the profile appear coarse, but there is no obvious need to
add an additional component to represent possible absorption at the A2
velocity. 
We estimate a plausible column density for Si~III based
on adding a fictitious A2 (unresolved) component and seeing how the blended profile
changes with $N$(Si~III).

A similar problem exists for defining a C~II column density limit for
A1, since any line at that velocity would be blended with the defining
A2 C~II component. We again estimate an upper limit to C~II at A1 by
blending a hypothetical component (this time with the $b$ value of the
broad Si~III component) with the existing C~II component.

\subsubsection{\lya}

The \lya\ absorption lines detected from NCC~1097 are complicated by
contamination by lines from two higher-redshift systems. One
contaminant is Ly$\lambda 930$ arising from a system at $z=0.31229$;
the H~I from this system is very well constrained however, since Ly$\beta$,
Ly$\gamma$, Ly$\delta$, Ly$\epsilon$, Ly$\lambda 923$ and Ly$\lambda
920$ lines are all present. Absorption 
can be fit with a
single component to give $\log N$(H~I)$\:=\:15.88$ and $b\:=\:38.9$~\kms
. With these values we can calculate a profile for Ly$\lambda 930$
(shown as a dashed line in Fig.~\ref{fig_stack_v20}) and remove it
from  the data.
There also likely exists a second contaminent to the \lya\ complex, 
an O~VI~$\lambda 1031$ line at $z\:=\:0.18272$. This system is
defined primarily by \lya , Ly$\beta$ and Ly$\gamma$ lines, and so
appears secure. There is a feature at 1227.2~\AA\ which is probably the
O~VI~$\lambda 1037$ line from this system. The O~VI absorption is weak
and simple, and a fit to the O~VI~$\lambda 1037$ line gives $\log
N$(O~VI)$\:=\:14.19$, $b\:=\:20.6$~\kms . We again model the corresponding
O~VI~$\lambda 1031$ line using these values and remove it from the
\lya\ lines from NGC~1097.  Both these interloping Ly$\lambda 930$ and
O~VI~$\lambda 1031$ lines are relatively weak compared to the strong
\lya\ lines from NGC~1097 and contribute little to the total EW.

The \lya\ profile clearly argues for at least 2 components, 
but determining the column density for A3
remains problematic; without including any more information from the metal
lines (apart from their velocity) values of
$\log N$(H~I)$=15.07$ and $b=18.9$ are obtained (Table~\ref{tab_fits_v20}), but 
the solution is not unique, and the same profile can be obtained for higher
$N$(H~I) and smaller $b$ values. An upper limit for these can be set by the
appearance of damping wings, which is not seen in  the data, and
which occurs when $\log
N$(H~I)$\:\ga 17.1$ and $b\la 11$~\kms .  

While a 2 component fit can adequately represent the observed \lya\
absorption, we noted in the previous section the possible presence of
an additional C~II component --- A2 --- which could have a
corresponding \lya\ component. There is no information in the \lya\
profile to constrain the physical parameters of a component matching
A2, but we can explore how badly the \lya\ $b$-values and column
densities of A1 and A3 might be affected by including A2. An upper
limit to $N$(H~I) for A2 can be set at $\log
N$(H~I)$\:\simeq 18.1$, when the damping wings of A2 are wider than
the entire profile.  The maximum $b$-value of A2 is given by the
$b$-value of the C~II component, which has $b=10.9$~\kms. If $b$(C~II)
was defined purely by turbulent processes, then $b$(\lya) would be the same as $b$(C~II), or
10.9~\kms. If $b$(C~II) was entirely a thermal width, then the corresponding
temperature would be $\sim 8.6\times 10^4$~K. The width of a \lya\
line at that temperature would be $b$(\lya )$\:=38$~\kms . We can
therefore use $b=10.9$ and $b=38$~\kms\ as a lower and upper limit to
$b$ for A2. We can then fix the velocity of all 3 components to those
of the metal lines, vary $N$(H~I) for A2 in fixed intervals, and then
refit, to see how $N$(H~I) and $b$ change for A1 and A3.

In fact, for either value of $b$ for A2, the values of $b$ and
$N$(H~I) change little for A1 and A3 until the $N$(H~I) of A2 reaches
its upper limit of $\log N$(H~I)$\simeq 18.1$. 
For $N$(H~I) less than this, most of the optical depth
at the velocity of A2 is always taken up by A1, so increasing
$N$(H~I) for A2 makes little difference to the optical depth at the
velocity of A2.
This suggests that $N$(H~I) for A1 and A3 may not be too
badly affected by the presence of even a moderately strong \lya\
component at the velocity of A2.

Finally, we note the presence of additional weak, higher velocity
components, which we label A4 \& A5 in Table~\ref{tab_fits_v20} and
Figure~\ref{fig_stack_v20}. The figure shows the \lya\ profile after the
contaminating O~VI~$\lambda 1032$ and H~I~$\lambda$~930 lines have been
divided out of the data.  We find values of $N$ and $b$ from the profile
fitting, but the lines are unresolved.  The limit of the COS resolution is
$\sim 15$~\kms\ FHWM (ignoring the non-Gaussian shape of the LSF) or
$b \la 9$~\kms . The EWs of both \lya\ lines lie on the linear part of the
Curve of Growth so long as $b\ga 7$~\kms , (i.e. $T \simeq 3,000$~K) with
$\log N$(H~I)$\: \simeq 13.1$. If $b$ is really less than 7~\kms , however,
then $N$(H~I) is a lower limit. As $b$ is unconstrained, we give $N$(H~I)
as a lower limit in Table~\ref{tab_fits_v20}.

It is useful to summarise the above discussion in terms of the {\it total}
$N$(H~I) along the sightline, a value we will use later in this paper. The
lower limit to $N$(H~I) is set by A1, and is $\log N$(H~I)$\:\simeq 15.6$
(Table~\ref{tab_fits_v20}).  The upper limit is set by a lack of a damping
wing on the red-side of A3, $\log N$(H~I)$\: \la 17.1$; however, the
possible existence of H~I corresponding to the C~II component A2 increases
this limit to $\log N$(H~I)$\: \la 18.1$, at which point damping wings
would be seen on both sides of the complex.

\subsection{Absorption Lines Towards   2dFGRS~S393Z082 \label{sect_v40fits}}

\subsubsection{Metal Lines}

\begin{figure}
\includegraphics[width=8.5cm]{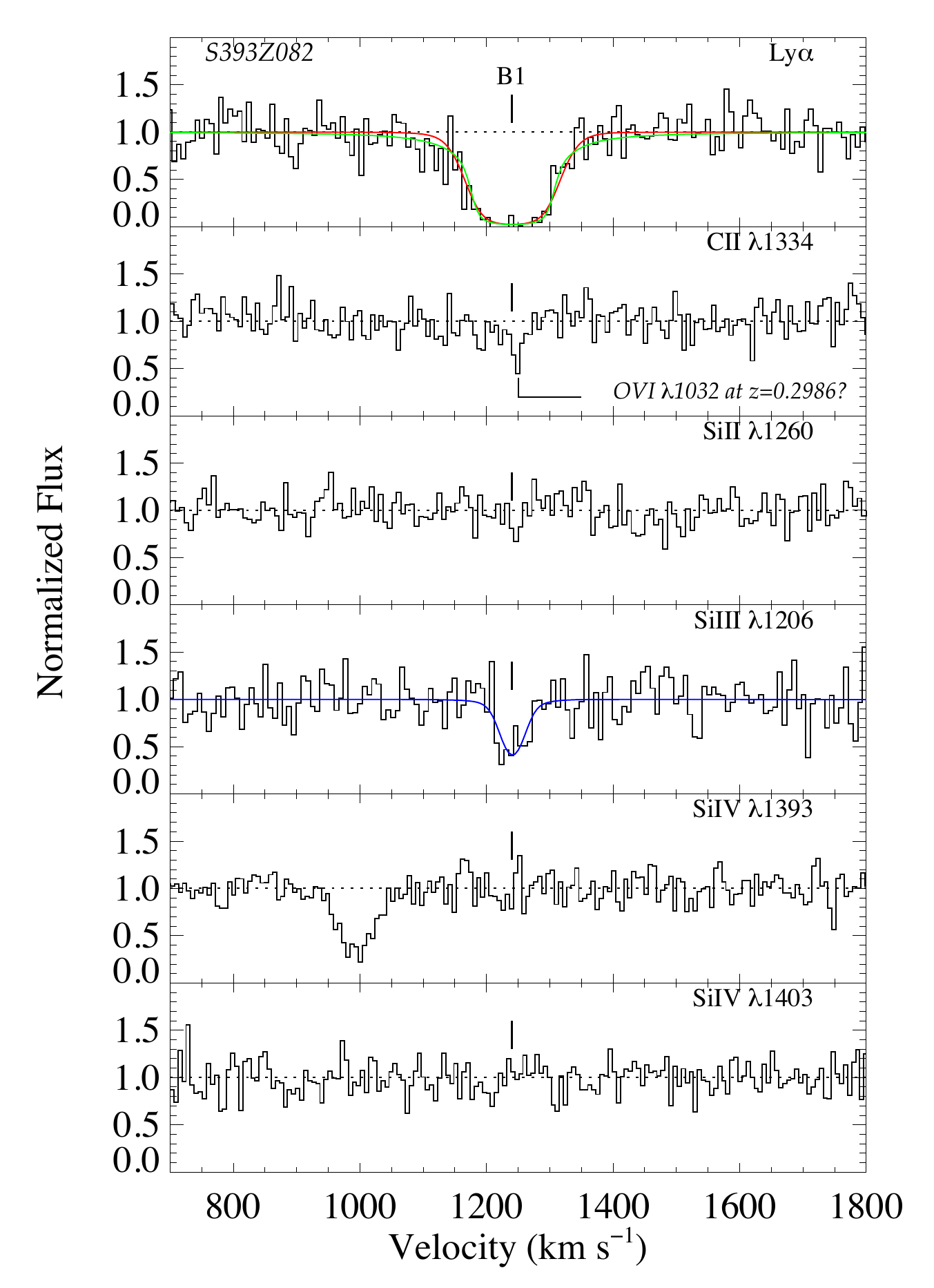}
\caption{Absorption lines found towards \vfour. The values of
  $N$(H~I) and $b$ are ill-constrained for \lya\ absorption from NGC~1097: two
  theoretical line profiles are shown in the top panel,  each
  representing nearly identical best-fits to the data:
  one has a
  high-$N$(H~I)/low-$b$ solution (green), the other has 
  low-$N$(H~I)/high-$b$ values (red). The two profiles are nearly
  indistinguishable from each other. There exists a possible
  detection of C~II~$\lambda 1334$ (second panel), 
  but the line is likely to be O~VI~$\lambda 1032$ at
  z=0.2986. Si~III~$\lambda 1206$ is the only metal line unambiguously
  detected from NGC1097 (fourth panel). \\
\label{fig_stack_v40} }
\end{figure}

\begin{figure}
\hspace*{0.75cm}\includegraphics[width=7.5cm]{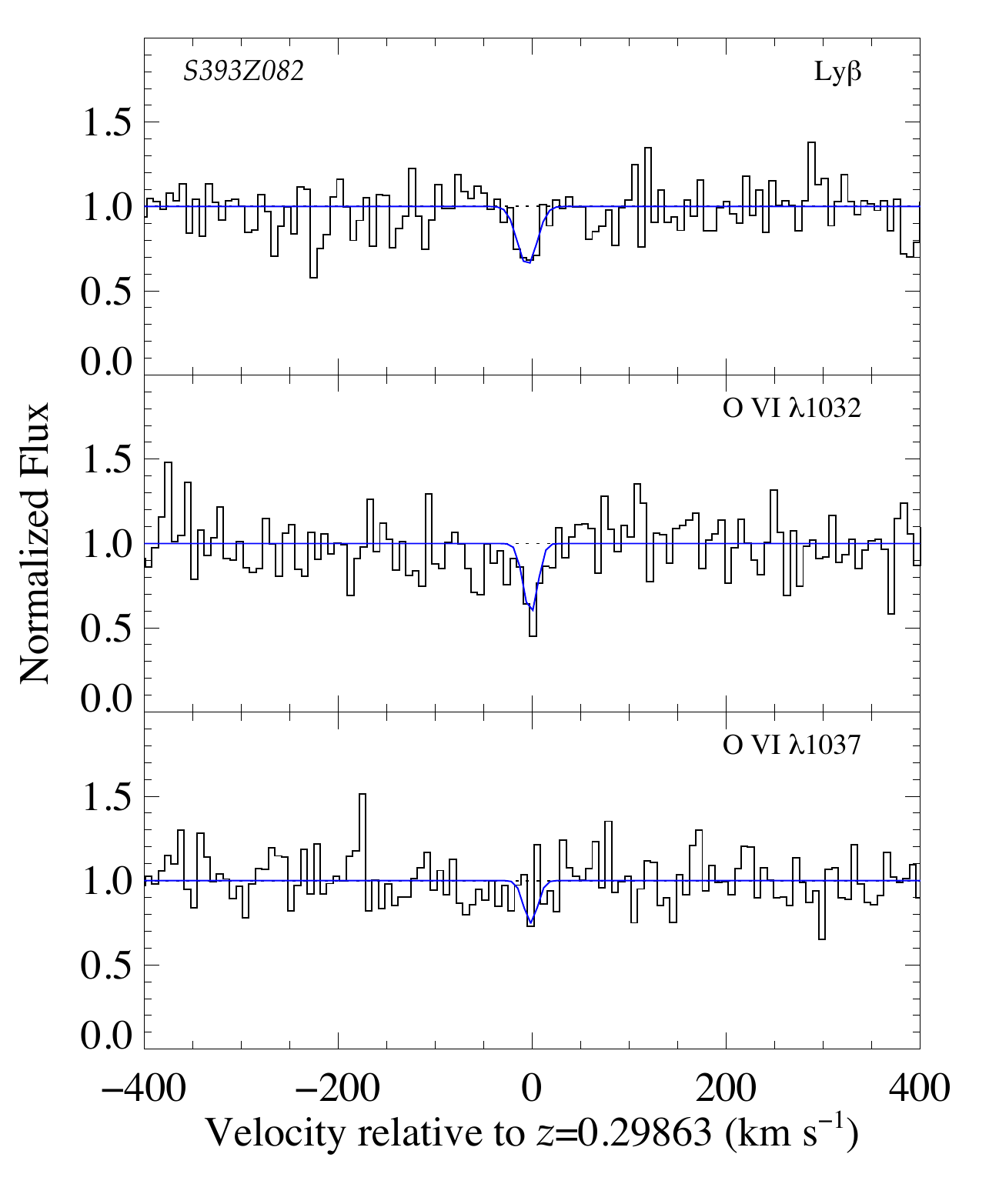}
\caption{While the line at 1340.1~\AA\ towards S393Z082 could be
  C~II~$\lambda 1334$ absorption from NGC~1097 (see
  second panel of Fig.~\ref{fig_stack_v40}) another
 possible interpretation of this line, along with another
 at  1332.0~\AA , is that they represent O~VI~$\lambda
 1032$ and Ly$\beta$ absorption at $z=0.29863$ instead. This figure shows these lines (as
 well as the expected absorption from O~VI~$\lambda 1037$) in
 that rest-frame, as well as theoretical profile fits for
 $b=13.8$~\kms, $\log N$(H~I)$\:=\:13.7$ and $\log N$(O~VI)$\:=\:13.5$.\\
\label{fig_badc2}}
\end{figure}

As shown in Figure~\ref{fig_stack_v40}, Si~III~$\lambda 1206$ is detected
unambiguously at 1240~\kms\ towards \vfour , a system we label B1.  A weak
feature is also detected at the wavelength where C~II~$\lambda 1334$ is
expected at the same velocity as B1, but the line could be
O~VI~$\lambda 1032$ at $z=0.29863$, because there is also a very weak
feature which could be Ly$\beta$ at that redshift.  The corresponding
O~VI~$\lambda 1037$ line is not detected at this redshift, but given the
weakness of the $\lambda 1032$ line, this is not unexpected for unsaturated
lines.  Fitting Voigt profiles to the \lyb, O~VI~$\lambda 1032$ and (to
provide an upper limit) O~VI~$\lambda 1037$, a solution of $b=13.8$~\kms,
$\log N$(H~I)$\:=\:13.7$ and $\log N$(O~VI)$\:=\:13.5$ can be found
(Fig.~\ref{fig_badc2}).  Although an O~VI/H~I absorber with such a narrow
line width is relatively rare, \citet{tripp08} have found such
systems. Further, the ratio of the column densities in our fit,
$\log$[$N$(H~I)/$N$(O~VI)]$=-0.17$ is consistent with the values found by
Tripp~\etal\ for the value of $\log N$(H~I)$=13.7$ that we find for the
system.  Given the precise alignment of the two lines, and the plausibility
of physical parameters obtained from the profile fits when compared to
other O~VI systems, we cannot be sure that the line at 1340.1~\AA\ is C~II
from NGC~1097 and refrain from using it in our analysis.  We note that if
the identification is wrong, and the line is indeed C~II~$\lambda 1334$,
then the profile fit to the line requires $\log N$(C~II)$\:\simeq\: 14.0$.

\subsubsection{\lya}

This is the simplest of the \lya\ absorption lines associated with
NGC~1097, but the physical parameters are not well constrained.
The data have the lowest S/N of the 4
sightlines, and there is some ambiguity over the number of components
to fit. A single component, which we label B1 in
Figure~\ref{fig_stack_v40}, appears to be adequate; there is a
slightly lower `shelf' at 1100~\kms , and adding a component to match
this feature changes $N$(H~I) for B1 by a factor of 4. The MC simulations used
to calculate the errors in $N$(H~I) and $b$-values, however, suggest that the
small amount of extra depth may well just be from noise, since in many
cases, the feature is not replicated in the synthetic spectra. We therefore
assume that no additional absorption occurs at this lower velocity.

The \lya\ line provides an example of the classic problem of measuring
$N$(H~I) and $b$ when its EW lies on the flat part of the Curve of Growth.
The line-profile fitting settles on two possible solutions, one with
$b\: =\: 46$~\kms\ and $\log N$(H~I)$\: =\: 14.8$, and a second one with
$b\:=\:19$~\kms\ and $\log N$(H~I)$\: =\: 17.6$.  The problem is
exacerbated by the low S/N of the data providing insufficient constraints
on the shape of the wings of the line. An analysis in which the width of
the line is determined by a \btherm\ and a \bturb\ component (see Appendix
\ref{sect_photoU}) shows that the change in $\log N$(H~I) from 14.8 to 17.6
occurs when $T>1\times 10^4$~K. Figure~\ref{fig_stack_v40} better shows the
problem, where theoretical lines profiles for both the high-$N$ fit (green)
and the low-$N$ fit (red) are drawn; the profiles for the two solutions are
almost identical.

\subsection{Absorption Lines Towards  Q0246$-$308 \label{sect_v60fits}}

\subsubsection{Metal Lines}

\begin{figure}
\includegraphics[width=8.5cm]{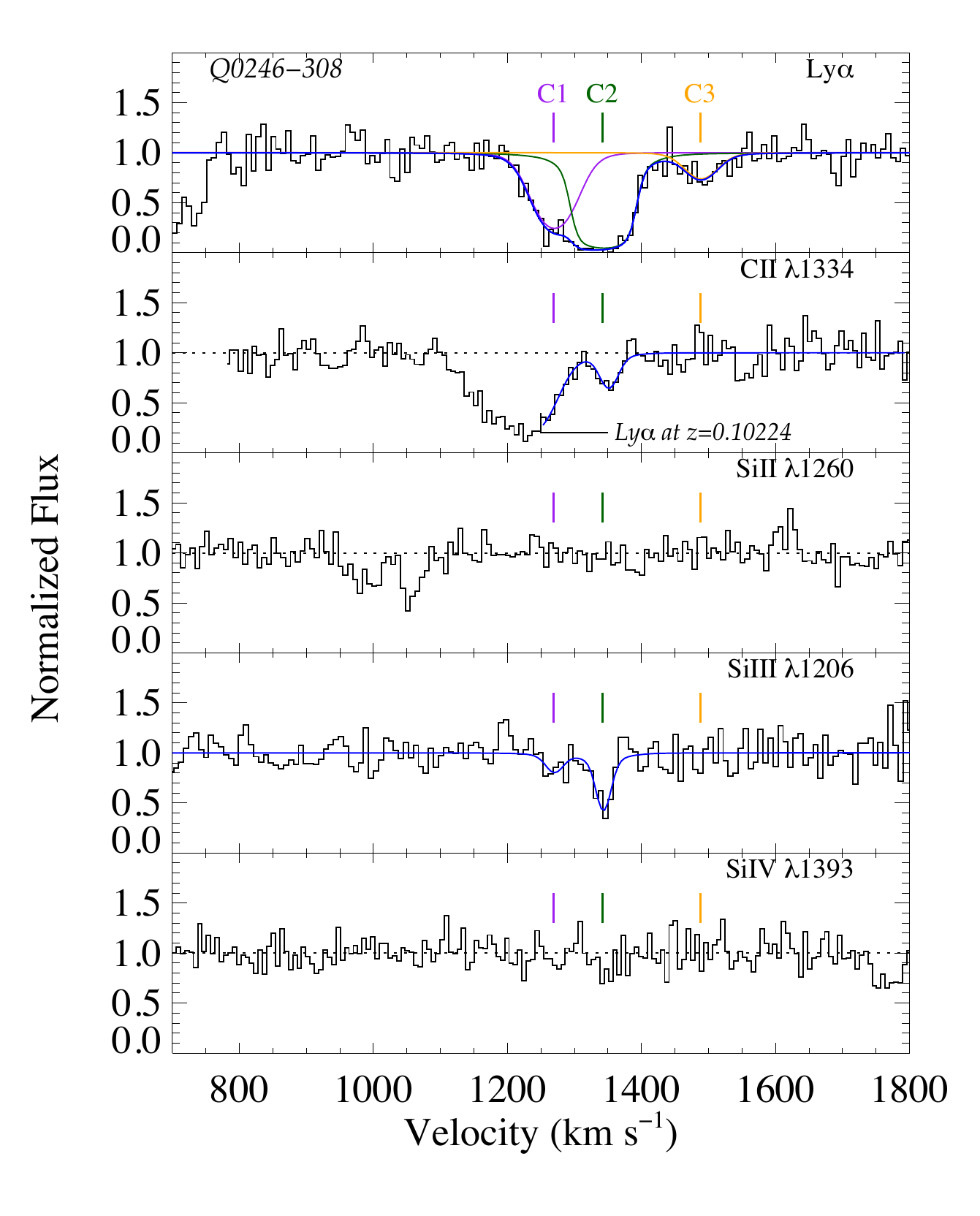}
\caption{Metal absorption lines found towards \vsix. The defining
  component of the system is labelled C2, seen in both Si~III, \lya,
  and possibly, C~II; weaker Si~III and \lya\ components exist at a
  more negative velocity,
labelled C1. No Si~II or Si~IV is detected from either C1 or C2. C3 is
identified only by \lya\ absorption. The
absorption at the expected velocities of the C~II components is complex, 
and the strongest feature is probably a \lya\ line at $z=0.10224$. Although we
identify C~II from system C2, it remains possible that this feature is part of
that higher-$z$ complex at 
$z=0.10272$.
 \label{fig_stack_v60}\\ }
\end{figure}

Along this sightline, Si~III appears to be detected at the redshift of
NGC~1097, and being shortward of \lya , the reliability of its
identification is strong. This component is labelled C2 in
Table~\ref{tab_fits_v60} and Figure~\ref{fig_stack_v60}.  There is an
additional weak Si~III component to the blue of C2, which is designated as
C1.

Any C~II~$\lambda 1334$ from system C1 is lost in a line that, given its
strength and width, is probably \lya\ at $z=0.10224$
(Fig.~\ref{fig_stack_v60}, second panel) although there are no other
corroborating lines.  C~II from system C2 is expected at 1340.5~\AA , and a
weak feature is seen very close to that wavelength, although it is nestled
next to the high-$z$ \lya\ line.  $N$(C~II) for the line is given in
Table~\ref{tab_fits_v60}, and while we consider the detection to be real,
it remains possible that the feature is simply an additional higher
velocity ($z=0.10272$) component that is part of the strong high-$z$ \lya\
line.

A feature is also seen at 1244.32~\AA\ which could be N~V~$\lambda 1238$
from NGC~1097, but there is no N~V~$\lambda 1242$ line to confirm
this. Fitting the N~V~$\lambda 1238$ line gives $\log N$(N~V)$\: =\: 13.50$
and $b=20$~\kms ; with these values, the N~V~$\lambda 1242$ line should be
detected, but is not. We therefore list only upper limits in
Table~\ref{tab_fits_v60}.

\subsubsection{\lya}

The \lya\ absorption from NGC~1097 towards \vsix\ is comprised of 3
components. Two of these well match the systems C1 \& C2 identified in
the metals; a third broader line is evident at a higher velocity,
which we label C3 in Figure~\ref{fig_stack_v60}, but which has no
corresponding metal-line absorption. The \lya\ component
in system C2 is another
example of a saturated \lya, and constraining $N$(H~I) is
challenging. For this spectrum, the S/N is higher than for \vfour,
which helps constrain the fits of theoretical Voigt profiles;
nevertheless, there is still a large uncertainty in $N$(H~I) which
is reflected in the errors listed in
Table~\ref{tab_fits_v60}.

\subsection{Absorption Lines Towards HE0241$-$3043 \label{sect_v70fits}}

\subsubsection{Metal Lines}

\begin{figure}
\includegraphics[width=8.5cm]{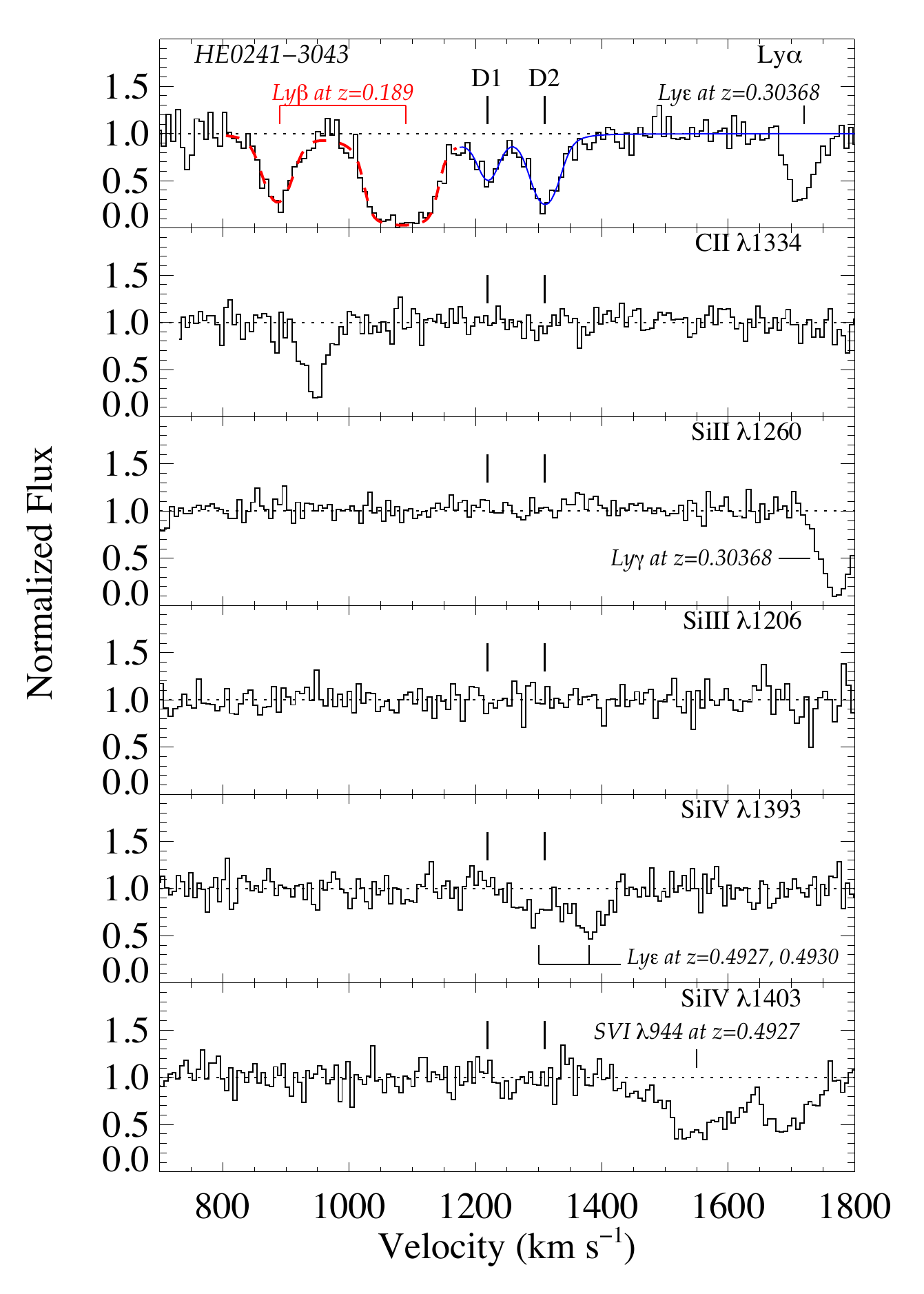}
\caption{Absorption lines found towards \vseven. Only \lya\ is
  detected from NGC~1097.
\label{fig_stack_v70} }
\end{figure}

No metal line absorption is detected from NGC~1097 along this sightline.
The lack of detections are shown in
Figure~\ref{fig_stack_v70}, and upper limits are listed in
Table~\ref{tab_fits_v70}. Absorption by Si~IV~$\lambda 1393$ from
NGC~1097 would be blended with a pair of Ly$\epsilon$ lines at
$z\simeq 0.493$, so column density limits are measured for
Si~IV~$\lambda 1403$ instead.

\subsubsection{\lya}

Though no metal-lines are detected, two \lya\ components are
seen from NGC~1097. These are labelled D1 \& D2 in
Figure~\ref{fig_stack_v70} and Table~\ref{tab_fits_v70}, and
lie close to Ly$\beta$ lines from a two component complex
at $z=0.18868$ \& $z=0.18945$.  D1 \lya\ from NGC~1097 is slightly
blended with the higher-$z$ of these two \lyb\ lines, as can be seen
in the figure. Unfortunately, any \lya\ components at velocities less
than that of D1 will be lost in the \lyb\ lines. Neither D1 or D2 
can, instead, be additional Ly$\beta$
components at $z\simeq 0.189$ because the corresponding \lya\
lines are not seen at longer wavelengths. We can find no other systems
for which D1 and D2 might be metal lines.

\section{Photo-ionization models of absorbing clouds }
\label{sect_photou_point}

In order to better understand the physical characteristics of the absorbing
clouds, we have attempted to reproduce the measured values of, or limits
to, the column densities for all ion species along the QSO sightlines using
photoionization models. The results depend on various assumptions, and as
the list is rather extensive, we present the details of our analysis in
Appendix \ref{sect_photoU}. Unsurprisingly, only components with detected
metal lines, those labelled above as A1, A3, B1, \& C2 give useful
constraints. In summary, the metallicities are probably sub-solar, $Z < 0.3\:
Z_\odot$ for A1, B1 \& C2. For A3, however, we can find no single-phase
solution, meaning that either the gas is not simply photoionized (or indeed
collisionally ionized, which was also considered) or that the 
component-fitting model used to model $N$(H~I) is inadequate.

\section{Constraints on Halo Gas Kinematics}
\label{sect_kinematics}

In this section we use the kinematics of the absorption lines toward
NGC~1097 --- the profiles of the lines and velocity centroids --- to
model the gas in the halo of the galaxy. We focus specifically on the
detected \lya\ lines as the H~I absorbing gas is the most sensitive
tracer of cool gas available from our observations, and is detected along
all sight lines.

\begin{figure*}[t]
\hspace*{2cm}\includegraphics[width=15cm]{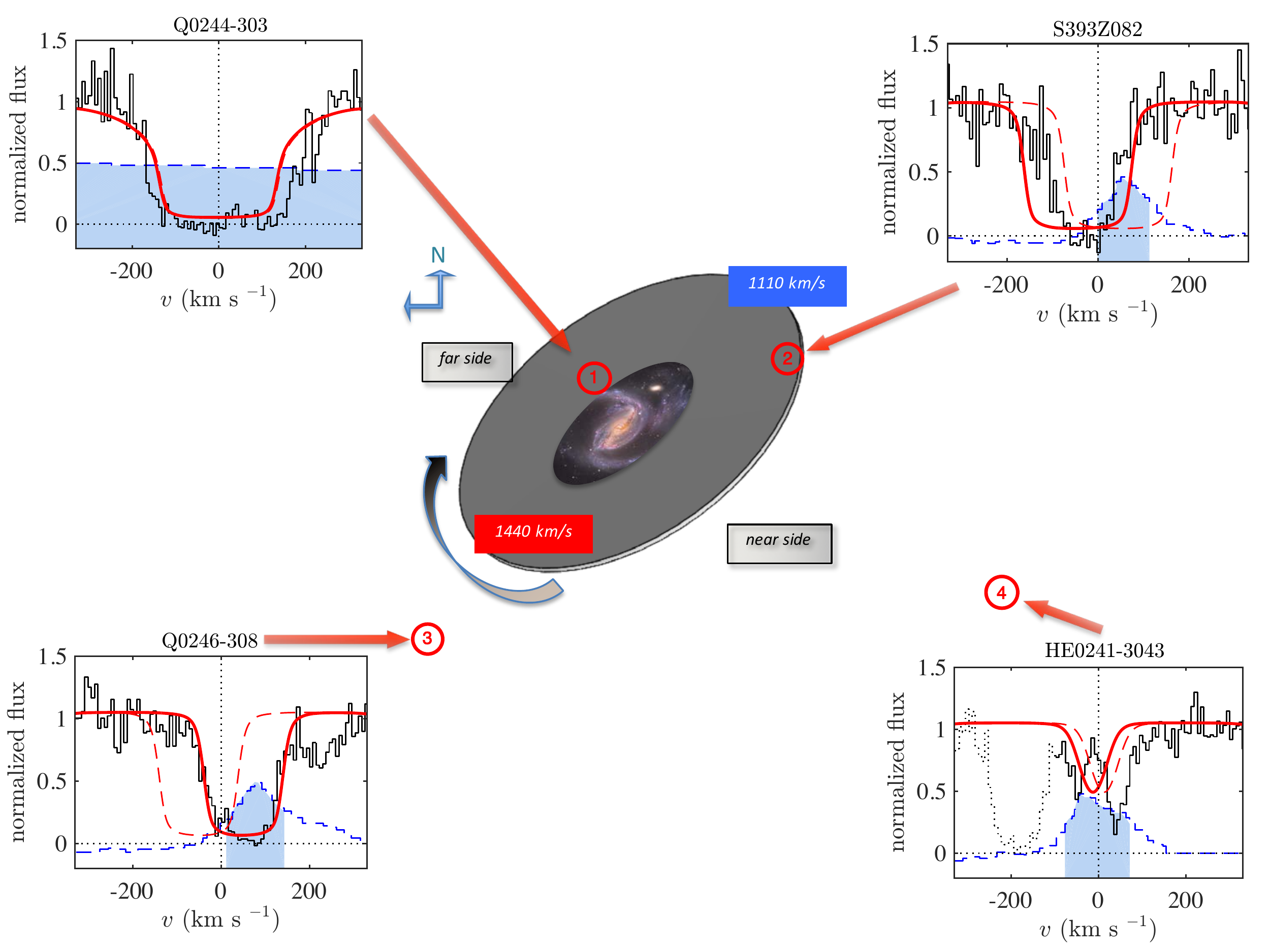}
\caption{Predictions of \lya\ line profiles (red lines) for QSO sightlines passing
  through a rotating geometrically thin disk model,
  compared to our COS data. The center figure is a simple schematic of the
  galaxy-QSO configurations, with a disk that has approximately the same inclination and
  PA as NGC~1097 shown in Fig.~\ref{fig_map}; the SW side
  of disk is
  assumed to be closer to us, with the spiral arms trailing the
  rotation. The positions of the QSOs are
  indicated by red circles;  they are arranged to roughly match their true
 configuration (Fig.~\ref{fig_map}),   
  and 
  indicate which structures of the model are intercepted along a given
  sightline. (The inset image of the image of the galaxy is not
  to scale, but is included to show the orientation of the spiral arms
  relative to the rotation.) The predicted line profiles are built from a
  model that follows the galaxy parameters in
  Table~\ref{tab_n1097}, as well as the positions of the QSOs on the sky, precisely.
The disk has rotation speed of
  $v_{\rm{rot}} = + 70$~\kms\ and a turbulent velocity of 20~\kms
  . The model is qualitatively consistent with the data towards \vtwo\
  and \vsix, although only one broad component is predicted for \vtwo\ when
  there are clearly two in the data; the model also over-predicts the strength of the absorption towards
  \vfour, and cannot reproduce the double line profile towards
  \vseven. Also shown are $f_{\rm{min}}$ curves  (blue dashed-line
  histograms) 
  that are used to estimate the allowed rotation speed
  interval for each sightline (as shown by the blue shaded regions). 
  For comparison, profiles are also plotted (red dashed lines) for an absorbing
  disk {\it counter}-rotating with respect to NGC~1097. Such
  counter-rotation would produce absorption at velocities quite
  different from that observed towards \vfour\ and \vsix .\\
\label{fig_disk}}
\end{figure*}

\subsection{Disk Models \label{sect_diskmodels}}

Simulations of galaxy structures in a $\Lambda$CDM universe predict that
the halos of galaxies should accrete gas from the IGM along cool gas
streams that possess their own angular momentum
\citep{danovich15,cen14,stewart13}.  
This leads us to question whether the
absorption seen towards NGC~1097 might trace similar structures.  We
consider two variations of gas rotation in  NGC~1097: {\it a}) a rotating
disk model, and {\it b}) a rotating-inflow disk model, where gas is flowing
into the disk (akin to an accretion disk). For reasons that we explain, we
will also consider a third option, that of the same rotating-inflowing disk model,
but with a conic outflow launched from the center of the galaxy,
perpendicular to the disk.  For all these models, we use the galaxy
parameters listed in Table~\ref{tab_n1097}.

\subsubsection{Rotating Disk Models}

In this model, we assume an inclined thin disk rotating at a fixed
azimuthal speed $v_{\rm{rot}}$.  A positive value of $v_{\rm{rot}}$
indicates clockwise rotation of the disk shown in Figure~\ref{fig_map} ---
the analysis by \citet{quillen95} suggests that the south-west (SW) side of
the galaxy lies nearer to us than the north-east (NE) side, and therefore
that the galaxy rotates with its spiral arms trailing the rotation.  The
column density of the disk is assumed to follow an exponential decline with
the de-projected \citep[e.g.][]{williams01} radius
$\tilde{\rho} \propto \exp (-\tilde{\rho}/\tilde{\rho}_d)$, where the disk
scale is $\tilde{\rho}_d = 11$~kpc.
To model the absorption we assume that
micro-turbulence in the gas produces lines with widths of
$v_{\rm{turb}} = 20$~\kms , which result in the observed line profiles
after convolution with the COS LSF.

Our aim is to use the parameters described to predict the line profiles
that would be expected along the four QSO sightlines. As there are many
parameters that can be varied to produce the predicted profiles, we refrain
from, e.g., attempting to minimize a $\chi^2$ statistic between the
theoretical profiles and the data --- there is no unique solution for the
number of parameters available. Instead, we look for more qualitative
agreements that might guide us toward finding a correct scenario for the origin
of the absorption.

So, for example, by adopting of value of $v_{\rm{rot}} \simeq +70$~\kms ,
we find that our predicted profiles show some agreement with the observed
data.  (This velocity is smaller than the circular speed of the inner disk
observed from 21~cm emission measurements, which is $\sim 200-300$~\kms , a
point we return to in the next section.)
Figure~\ref{fig_disk} shows the predicted profiles in comparison to the
observed profiles. While the impact
parameters of the two outer sightlines differ by only 15 kpc, the
\lya\ line profiles of each are markedly different, with the sightline
towards \vseven\ showing none of the saturated absorption seen towards
\vsix. As the former sightline lies closer to the minor axis of the
disk, a lower opacity would indeed be expected. Further, the absorption
velocities towards \vtwo\ and \vseven\ are centered around the
systemic velocity of NGC~1097, as would be expected for sightlines at
opposite ends of the galaxy's minor axis.

In contrast, the sightlines towards \vfour\ and \vsix\ lie close to the
opposite ends of the major axis and the absorption ought to be shifted away
from the galaxy's systemic velocity, and with the velocities in opposite
directions. The implied sense of rotation is consistent with that assumed
for the inner disk; for comparison, Figure~\ref{fig_disk} also shows the
results from a disk model where the gas counter-rotates with the galaxy
($v_{\rm{rot}}=-70$~\kms ): here the predicted absorption clearly fails to
match the data. As expected, the discrepancies are largest towards the
sightlines along the major axis where differences in $v_{\rm{rot}} $ would
be most noticeable.

To estimate the uncertainties associated with $v_{\rm{rot}}$, we
define a minimum fraction of the observed line profile which lies on
either side of the centroid velocity of the predicted absorption
$v_c$, i.e.
$f_{\rm{min}}(v_{\rm{rot}}) = {\rm{min}}[{\rm{EW}}(v>v_c(v_{\rm{rot}})), {\rm{EW}}(v<v_c(v_{\rm{rot}}))]/$EW.
Hence a model whose predicted velocity is centered around the observed
absorption would have $f_{\rm{min}} \simeq 0.5$. For predicted line
profiles whose $v_c$ significantly deviates from the observed centroid
velocity, $f_{\rm{min}} \ll 1$. To estimate the uncertainty of
$v_{\rm{rot}}$, we calculate $f_{\rm{min}}$ for a range of
$v_{\rm{rot}}$ values, and take the profile to be an acceptable replica of
the data if
$f_{\rm{min}} > 0.2$. This variation of $f_{\rm{min}}$ as a function
of $v_{\rm{rot}}$  is shown as a dashed line at the bottom of each
panel in Figure~\ref{fig_disk}. 
As noted above, some sightlines are more
sensitive of the disk dynamics than others, depending on where they
probe the disk.  For example, the sightline towards \vtwo\ is largely
insensitive to disk rotation (top left panel of Fig.~\ref{fig_disk}), 
because its sightline passes close to the
minor axis of NGC~1097, and hence $f_{\rm{min}}$ changes little for
any value of $v_{\rm{rot}}$ ($f_{\rm{min}} \simeq 0.5$ for $-500$~\kms $<
v_{\rm{rot}} < 500$~\kms). The other sightlines provide more stringent
constraints on $v_{\rm{rot}}$, as Figure~\ref{fig_disk} shows, and the similarity
in the velocities at which
 $f_{\rm{min}}$ peaks in the three other panels, $\sim 70$~\kms,
 suggests that this is a good estimate of $v_{\rm{rot}}$.

\begin{figure*}[t]
\hspace*{1cm}\includegraphics[width=16cm]{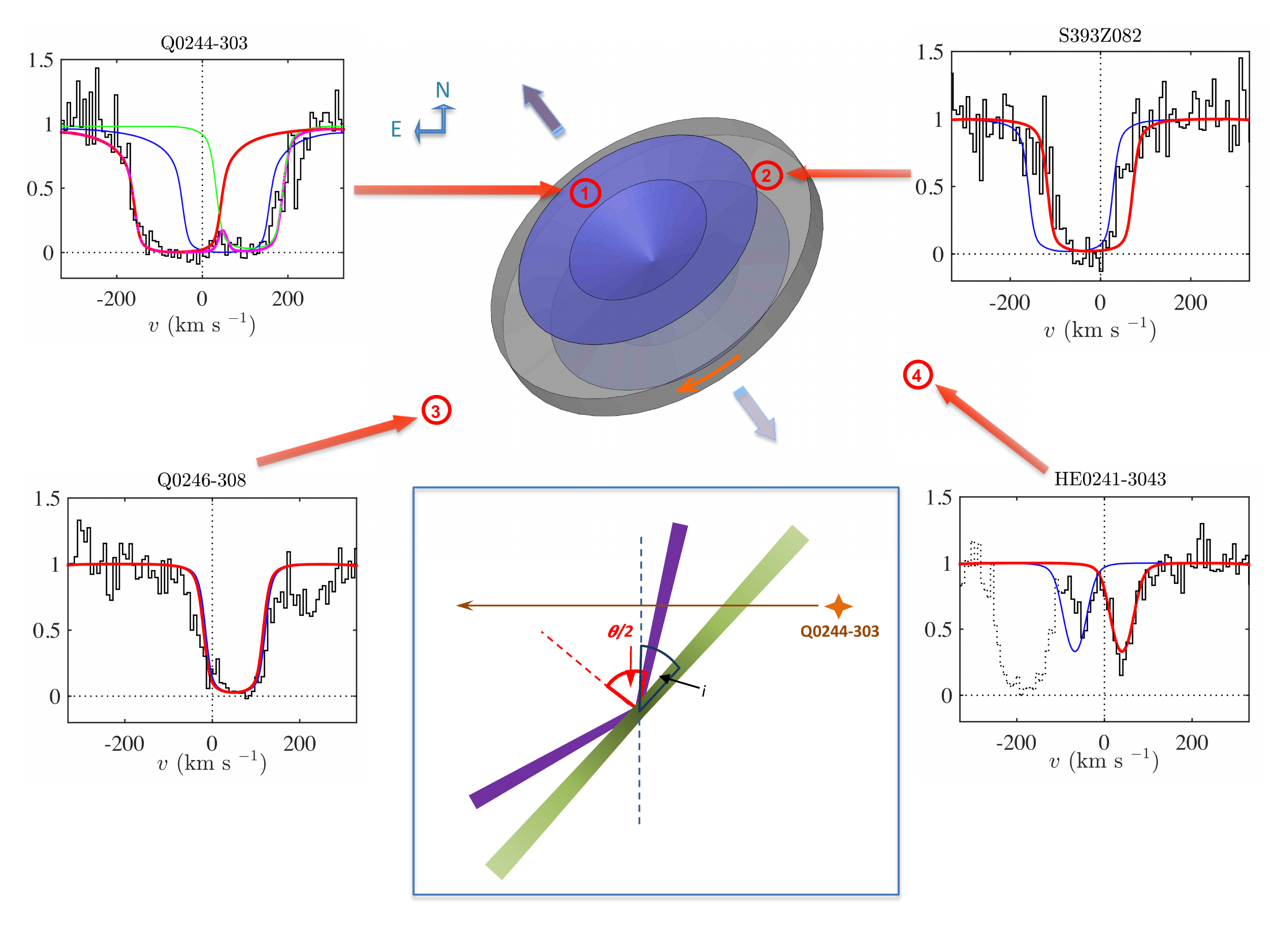}
\caption{Predicted \lya\ line profiles (red lines) for QSO
  sightlines passing through gas accreting onto NGC~1097 via a
  rotating geometrically thin disk model,
  as compared to our COS data. The disk
  has $v_{\rm{turb}} = 20$~\kms, a rotation speed of
  $v_{\rm{rot}} = 70$~\kms, and an inflow speed along the disk's
  (inward) radial direction of $v_r = 70$~\kms .   For
  comparison, the predicted line profiles for gas flowing away
  from the disk at a speed of $v_r = 70$~\kms\ are shown as blue lines.  
  Compared to the thin-disk model shown 
  in Figure~\ref{fig_disk}, this model better reproduces
  the absorption seen towards \vfour\ and \vsix , component A1 towards
  \vtwo , and D2 towards \vseven. However, it fails to
  account for components A3 towards \vtwo\ and D1 from \vseven .
One way
  to explain component A3 towards \vtwo\ is to add a conic outflow
  (purple cone in the middle schematic) with an outflow velocity of 300~\kms\ subtending an angular
  range of $65\degr < \theta < 70\degr$ and extending out to at least
  50~kpc above the disk. Such a model leads to the green line in the
  top-left panel, with the blend of the profiles from the two models shown in
  magenta. The inset diagram (middle, framed bottom panel) represents a cut
  through the model at the position of \vtwo , that shows further how
  positive velocity absorption (i.e. component A3) can be obtained
  from a wide conic outflow (purple regions) as it emerges from the (green)
  disk. So long as $\theta \geq 90-i$, the edge of the outflow will
  have a velocity greater than the systemic velocity. In this particular scenario,
  outflowing gas occurs on only one side of the galaxy.
\\
 \label{fig_flow}}
\end{figure*}

Although this model reproduces some of the absorption line features, 
there are notable discrepancies between the rotating disk model and
the data. For example, the model fails to predict the two-component
absorption seen towards \vseven. This might be the easiest discrepancy to
understand. One possibility is that a thin disk
model might no longer apply at such large radii. In fact, at a radius of
$\sim 200$~kpc, the dynamical time 
$\sim 2\pi \tilde\rho / v_{\rm{rot}}$ is comparable to the
Hubble time.  Alternatively, the two-component absorption may mean that
at large radii, H~I clouds no longer form a homogeneous thin disk.
Additionally, discrepancies associated with under- and over-prediction
of the opacity along particular sightlines can be resolved if either
$v_{\rm{turb}}$ or the column density profile deviate with radius
\citep[see, e.g., the variations in the rotation curves around a mean
value in][]{ondrechen89, hsieh11}. More worryingly, the model predicts only
one absorption component towards \vtwo, whereas at least two are
observed; and the absorption towards \vfour\ is somewhat narrower than the
model predicts.

\subsubsection{Rotating-Inflowing Disk Models with a Conic
  Outflow \label{sect_fulldisk}}

As noted above, the rotation speed of an extended disk with
$v_{\rm{rot}} \simeq 70$~\kms\ is significantly smaller than the
$200-300$~\kms\ rotation speed of the inner disk \citep[][see also
\S\ref{sect_best}]{ondrechen89}. This fact, combined with results from
$\Lambda CDM$ models that imply the existence of cold-flow accretion,
lead us to consider a model in which the absorbing gas has an inward radial
velocity $v_r$.   

The results of such a model are shown in
Figure~\ref{fig_flow};
data
towards sightlines that lie in the direction of the minor axis are
more sensitive to $v_r$, and imply $v_r < 100$~\kms .
We find that a model with $v_{\rm{rot}} \simeq v_r \simeq 70$~\kms\
improves the match between data and model towards some of the sightlines.
Towards \vtwo , the blue component, A1 is better reproduced by the model,
but at the expense of A3, which is not predicted at all. The predicted
absorption towards \vfour\ is improved, while the predicted absorption
towards \vsix\ is unchanged.
One of the components towards \vseven , D2, now matches the
predictions, but the other component, D1, is still not reproduced.

The values of $v_{\rm{rot}}$ and $v_r$ are not entirely
unique and similar (but not superior) fits can be obtained
for other $v_{\rm{rot}}$ and $v_r$ combinations. This model offers an
improvement over 
the simple rotating disk presented above, but still fails to match all the
observed features.

NGC~1097 shows moderate amounts of star formation at its
center \citep{calzetti10,hsieh11} so it is reasonable to consider the
possible effects of an outflow in our models. In several studies where
outflows have been traced by gas and dust emission (most notably
towards AGN) a hollow cone geometry has been inferred whose axis lies
along the disk's angular momentum vector \citep{veilleux01,veilleux05}
and where the wind may be detected in emission out to several kpc
above the disk, and potentially out to even larger scales in
absorption \citep{murray11}.

We consider the effect of adding a hollow-cone model
to our rotating-inflowing model by including gas that flows out of the galaxy
along radial trajectories at a speed $v_{\rm{rad}}$. This addition is also shown 
in Figure~\ref{fig_flow}. The most notable result
from adding an outflow component is that we can replicate the additional
A3 absorption component (Table~\ref{tab_fits_v20}) towards \vtwo,
using a model that has a cone with 
 $v_{\rm{rad}} \simeq 300$~\kms\ and 
that is so wide --- $65\degr < \theta < 70\degr$ --- that part of the cone wall
is still receding from us, providing an absorption component that
has a positive velocity (redshift) with respect to the rest of the 
galaxy. This scheme is highlighted in the inset diagram in
Figure~\ref{fig_flow}. 
Wide angle outflow cones have been seen in the inner regions of some galaxies, e.g.,
towards ESO~097-G013 \citep{greenhill03}, so a weak outflow
that contributes to the absorption is
plausible. A wide, amorphous outflow more like the superwind seen towards
M82 might be the best analog  \citep[e.g.][and refs. therein]{leroy15}.

\subsubsection{Contributions from Warped Disks}  

Galaxy disks are known to exhibit warps, especially in systems
undergoing interactions with other galaxies.
If the inclination angle of the disk changes with
radius, for example, and/or the position angle of the major axis
changes with radius, absorption line profiles could be replicated to
better match the data with these added variables. For example, the
simplest case is where the inclination angle varies with radius, which
would allow for a wider range of $v_{\rm{rot}}$ at different
radii. Among the various possibilities for geometrically thin disks we
mention only two: good agreement between data and line profiles can be
found when the outer disk has $v_{\rm{rot}} = 250$~\kms\ and
$i=10\degr$ (i.e. nearly face-on), or, $v_{\rm{rot}} = 60$~\kms ,
$i=75\degr$, and a very large disk scale of $\sim 30$~kpc.  So, given the
inclination of NGC~1097 ($40\pm5\degr$), substantial warping would
be required to account for the observed absorption.  There
is some indication from the 21~cm H~I maps of \citet{higdon03} that
the outer spiral arms of NGC~1097 are disturbed by tidal interactions
(which we discuss later), so distortions in the structure of any
extended disk are a possibility. However, far more additional data on
the nature of the warps in NGC~1097's disk would be needed before they
could be reliably added to our model.

\subsection{Outflow Models \label{sect_outflowmodels}}

While a conic outflow can be added to the disk model to explain the
additional absorption (component A3) seen towards \vtwo , its inclusion is
somewhat ad hoc, and leads us to examine whether, in fact, outflow models
alone --- with no contribution from an absorbing disk of gas --- could
adequately reproduce the observations.  
We assume bi-conic outflows whose symmetry axis is parallel to the
angular momentum vector of the inner disk. Geometrically, the outflow is
defined by an inner and an outer polar angle, as well as an inner and outer
radius. This parameterization allows us to construct both
spherical and polar outflows. By specifying an inner and outer
radius to the outflow, intermittent flows or shells --- rather than
continuous winds --- can also be modeled. 
We set the
outer radius to arbitrarily large values; whether outflowing gas ejected
from the center of a galaxy can travel well into, and survive within, the
outer regions of a galactic halo, is a separate issue, and beyond the scope
of this paper. For the sake of brevity, we do not reproduce a figure
similar to Figures~\ref{fig_disk} and \ref{fig_flow}, but simply describe
the results from our analysis.

We assume that the outflows are coasting at a constant radial outflow
velocity, $v_{\rm rad}$. If mass is conserved along outflow lines
(e.g., no mass loading), then the gas density is $n\propto r^{-2}$, where $r$ is the
radial coordinate. For continuous
spherical winds, this also implies that the absorbing column density
declines as $1/\rho$. In contrast, for thin shell geometries whose
size is larger than the impact parameter range in question, the column
density is independent of $\rho$. Clearly, significant deviations from
the above scalings could occur if $v_{\rm rad}$ varies with $r$, or if
there are multiple ionization phases within the outflow 
 \citep{rupke13,scannapieco15}.
Such models cannot be constrained by the current data
set. 

\subsubsection{Spherical Outflow Models \label{sect_spherical}}

By virtue of isotropy, the absorption line profiles predicted from a
spherical outflow model are centered symmetrically around the systemic
velocity. For example, expanding shells would generally produce
two-component line profiles as the sightlines towards the background QSOs
intercept the front and back sides of the outflow. This well matches the
double component profile (D1 and D2) towards
\vseven, and we used these components to calibrate a thin-shell model with
$\delta r \ll r$ \citep[a similar approach was adopted
by][]{tripp11}. Specifically, by matching the model to the D1 and D2
components, we find that the mass in the outflowing
shell is $>3\times 10^5\eta\,{M_\odot}$, where $1/\eta$ is the mass
fraction of the gas in the form of H~I, a potentially small number related
to the uncertainties in the ionization corrections and/or the multiphase
nature of the outflow.  The mass-loss rate is then of order
$4 \pi r^2 \eta N({\rm H~I})m_p (v_{\rm rad}/r) \simeq 10^{-4}\eta
(r/200\:{\rm kpc})(N({\rm H~I})/10^{14}\,{\rm cm}^{-2 })(v_{\rm rad}
/40\,{\rm km~s^{-1}})\,{M_\odot~{\rm{yr}}^{-1}}$,
which is probably much smaller than the current rate of
star-formation. Irrespective of this argument, thin-shell models fail to explain the
decreasing gas opacity with impact parameter --- the predicted column
densities are always too low to produce the strong absorption seen along
all the sightlines. Moreover, the asymmetric absorption around the systemic
velocity along sightlines at opposite ends of the disk's major axis
(i.e. towards \vfour\ and \vsix) cannot be reproduced.

We find that the simplest models of spherical outflows seem to be inadequate for
reproducing the observed absorption, and given the limited set of
observational constraints, a more detailed analysis seems unwarranted.

\subsubsection{Polar Outflow Models}

In \S\ref{sect_fulldisk} we showed how adding a polar outflow to the
rotating disk models could account for the A3 component seen towards
\vtwo.  In general, however, models which include {\it only} polar outflows
and no disk component run into difficulties explaining the absorption
along the sightlines which lie in the direction of the major axis and
are at large impact parameters, unless the opening angle is large or
the flow extends far enough above the disk. For example, assuming an
outflow that extends out to $R_v$, and opening angle of $>50\degr$ is
needed to produce absorption towards \vsix . 

The model shown in Figure~\ref{fig_flow}
assumes outflow from only one side of
the galaxy, but a bipolar outflow would produce additional
positive velocity absorption. Outflowing gas from the
backside of the galaxy
would need to extend further than the front-side flow to intercept the
sightline. We do indeed see weak components  --- A4 and A5 --- at
higher velocities, which might fit such a scenario. Whether the extent and
covering factor of the outer regions of an outflow could account for such
weak absorption is unclear.

We should note that
gas outflowing at the observed velocities would need to
travel over timescales that are comparable to the dynamical time of
the halo, which would most likely invalidate the simple
conic geometry. For example, gas may be dragged perpendicular to
the outflow if the halo is rotating, or minor mergers may disrupt the
geometry of the escaping gas. Again, modelling the absorbing properties of
such disrupted outflows is beyond the scope of this paper.

\bigskip

\section{Global Properties of the absorption }
\label{sect_global}

We have observed 4 QSOs with impact parameters of $\rho = 48-165$~kpc from
NGC~1097 and searched for absorption from the galaxy along each sightline.
The lowest quality spectrum has a 2$\sigma$ equivalent width limit of
35~m\AA\ at short wavelengths, defining the limiting EW of our survey for
\lya\ and Si~III~$\lambda 1206$. These limits correspond to column
densities of $\log N$(H~I)$\:=12.9$ and $\log N$(Si~III)$\:=12.3$ (for
Doppler parameters of $b\geq 7.1$~\kms , or thermal temperatures
$\geq 3000$ K). We have seen that while the EWs of all the detected lines
are measured precisely, $N$(H~I) derived from the \lya\ lines for 3 of the
sightlines are less well defined, because the lines have equivalent widths
where $b$ and $N$(H~I) are degenerate.  For the four sightlines, our most
basic statistic suggests that the covering fraction of H~I and Si~III is
100~\% and 75\%, respectively, at a limit of $\rho = 165$~kpc.

\begin{figure*}
\centering
\includegraphics[width=7cm]{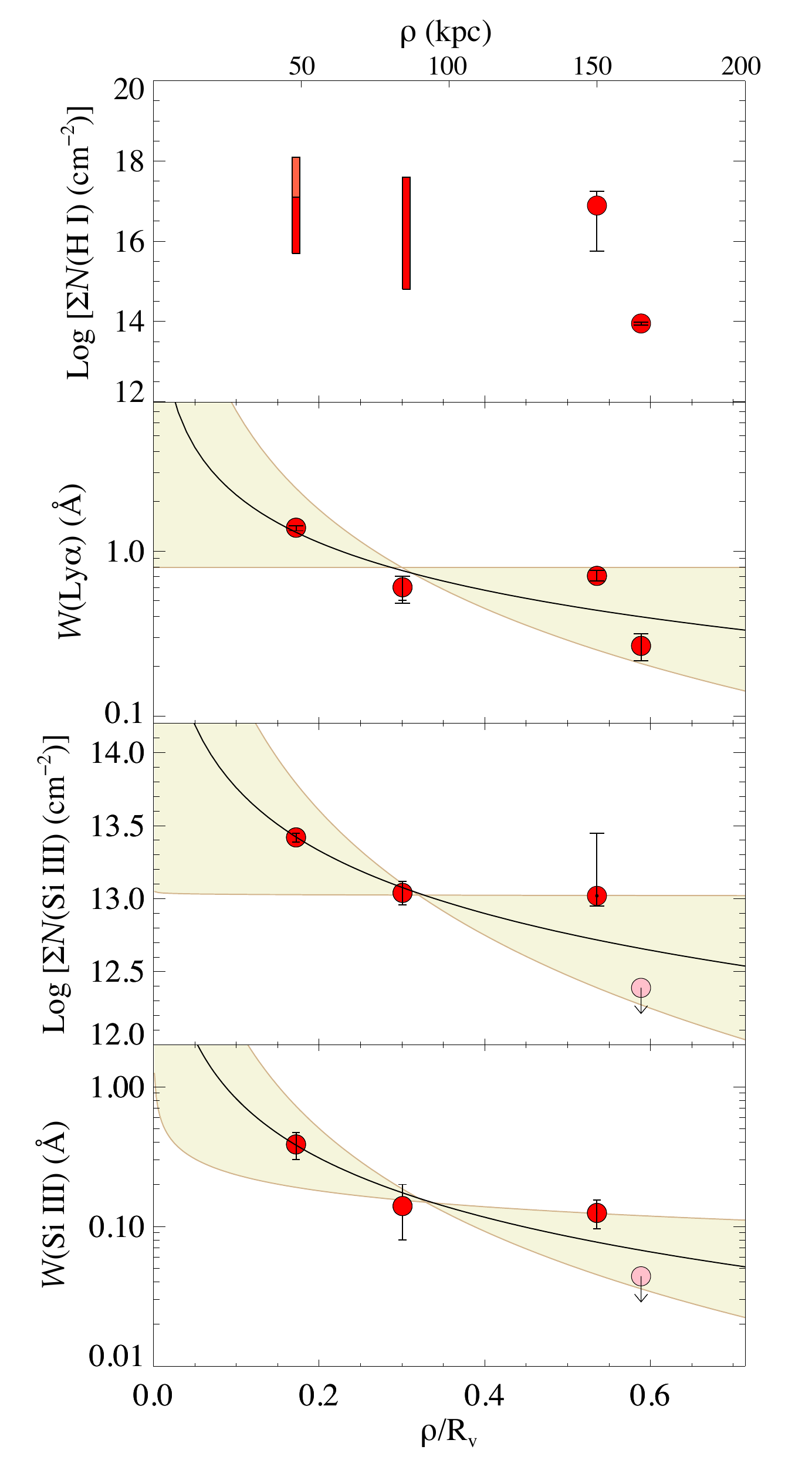}
\includegraphics[trim=0 -1cm 0 0,width=4.1cm]{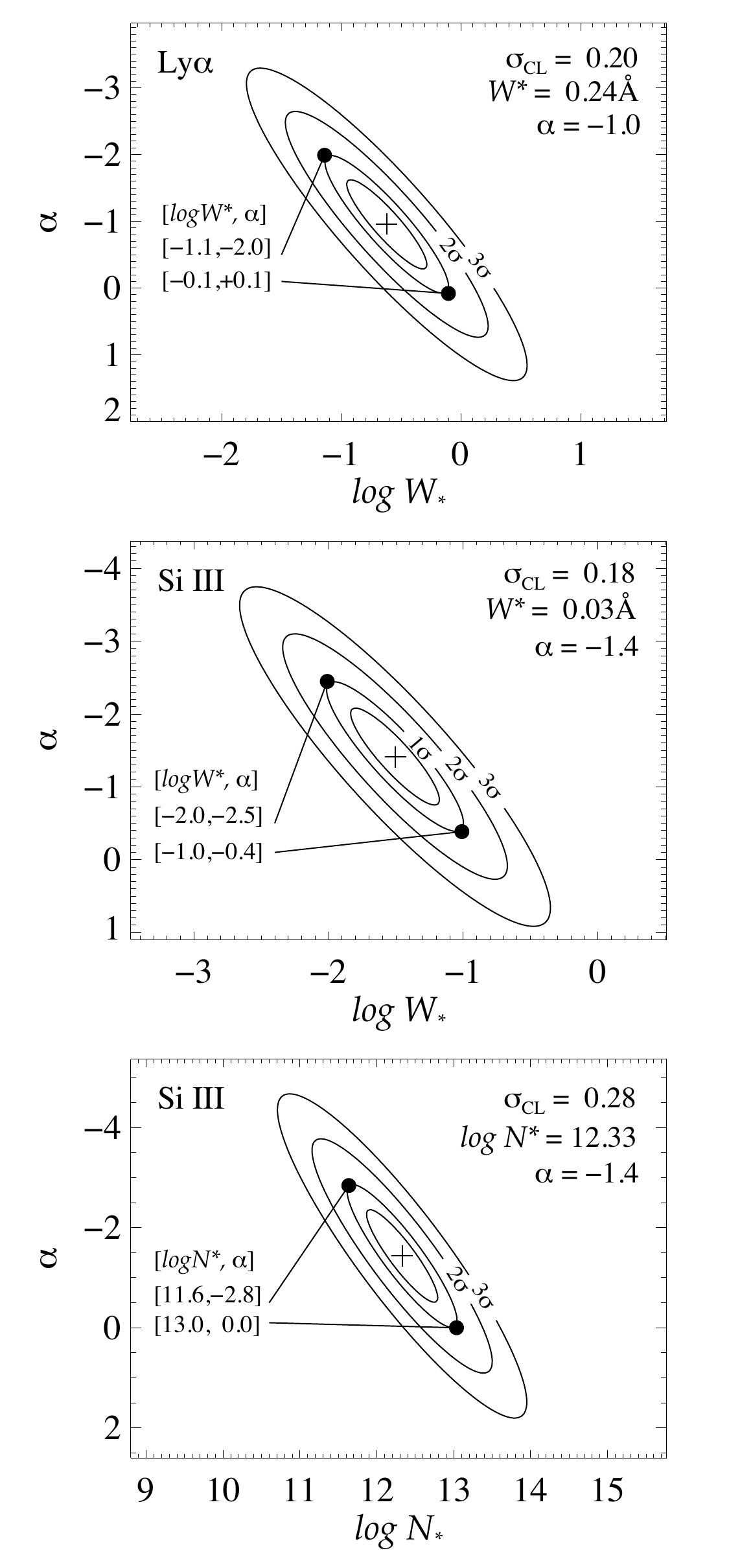}
\caption{
{\bf Left:} Distribution of total $N$(H~I) summed over all components (top panel),
\lya\ EW (second panel), total $N$(Si~III) (third panel), and
Si~III~$\lambda 1206$ EW (bottom panel) towards NGC~1097 as a function
of impact parameter normalized by virial radius. Our data are shown as red
circles for detections, and pink circles for upper limits. For the
$N$(H~I) values in the top panel, our measurements are
uncertain for the sightlines with the two smallest $\rho$, and we
represent the range in possible $N$(H~I) as two red/orange vertical bars. 
For \vtwo , there are two possible upper limits to $N$(H~I), one set by the maximum
possible from component A3, which is shown in red, and one set by the
maximum towards A2, extended in orange (\S\ref{sect_v20fits}).
For \vfour, the range is set by the two possible values of $N$(H~I)
that arise from two line profiles that are nearly identical (\S\ref{sect_v40fits}). 
For the three lower panels, power-law fits to the 4
points are shown
as black lines, and the beige regions indicate errors obtained from
$\chi^2$ confidence intervals (see text). 
{\bf Right:} 
Contour plots of $\chi^2$ centered on the minimum values
  (shown with crosses) found
  for power-law fits to \lya\ EWs (top panel), Si~III EWs (middle
  panel) and $N$(Si~III) (bottom panel). The best fit values are
  listed in the top right corner of each panel. Sets of $\alpha$ and
  $W^*$ or $N^*$ at the extreme deviations of the $1\sigma$ confidence
  intervals (black circles) are also indicated, and these two sets of
  values are the ones used to draw the shaded beige regions in the
  left-hand side of the plot.
\label{fig_corr1}}
\end{figure*}

The values of \lya\ and Si~III EWs, and H~I and Si~III column densities,
plotted against their impact parameters normalized by the virial radius
$R_v$, are shown in the left-hand column of Figure~\ref{fig_corr1}.  The
gas in the halo of the galaxy does not appear to be smoothly distributed.
While a rank correlation test suggests that $\rho/R_v$ and $W$(\lya) are
significantly anti-correlated, a Pearson's product-moment analysis shows
that the correlation is not significant (the probability of the null
hypothesis is $p=0.18$, well above the 0.05 significance level often used
to define a significant correlation).  The same test for $W$(Si~III) with
$\rho/R_v$ actually finds more significance for anti-correlation.

Irrespective of these rank correlation tests, however, we would like to
measure how $\rho/R_v$ and $W$(\lya) vary if the underlying relationship is
a power-law, in part because a such a relationship is often suggested for
higher-$z$ collations between EW and impact parameter in
single-galaxy/single-QSO-sightline systems (see \S\ref{sect_others}).  Even
though we have only 4 sightlines through NGC~1097, the sampling of multiple
points in a single galactic halo is unique, and we therefore examine the
results of fitting a power-law to the NGC~1097 data.

Both EWs and column densities are assumed to be related as

\begin{equation}
W = W^{*}  \left( \frac{\rho}{R_v} \right) ^{\alpha},
\:\: N  = N^{*}  \left( \frac{\rho}{R_v} \right) ^{\alpha}  .
\label{eqn_pw}
\end{equation}

\noindent We fit  the \lya\ and Si~III EWs, and the distribution of
Si~III column densities; we do not fit the H~I column densities
because we only have a range of values for 2 sightlines.
We also take the Si~III EW and column density upper limit
towards \vseven\ to be a detected value.

If a power-law fit to the data is justified,
then the distributions do not vary
smoothly  with $\rho/R_v$. We can fit a power-law 
by varying
$W^*$ or $N^*$, and $\alpha$, and minimizing the $\chi^2$ statistic
between the NGC~1097 data and values expected from equation \ref{eqn_pw},
but the reduced $\chi^2$ is 
$ \chi^2_\nu \gg 1$ given the small errors relative to the overall dispersion in $N$
and $W$. To account for this, we introduce an additional ``error''
which accounts for the clumpiness in the distribution. This approach
is the same as that used to account for patchiness in absorption from
O~VI in the Milky Way 
described by \citet{bowen08} [after the work of \citet{savage90}]. 
For each value of $W$ measured towards the QSOs
behind NGC~1097 we introduce an
additional {\it relative} error of $ \sigma_{\rm{CL}}$ beyond that of
the equivalent width error $\sigma(W)$ described in \S\ref{sect_ews}:

\begin{equation}
\sigma^2(W)_T = \sigma^2(W) + \sigma^2_{\rm{CL}}.
\end{equation}

\noindent For the given points, $\sigma_{\rm{CL}}$ can be varied until
$\chi^2_\nu =1$.  The values of $\sigma_{\rm{CL}}$ that we find are listed
in the right hand panels of Figure~\ref{fig_corr1}; for both the \lya\ and Si~III EW,
$\sigma_{\rm{CL}}$ is similar, around 0.2 dex (top two panels), which
might be expected if the \lya\ and Si~III absorbing regions are the
same. For $N$(Si~III), $\sigma_{\rm{CL}}$ is a little higher,
$\simeq 0.3$~dex. Of course, $\sigma_{\rm{CL}}$ need not be the same
for EW and column density measurements, for while the two parameters
are obviously related, the EW of a line represents both the column
density of absorbing clouds as well as the dispersion between
components comprising the line. In all three cases,
$\sigma_{\rm{CL}} \gg \sigma(W) $, as expected.  With
$\sigma_{\rm{CL}}$ fixed to the values that gives the $\chi^2_\nu =1$
value, the corresponding values of $W^*$, $N^*$ and $\alpha$ are
considered to be the best-fit values. These are again listed given in
the right hand panels of
Figure~\ref{fig_corr1}, and the resulting curves are shown in the
bottom three panels of the left hand column of Figure~\ref{fig_corr1} as black solid
lines. The values of $\alpha$ for $N$(Si~III) are identical, 
$\alpha = -1.4$, which may reflect a more direct relationship between
the simple, weak Si~III EWs and their column densities.

The errors in $W^*$ or $N^*$, and $\alpha$, can be estimated by
calculating $\chi^2$ as these values move aways from a minimum
$\chi^2$. Full details are given in \citet{bowen08}, which followed an
identical procedure. To summarise: as $W^*$ or $N^*$, and
$\alpha$, vary, contours in $\chi^2$ can be selected at particular
confidence levels;  
 in the right hand column of Figure~\ref{fig_corr1} we
show the confidence levels at 0.38, 0.68, 0.95 and 0.997, which
resemble the confidence intervals applied to a normal distribution. In
that figure, we also list the values of $W^*$ or $N^*$, and $\alpha$,
that correspond to the extreme deviations of the $1\sigma$ confidence
interval for both parameters (the end point of the major axis of the
contour); and for these pairings, we plot envelopes to the deviations
in the left hand panels of Figure~\ref{fig_corr1} (shown in beige). Unsurprisingly, given only 4
data points, the errors are large. In fact, for
one of the $1\sigma$ confidence intervals, the deviations are consistent
with no correlation at all for the \lya\ EW and $N$(Si~III).

\subsection{Comparison to Other Datasets \label{sect_others}}

In this section, we compare the properties of the absorption lines
seen towards NGC~1097 with the properties of absorption lines seen
towards galaxies in three other low-redshift studies. We first review the
samples and galaxy types in \S\ref{sect_cfch}$-$\S\ref{sect_cflc}, and
draw comparisons between the published results and NGC~1097 in
\S\ref{sect_typical}. Our goal is to determine whether NGC~1097 is a
{\it typical} QSOALS.

\subsubsection{Comparison to \CH \label{sect_cfch}}

In Figure~\ref{fig_corr1} of \S\ref{sect_global} we showed how \lya\
and Si~III column densities and EWs changed with impact parameter. We 
also introduced a comparison set of data, taken
from the \CH\ program, and in this section we discuss these results further. The properties of
\lya , and of metal lines, from the absorbers were studied by
\citet{tumlinson13}, and \citet{werk13a}, respectively.
To compare with our NGC~1097 data, we need to 
derive galaxy virial radii  in the same
way as we have for NGC~1097. 
Tumlinson~et~al list stellar masses for
all their galaxies, derived from SDSS colors, from which we can
estimate halo masses, and hence $R_v$, again using the relationships between the two
described by \citet{behroozi13}.

\begin{figure}
\hspace*{-0.5cm}\includegraphics[width=9.6cm]{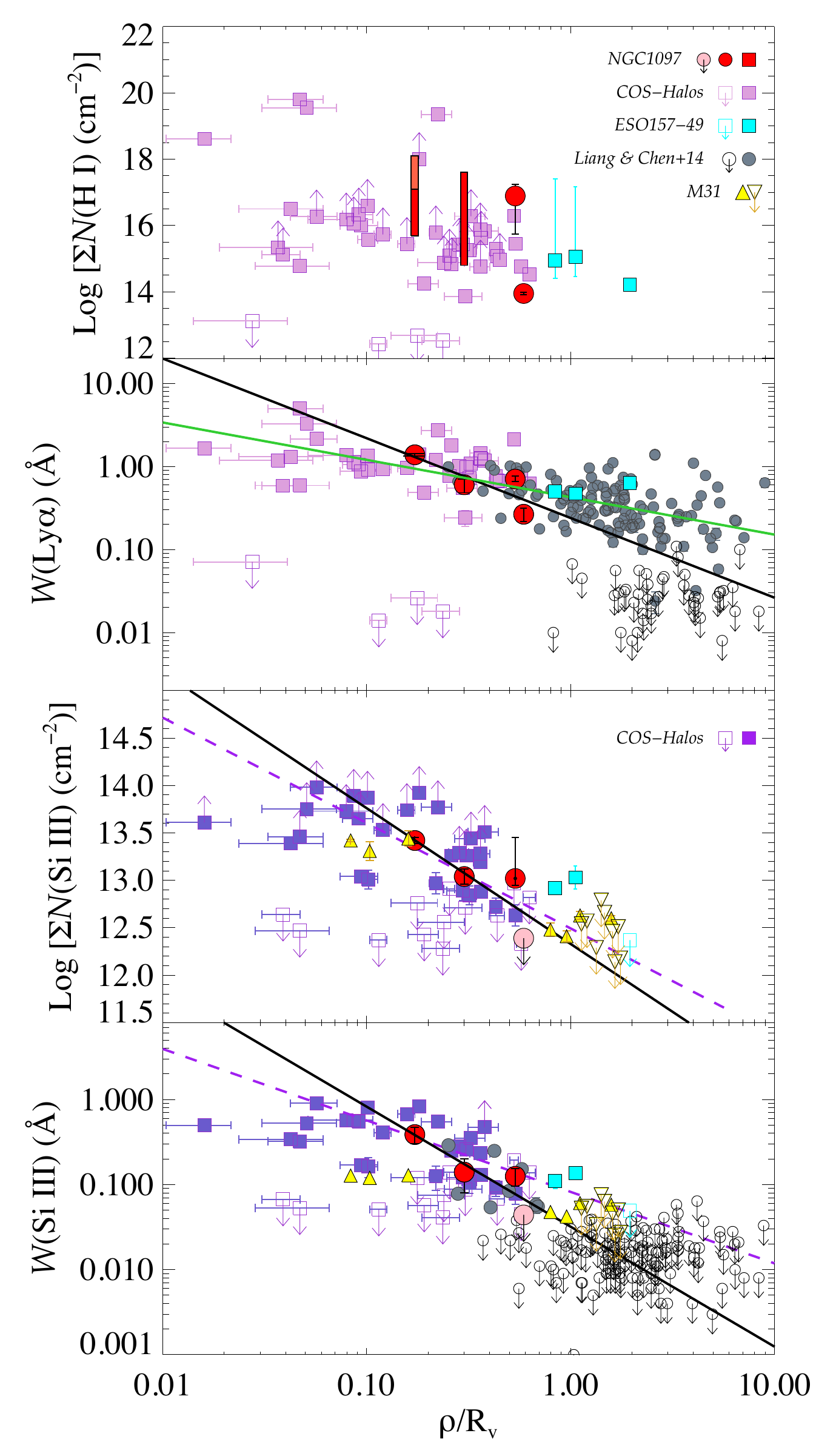}
\caption{Same as Fig.~\ref{fig_corr1}, now with the addition of values from 
  published datasets, and with the x-axis extended beyond $1R_v$.
  Data from \CH\ are shown as
  light pink squares for H~I \citep{tumlinson13} and purple squares for Si~III
  \citep{werk13a}.  
  The black lines in the bottom three panels show the power-law fits
  to the NGC~1097 data alone from Fig.~\ref{fig_corr1}; the
  dashed purple lines for Si~III show the fits from
  \citet{werk13a} [where the slopes were found to be 
  $-0.84\pm 0.24$ and $-1.11\pm0.29$ for $W$(Si~III)
  and $N$(Si~III), respectively].
  The green line in the second panel represents a least-squares fit to the
  \CH\ and LC14 data, ignoring their upper limits. 
 \label{fig_compare_others} }
\end{figure}

In Figure~\ref{fig_compare_others}
the \CH\ data are shown as light pink
squares, while the Si~III values are purple squares. 
The H~I column densities and EWs show a large
dispersion, and include a set of sensitive non-detections at very
small impact parameters which Tumlinson et~al suggested was due to the
lack of a cool halo around early type galaxies. These non-absorbing
galaxies are only a small fraction of the total sample, and the bulk
of the absorbers show unity covering fractions out to the
$\sim 0.6\:R_v$ limit of the survey.  
Our range of
$N$(H~I) values, and the inferred covering fraction, 
appears consistent with the \CH\ sample.

For Si~III lines, Werk~et~al fitted a power law to the distribution
of $N$(Si~III) and $W$(Si~III) with $\rho/R_v$
(ignoring both upper and lower limits).
These fits are reproduced in the
bottom two panels of Figure~\ref{fig_compare_others};
their fits to $N$(Si~III) and $W$(Si~III)
are close to what we find for the four sightlines towards NGC~1097.  
They also calculated a covering fraction for Si~III of
$73^{+9}_{-14}$\% over a radius of $0-160$~kpc for galaxies with
stellar masses $\log M_* (M_\odot)   \ge 10.5$. This again is 
consistent with the 3/4 detections in the halo of NGC~1097 at similar
radii and at the same stellar mass limit.

\begin{figure}
\hspace*{-0.5cm}\includegraphics[width=9.6cm]{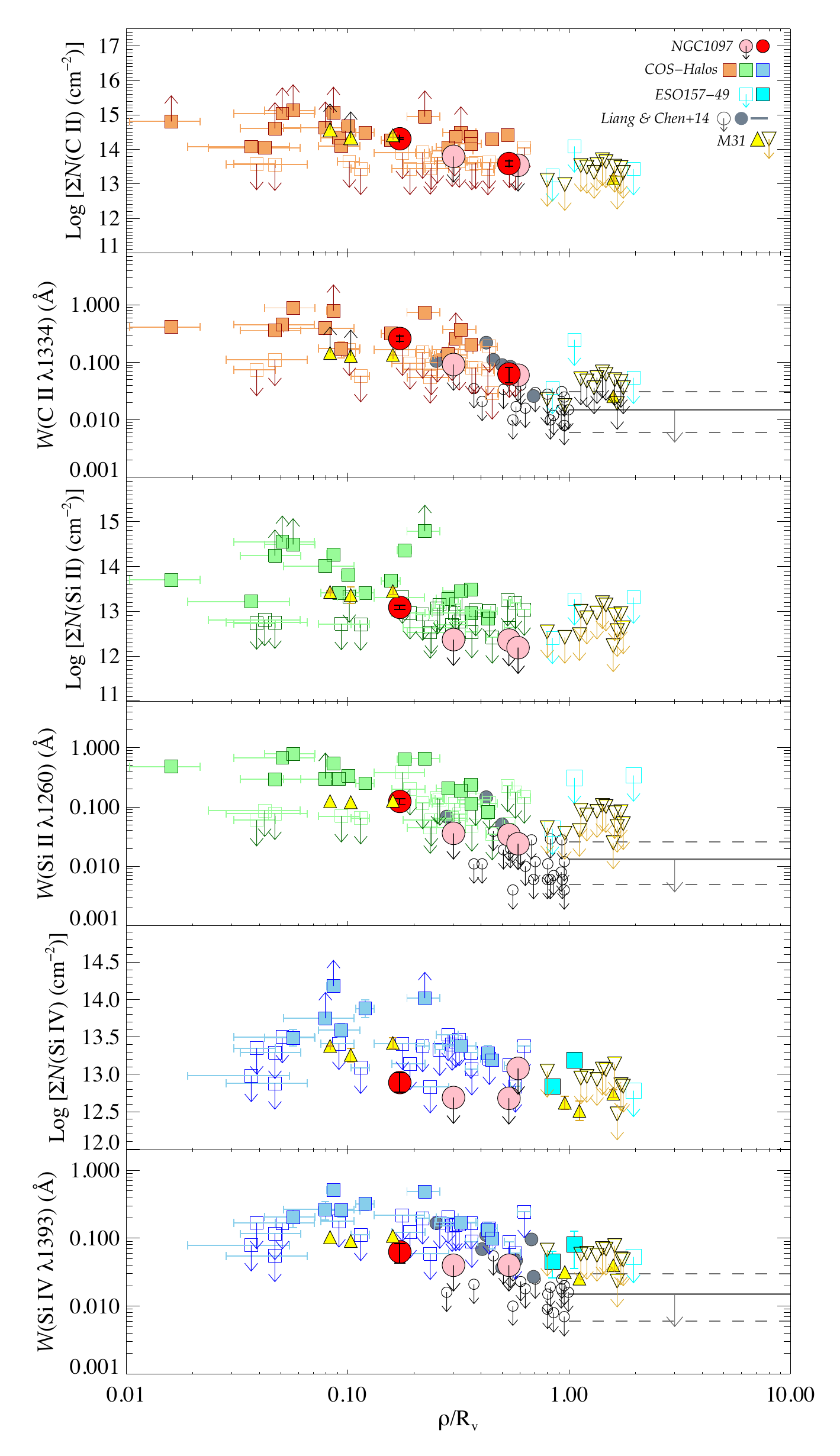}
\caption{Distribution of total column densities summed over all
  components, and equivalents widths, for C~II, Si~II and Si~IV,
as a function
of impact parameter normalized by virial radius. Our data towards
NGC~1097 are shown as red
circles for detections, and pink circles for upper limits.   
Several other datasets are included for comparison, and their origin
is shown in the legend top right of the first panel.
Unfilled symbols
represent upper limits, while lower limits (usually from mildly
saturated lines) are indicated as filled
symbols with the appropriate upward pointing arrow. Liang \& Chen's data contain a large number of
EW upper limits beyond $1R_v$, and to avoid confusion, we draw the
median of those upper limits with a single solid grey line and indicate the
range of their limits with two dashed grey lines. Individual upper limits at $< 1R_v$
are still plotted. Errors in $\rho/R_v$ are shown for the \CH\ data,
but to avoid confusion, they are omitted for the rest of the data sets. In
most cases, uncertainties in $R_v$ are similar.
 \label{fig_metals} }
\end{figure}

In Figure~\ref{fig_metals} we introduce the detections and limits for
C~II (orange points), Si~II (green points) and SI~IV (blue points) for
the \CH\ galaxies.  With C~II, Si~II and Si~IV detections towards only
\vtwo, and a possible C~II detection towards \vsix \ (which we
include) comparisons with NGC~1097 are tenuous.  We note simply that
the strengths of the metals towards NGC~1097 are similar to the \CH\
sample. They lie in regions where the covering fraction of the metals
in the \CH\ sample are less than unity, and our non-detections are
compatible with those low values.

\subsubsection{Comparison to ESO~157-49 \label{sect_cfeso}}

Three background probes were used by \citet{keeney13} to probe the edge-on
galaxy ESO~157$-$149, a much less luminous ($L/L^* = 0.12$) and weaker
star-forming galaxy ($0.2-1.1$ $M_\odot$~yr$^{-1}$) than NGC~1097.  We
assume a viral mass of $\log M(M_\odot) = 10.6$ for the galaxy (the lower
limit set by Keeney~\etal ), which corresponds to $R_v = 88$~kpc. This puts
their sightlines at impact parameters of $\rho/R_v \simeq 0.9, 1.1$, and
2.0, distances larger than those probed for NGC~1097. We adopt the sum of
the 3 $N$(H~I) components towards the outermost sightline, HE~0435$-$5304,
to be $\log N($H I) = $14.22\pm0.11$ and calculate the total EW by
reconstructing the line profile from their Voigt profile fits, $W$(\lya) =
0.63~\AA . The \lya\ lines towards the inner 2 sightlines are of similar
strength to the inner two sightlines towards NGC~1097, and Keeney~\etal\
had the same problem in determining $N$(H~I) as we have had, with
moderately strong $W$(\lya) yielding ambiguous column
densities. Figure~\ref{fig_compare_others} shows these points compared to
the NGC~1097 values; although the errors are large, $N$(H~I) (filled cyan
squares) appears to continue to decline with $\rho/R_v$ beyond 0.9$R_v$.
Si~III is detected towards the inner two ESO~157$-$49 sightlines, beyond
the last limit from NGC~1097.  No C~II or Si~II is detected towards any of
the sightlines (open cyan squares) but Si~IV is seen at distances beyond
where the \CH\ or the NGC~1097 data probe.

With the absorbers at either side of the major axis of the edge-on
galaxy {\it both} having negative velocities relative to the galaxy,
Keeney~\etal\ concluded that the absorption (at least for one of the
sightlines) was unlikely to be associated with any galaxy
rotation, and that it was more likely to have arisen in
material once ejected from the galaxy. This interpretation is clearly
quite different from the ones discussed in \S\ref{sect_kinematics} 
designed to account for the absorption towards
NGC~1097.

\subsubsection{Comparison to M31 \label{sect_cfm31}}

M31 offers an interesting comparison with NGC~1097 in that M31 has a
luminosity and halo mass close to that of NGC~1097
\citep{marel12,marel12b}.  The one important difference is that M31 is
interacting with another galaxy of comparable mass only 0.8~Mpc away, with
the same halo mass and luminosity --- namely the Milky Way; as discussed in
\S\ref{sect_lss}, there are no galaxies as bright as NGC~1097 within
$4-5$~Mpc.  Although there are a large number of QSOs that can be found at
interesting impact parameters behind M31, the low velocity of the galaxy
makes it difficult to study in absorption (\S\ref{sect_intro}).
Nevertheless, \citet{rao13} found no absorption beyond $0.2\:R_v$ along the
major axis of the disk (assuming $R_v = 300$~kpc for M31), but
low-ionization absorption from 4 of 6 sightlines below $0.1\:R_v$. These
data are hard to incorporate in this study because {\it i}) the low
resolution of the COS G140L and G230L observations make it difficult to
fully separate out absorption from the strong MW ISM absorption lines, {\it
  ii}) understanding the relationship between the velocities of the
detected lines and the velocity of MS+HVC gas (if any) at the positions
probed is beyond the scope of this paper; and {\it iii}) both the \lya\ and
Si~III~$\lambda 1206$ line which are integral to our study are not
available in their low resolution data. For these reasons, these data are
not plotted in Figures~\ref{fig_compare_others} and
\ref{fig_metals}. Nevertheless, Rao~\etal\ drew an important distinction in
claiming that M31 is not a {\it typical} Mg~II absorbing galaxy compared to
higher-redshift samples.

On the other hand, \citet{lehner15} used COS G130M and G160M Archival data
of 18 sightlines passing within $2R_v$ of M31. They distinguished a
specific velocity range over which to measure absorption that they
considered sufficient to avoid contamination of MW ISM absorption and the
MS. These data are included in Figures~\ref{fig_compare_others} and
\ref{fig_metals} (shown as triangles).  Lehner~\etal\ did not provide EW
measurements from their data, and so we estimate EWs directly from the
column densities assuming the lines are optically thin. These represent
lower limits to the true EWs because not only may some of the lines be
mildly saturated, but also because the wings of the absorption profile are
often lost in the blended MW ISM and the MS lines either side of the
absorption from M31.

\subsubsection{Comparison to Intermediate-Redshift Sample \label{sect_cflc}}

Like the \CH\ sample, the collation by \citet{liang14} is of
absorption along single sightlines through many
galaxies. Nevertheless, the sample is useful in that the sightlines
probe beyond the impact parameters of the \CH\ galaxies, and that
measurements of many species of metal lines, as well as \lya , are
provided. This sample of galaxies (hereafter designated as LC14) has a
median redshift of 0.02, and samples galaxies with stellar masses of
$10^7 - 10^{11}\: M_\odot$. Their derivation of $R_v$ is the same as
described above, so we use their tabulated values.  Only EWs are given
by Liang \& Chen, and since these cannot be converted to column
densities with any reliability, column densities are not plotted. In
Figure~\ref{fig_compare_others} all the data for \lya\ and
Si~III~$\lambda 1206$ are included, but in Figure~\ref{fig_metals},
most points at $\rho/R_v > 1.0$ are only upper limits, and not
sensitive enough to show any significant decline in EW above those
radii. To avoid confusion with points from the other datasets, we 
show only the median EW limit at $\rho/R_v > 1.0$ as a grey horizontal
bar.

\subsubsection{Is NGC~1097 a typical absorber? \label{sect_typical}}

Our fitting of a power-law to the data obtained for NGC~1097 shows that
the evidence for a decline in H~I and Si~III column densities and EWs
with $\rho/R_v$ --- based only on the 4 sightlines --- is weak.
Tumlinson~\etal\ reported no decline in $N$(H~I) or $W$(\lya ) with
$\rho/R_v$ in the \CH\ sample, which we see recreated in the top panel
of Figure~\ref{fig_corr1}. However, Figure~\ref{fig_compare_others}
shows that when we combine the \CH\ results with the LC14 data it
seems more plausible that \lya\ and Si~III decline with impact
parameter, and indeed that NGC~1097 follows the same decline.  If we
exclude the upper limits in the \CH\ and LC14 datasets, then a linear
regression analysis of the two larger samples confirms a strong
correlation with a high probability of being able to exclude the null
hypothesis that there is no correlation.  A least-squares fit to the
$W$(\lya ) data (shown as a green line in the second panel of
Fig.~\ref{fig_compare_others}) has $\log W^* = -0.37\pm0.02$~\AA ,
$\alpha = -0.45\pm0.04$; but more importantly, the standard error in
the regression, $s$ is 0.27 dex, while for NGC~1097, $s=0.21$~dex. A
similar result is found for $W$(Si~III). This standard error is
the average distance that the observed values fall from the fitted
line, and conventionally, 95\% of all points should fall within
2$s$. The fact that the values of $s$ for NGC~1097 and the two samples
of \CH\ and LC14 are similar suggests that the overall variability or
clumpiness within the halo of NGC~1097 is largely the same as the
dispersion seen from 
the galaxies in the larger single-probe datasets.

Of course, while the \lya\ and Si~III absorption in the halo of NGC~1097
may have much the same characteristics as the other datasets, we only know
this to be true for impact parameters $<0.6R_v$. Beyond $1R_v$, LC14 have
shown that the number of non-detections of \lya\ increases rapidly,
i.e. the covering fraction declines to $\sim 60$\% at an EW threshold of
0.05~\AA\ (their Fig.~11). Keeney~\etal 's observations of ESO~157$-$49
does not show such a decrease (and remains as flat as the values for
NGC~1097 at smaller $\rho/R_v$). Whether this represents the way H~I is
distributed around all galaxies, or whether the detection/non-detection of
\lya\ points to some other property of a galaxy that influences the extent
of H~I around its CGM, is not clear. Observations of additional probes
beyond $0.6 R_v$ for NGC~1097 would help explain how the covering fraction
really declines in a single halo at large radii.

We conclude that based on the absorbing column densities and EWs alone,
NGC~1097 --- and probably M31 and ESO~157$-$49 --- are `typical' absorbers
compared to the \CH\ and LC14 samples. This is curious, because the
structures which have been suggested as origins of the absorbing gas (an
isolated galaxy for NGC~1097, an interacting group for the MW-M31-MS
system, and post-ejection products from ESO~157$-$49) are quite
different. It may be that while galaxies do indeed possess definable CGMs
which decline with radius, some absorption line properties ---- in this
case $N$ and EWs --- are relatively insensitive to the structures that
contain the baryons.  For NGC~1097 we have also used kinematics to attempt
to better understand the origin of absorbing gas, leveraging the use of
{\it multiple} probes of a single galaxy to constrain a model (an approach
that is far more ambiguous for single sightlines through single
galaxies). Again, though, a similar consideration of the kinematics of the
three sightlines towards ESO~157$-$49 suggest an origin quite different
than that for NGC~1097, and it remains to be determined if either set of
results is universal for galaxies.

\section{Discussion}
\label{sect_discussion}

\subsection{Likely Origin of the Absorbing Gas \label{sect_best}}

\begin{figure}
\includegraphics[width=8cm]{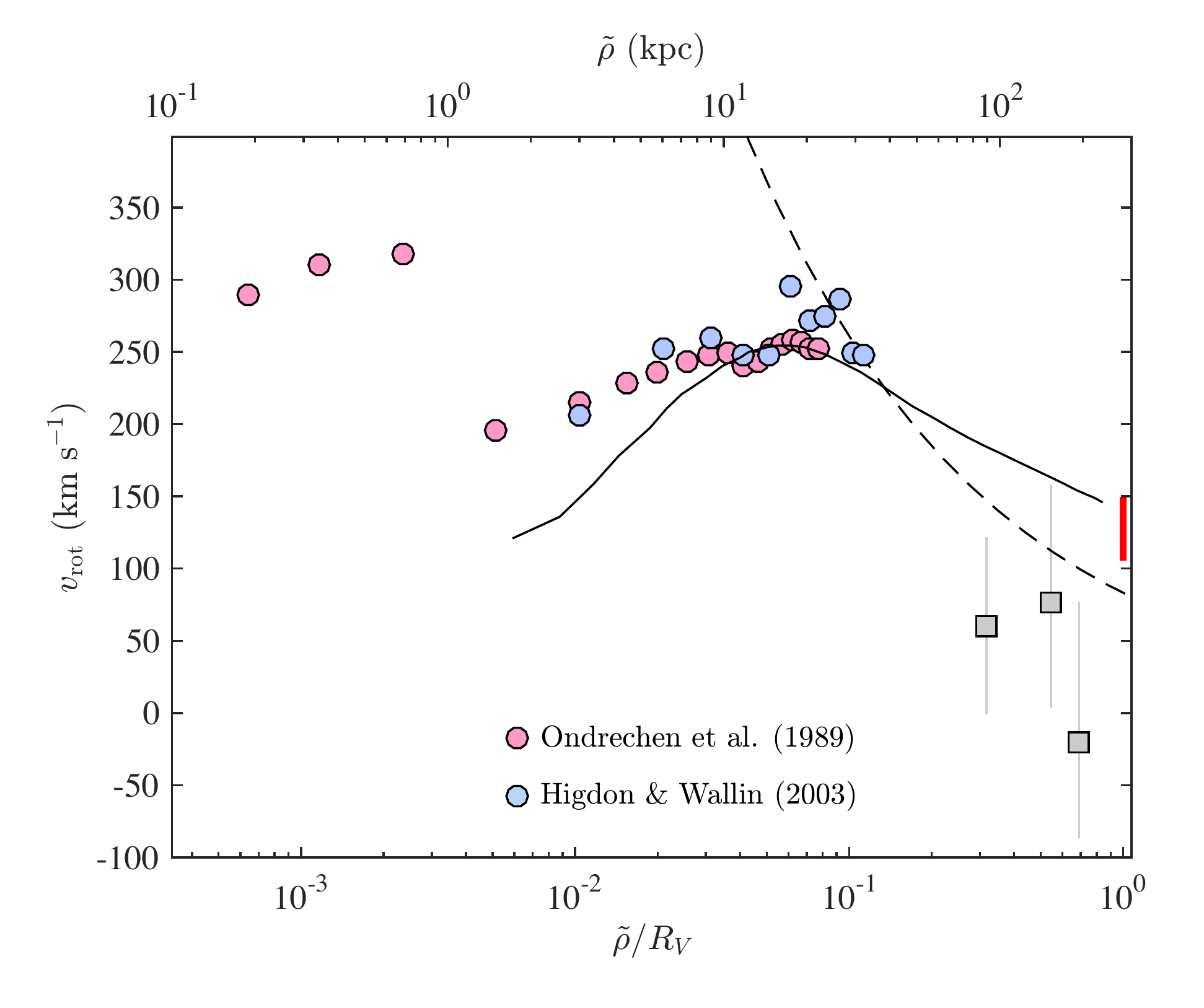}
\hspace*{0.2cm}\includegraphics[width=7.8cm]{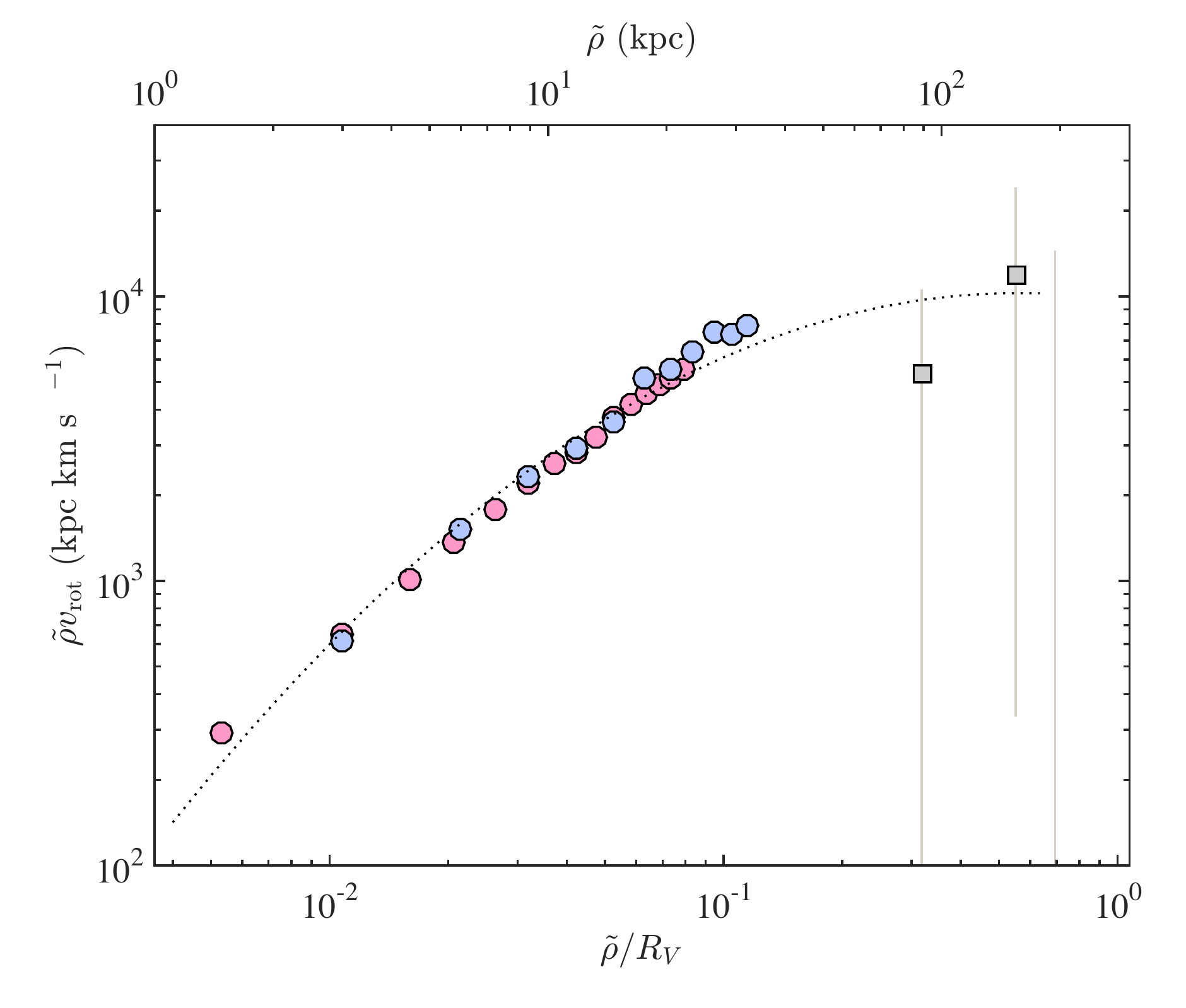}
\caption{  {\it Top:} Rotation
  curve for NGC~1097's inner disk from 21~cm data (pink and blue
  circles), compared to the absorbers in the CGM (grey squares)
  assuming that the absorbing gas is in the form of a disk and has the
  same inclination and position angle as the inner disk.  All values
  are corrected for inclination effects. The sightline towards \vtwo\
  leads to constraints on $v_{\rm{rot}}$ that are larger than the scale plotted on
  the y-axis, (because the sightline lies along the minor axis so has
  only a very small Doppler shift, and because the absorption is very
  broad) 
  and so the sightline is excluded. The dashed line shows
  the expected $v_{\rm{rot}}$ if there were no dark matter beyond
  25~kpc, and the black solid curve is a rescaled version of the
  \citet{seljak02} galaxy-halo rotation curve. The vertical red line
  marks the expected virial speed for NGC~1097.
 If the absorbing gas
  really is in the configuration of a disk, the fact that \vrot\
  is below the DM model expectations 
 may imply that
  the gas is either infalling or that the disk is warped at large $\tilde\rho$. 
  {\it Bottom:} Specific angular momentum of the gas. The dotted line
  shows a polynomial fit to the data.
    \label{fig_vrot}}
\end{figure}

We have shown in \S\ref{sect_diskmodels} that most of the absorption line
profiles can be reproduced by a model in which gas is distributed in an
extended rotating disk whose geometry matches those of the inner H~I disk
of NGC~1097. This implies that we are able to trace the rotation curve of
the galaxy out to distances of $0.6 R_v$. Figure~\ref{fig_vrot} shows the
21~cm rotation curves obtained by \citet{ondrechen89} and \citet{higdon03}
as a function $\tilde\rho/R_v$, in comparison to the \lya\ lines towards
the background QSOs. Figure~\ref{fig_vrot} shows how velocities would
decline for a Keplerian rotation; an extended disk model is only marginally
consistent with being rotationally supported if there were no mass exterior
to the inner H~I disk (i.e. beyond $\sim 40$~kpc).  Obviously, this is at
odds with $\Lambda$CDM theory in general, and specifically, with the
NGC~1097 system, where a large DM halo is required to produce the stellar
streams \citep{higdon03,amorisco15}. Instead, we are likely seeing
absorption from an extended disk with a cool gaseous component, probably of
low metallicity, falling into the inner disk of
NGC~1097. Geometrically-thick quasi-disk gas distributions that are
accreting on to a central object from large scales have been shown to occur
naturally in simulations of structure formation \citep[][e.g. their
Fig.~13]{kaufmann06, stewart13, danovich15}. There is evidence that such
cold accretion inflows are now being detected at high redshift
\citep{martin15_web}.

Figure~\ref{fig_vrot} also shows that the specific angular momentum of the
absorbing gas along the axis perpendicular to the inclined disk
($\tilde\rho\, v_{\rm{rot}}$) is consistent with that deduced from the
21~cm observations at the outer edges of NGC~1097's disk. This implies that
the accreting gas is infalling and does not lose angular momentum until it
joins the inner halo at $\sim 0.3 R_v$. This agrees with initial
semi-analytic works on galaxy formation \cite[e.g.][]{fall80} as well as
the more recent simulations of \citet{danovich15} (see their Fig.~20) that
explain the origin of angular momentum of galaxies by the deposition of gas
from cold inflowing streams.

Our interpretation suggests that the spin parameter of the extended
inclined disk \citep[][and refs.~therein]{dutton12} is
$\lambda_{\rm{extended-disk}} \simeq 0.15 (\tilde\rho / 100 \:{\rm
  kpc}) (v_{\rm{rot}} / 70 \:$\kms ).  This is $\sim 4$ times larger
than the typical spins values of DM halos \citep{dutton12, genel15}
and is consistent with the theoretical expectations for the spin of
cold dense gas which originates from larger scales and accretes onto a
halo.

Conversely, there is little obvious evidence to suggest that the
absorption seen along the 4 QSO sightlines arises in outflowing gas from
NGC~1097. Our attempts to replicate the kinematics of the
absorption do a poor job when only outflowing gas is included,
compared to the inflowing/rotating thick disk model. Nevertheless,
{\it adding} an outflow to the thick disk model helps account for the A3
component towards \vtwo . 
The primary constraint on such an outflow is that
the opening angle must be large.

Moreover,
galactic outflows are expected to have metallicities of order the
solar value, while our investigations suggest that the absorbers might have
metallicities less than this (Appendix \ref{sect_photoU}). 
 The A3 component towards \vtwo,
again, is a possible exception, since we are unable to constrain its
metallicity from photo-ionization considerations.

Finally, there is the issue of tidal debris around NGC~1097.
As discussed in \S\ref{sect_local}, NGC~1097 is interacting with an
LMC-like galaxy (NGC~1097A), and has seen the passage of another, now
destroyed, satellite galaxy which has given rise to the dog-leg
stellar trail shown in Figure~\ref{fig_pretty}.  
It has been suggested that galaxy mergers and interactions are an
effective way of distributing gas out to large distances and
relative velocities from galaxies, and that complex, multi-component
absorption line systems found in QSO spectra could be probing such
debris \citep{bowen_93j, keeney11,rosenberg14,fox14}.  The three outer
QSO sightlines towards NGC~1097 show no obvious evidence for complex
multicomponent structures spanning many hundreds of ~\kms , which we might
use to argue for the presence of tidal debris.  The sightline towards
\vtwo, however, lies only $\simeq\:$15~kpc from Knot-A (at least, in
projection), and only $\approx 10$~kpc from the limit of the trail
discernable on the image. 
Components A1$-$A3 span only 157~\kms , but if the very
weak positive-velocity \lya\ components (A4 and A5) are included, the
absorption covers at least 260~\kms .  Indeed, the spread in velocity
from components arising in tidal debris will depend strongly on the
orientation of the debris structure. Structures that are roughly
confined to a plane that is parallel to our sightline will show
components that have velocities determined only by internal random
motions rather than the bulk flow. Hence complex
absorption lines may not be a necessary signature for tidal
debris, particularly for minor mergers, where the disk of the more
massive galaxy might be less perturbed.

\citet{higdon03} found no H~I associated with the streams to a limit of
0.06 $M_\odot$~pc$^{-2}$ or $8\times10^{18}$~\pcm ; 
at the position of the right-angle of the dog-leg trail, they set a
limit of 0.021~$M_\odot$~pc$^{-2}$ or $3\times10^{18}$~\pcm\ in a 4
arcsec aperture. [This is consistent with the value of
$N$(H~I) found towards \vtwo .] 
Higdon \& Wallin's $N$-body simulations suggest that X-shaped tidal
tails develop only after a second or third pass of the accreting
galaxy through the host ($1.7-3.8$~Gyr from the first collision), and
that the dwarf's ISM will have been destroyed by ram pressure from the
disk. In that case, there may be no remnants of cool gas associated
with the dwarf or its interaction with the host galaxy.  Our
absorption line detections towards all four QSO sightlines, however,
are sensitive to the much lower column densities that may arise in more
highly ionized gas, and it remains possible that the absorption we see
is related to the fate of the gas from the cannibalism of the
dwarf galaxy.

We have seen in \S\ref{sect_fulldisk} that while a rotating, infalling
disk can explain most of the kinematics of the absorption lines from
NGC~1097, an additional model component of an conic outflow is needed
to explain the A3 component to the absorption. It is possible that
instead, this component --- and the weak A4 and A5 \lya\ lines ---
come from tidal debris associated with the stellar trails. This may
help explain the difficulty in finding suitable solutions for
photoionized gas in \S\ref{sect_A3} for A3.  The models of the debris
constructed by Higdon \& Wallin (their Fig.~13) show that stellar
material ought to be distributed out to the sightlines of \vtwo\ and
\vfour , but, as expected, the covering fraction would probably depend
quite strongly on the galaxy's orientation to the line of sight.
Modelling the evolution of gas associated with tidal
mergers is beyond the scope of this paper. While the kinematics of the
gas and the arguments made above show consistency with the models
described, it is worth emphasizing that with only 4 sightlines, the
correspondence could be fortuitous.

\begin{figure*}
\includegraphics[width=18cm]{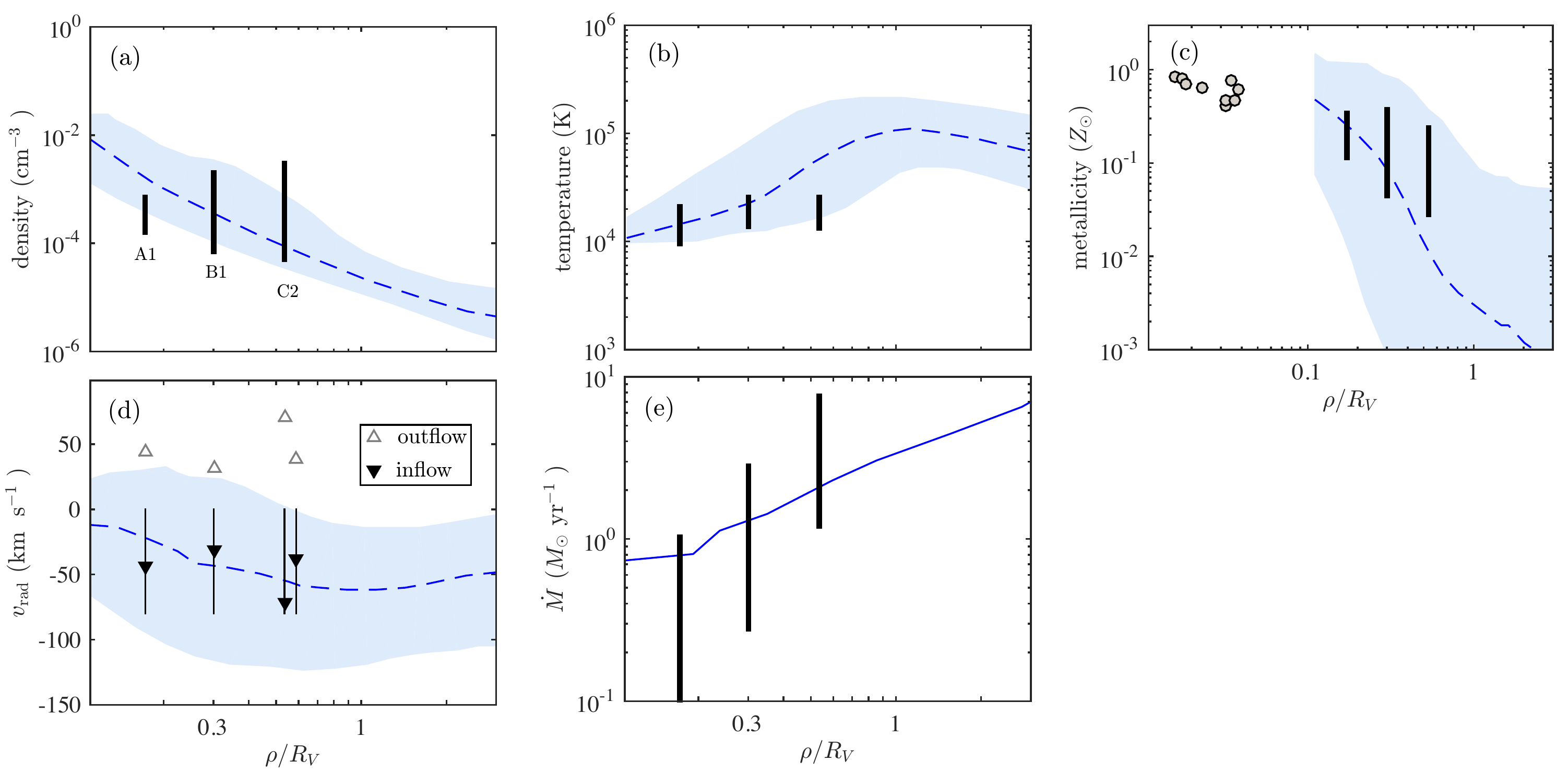}
\caption{ Comparison between the results for the absorbers towards
  NGC~1097 and those predicted by the simulations of cold-mode gas
  accretion at $z=0$ made by \citet{vandevoort12}. The dashed blue
  lines mark the median of the property from their simulations, while
  the blue shaded regions mark the 16th to 84th percentiles around the
  median. Triangles mark upper or lower limits, while
  vertical bars mark the range of values, found from our
  analysis of the absorbers. The simulations calculate distances in
  real (3D) space, while we only probe projected impact parameters $\rho$; as
  shown in previous plots, however, $\rho$ and the de-projected values
  $\tilde\rho$ are similar for all but the outermost sightline, and we
  plot our data in terms of $\rho/R_v$. Panels ({\it{a}}) to
  ({\it{e}}) show the gas density, temperature, metallicity, inflow
  velocity and mass accretion rate assuming azimuthal symmetry in the
  plane of the disk. In panel ({\it{c}}) we plot inner disk
  metallicities of NGC~1097 (grey circles) taken from
  \citet{moustakas10} [using the normalization from
  \citet{pilyugin05}].  For the mass accretion rates 
  in panel ({\it{e}}), van de Voort \& Schaye do not quote estimates
  in the scatter of $\dot{M}$, so no blue shaded region is drawn. In
  panel ({\it{d}}) two sets of limits exist, either for inflowing or
  outflowing gas, while the vertical bars indicate the allowed inflow
  velocity range based on the rotating-inflowing disk model
  (\S\ref{sect_fulldisk}). Overall, there is good agreement between
  the absorber properties deduced in this paper and these particular simulations.
\\
  \label{fig_vvs}}
\end{figure*}

\subsection{Comparisons with simulations \label{sect_sims}}

As argued above, there are good reasons to believe that most of the
absorbing gas we see
in the CGM of NGC~1097 comes from material that is
being accreted onto the galaxy from the outskirts of the
halo. 
We conclude by comparing our data with one further set of simulations that
evolve galactic structures down to $z=0$, those of 
\citet{vandevoort12}. Their calculations
include radiative cooling, sub-grid star-formation and feedback
prescriptions. Approximately 500 simulated halos are used to
quantify the statistical properties of the halo gas. Conveniently,
their results include a halo mass that is consistent with that of
NGC~1097. Two `modes' of gaseous structures are identified, a ``cold-mode''
component in the form of cool $<10^{5.5}$~K gas filaments, and
a ``hot-mode'' component which contains a volume-filling gas
component. Our data imply gas temperatures of $<10^5$~K, and so
would be related to the cold-mode structures in the halo.

Figure~\ref{fig_vvs} compares the results from the simulations with our
analyses of the absorption systems. We find good agreement
with many of the variables recorded in the simulations as they vary with
$\rho/R_v$: the surface density profile is consistent with the model within
the (non-negligible) uncertainties; the deduced gas temperatures coincide
with the simulated values; and the absorbing gas metallicities also show
good correspondence with the model.

The comparison between gas kinematics is shown on the bottom panel of the
first column of Figure~\ref{fig_vvs}.  It is impossible to
distinguish an outflow that is launched away from us from the more distant side of
a disk, from an infall of gas towards the side of a disk nearest to
us. If we {\it assume} that the absorbers are falling towards the
center of NGC~1097, then the absorption velocities are consistent with
values in the simulation. We note 
however, that we only measure line-of-sight velocities, with a radial
velocity component being potentially larger. 

We can crudely estimate the mass infall rate from our data
as a function of galactocentric radius, by assuming azimuthal
symmetry:  $\dot{M}(\tilde\rho) \simeq \pi \tilde\rho^2 N$(H) $(\tilde\rho)
(v_{\rm{rad}}/\tilde\rho )$ . If $v_{\rm{rad}}$ is independent of
$\tilde\rho$, we find $\dot{M}(\tilde\rho)$ rises monotonically with $\tilde\rho$, as
predicted in the simulations. Specifically, if we take a fiducial value of
$v_{\rm{rad}}  = 70$~\kms, 
there is qualitative agreement between a mass infall rate of
$\simeq 1 M_\odot$~yr$^{-1}$ at small $\tilde\rho/R_v$ and the current
star-formation rate, despite the $\ga 1$~Gy dynamical timescales
implied.

Figure~\ref{fig_vvs} suggests good agreement between theoretical
 simulations and the absorption line data for
NGC~1097. Intriguingly, van de
Voort \& Schaye noted that in their simulations, the probability of
detecting these cold streams was low. Similarly, \citet[][their
Fig.~4]{goerdt12} calculated a covering fraction of only $\sim 20$\%
for cool gas within
$1R_v$ at $z=1.4$, with the value dropping with time. In the case of
NGC~1097, and despite the small numbers of sightlines used, the
covering fraction at $\log N$(H~I)$ \ga 14$ is of order unity out to
$\simeq 0.6 R_v$. While it is possible that NGC~1097 may be  unique in
some way, there
are no clear indications from the absorption line data that this is so. In
fact, in \S\ref{sect_typical}, we argued that NGC~1097 was a 'typical'
absorber, in comparison to the ensemble of higher-redshift
absorbers. If our results are not entirely due to a fortuitous
combination of sightlines, then the apparent discrepancy
could be related to problems with the simulations, a topic that is beyond
the scope of this paper.

\bigskip

\section{Summary}
\label{sect_summary}

In this paper, we have observed 4 QSO whose sightlines pass through
the halo of NGC~1097: Q0244$-$303 (referred to as \vtwo\
in the text), 2dFGRS~S393Z082 (\vfour),
Q0246$-$308 (\vsix) and HE~0241$-$3043 (\vseven). These sightlines lie
at impact parameters of 48, 84, 150, and 165~kpc,
or $\approx 0.2, 0.3, 0.5$ and 0.6 times the virial radius of the
galaxy, which we take to be $R_v = 280$~kpc. We can summarise our
results as follows:

\begin{enumerate}

\item We detect \lya\ absorption towards all four sightlines. The
  total equivalent widths range from $\simeq 0.3-1.4$~\AA , but the
  column densities towards the inner two sightlines are difficult to
  determine due to saturation of the lines.  Nevertheless, they all
  have $N$(H~I)$\: \geq 14.0$, which implies that the cross-section of
  NGC~1097 is unity out to 165~kpc ($0.6\:R_v$) at that limit. Si~III~$\lambda
  1206$ is detected towards the 3 inner sightlines, with $W = 0.13 -
  0.39$~\AA , all lines having $\log N$(Si~III)$> 13.0$, again
  suggesting a covering fraction of unity at 150~kpc ($0.5 R_v$) at this
  column density.

\item The strongest \lya\ absorption, and the one that shows the most metal
  lines (including C~II, Si~II, Si~III and Si~IV lines) occurs along the
  sightline to the inner-most sightline, Q0244$-$303 at 48~kpc ($0.2 R_v$),
  and is comprised of a minimum of 2 components. A lower limit for the
  integrated column density of $\log N$(H~I)$\:\geq\:15.7$ is robust, but
  the upper limit depends on whether a third component also exists, which
  is defined by the presence of a single C~II component and no other metal
  lines. For a 2 component model, we establish $\log N$(H~I)$\:\leq 17.1$,
  but if a third component exist, we cannot rule out a column density as
  high as $\log N$(H~I)$\la 18.1$. It is therefore possible that the
  absorption along this sightline is equivalent to a higher redshift Lyman
  limit system. This may also be true for the second inner-most sightline,
  2dFGRS~S393Z082 ($\rho \simeq 0.3\:R_v$), because although the \lya\ line
  appears to consist of a single component, a high-$N$(H~I)/low-$b$ profile
  fit is indistinguishable from a low-$N$(H~I)/high-$b$ solution.

\item As noted above, Si~IV is detected only along the inner-most
  sightline, with $\log N$(Si~IV)$=12.9$. Beyond this, Si~IV is absent
  at a level of $\log N$(Si~IV)$\simeq\: 12.7-13.1$,  implying a small
  covering fraction at these limits beyond 48~kpc ($0.2 R_v$). 

\item There is only marginal evidence that  $N$(H~I) and $N$(Si~III)
  decline with galactocentric radius, and the distribution of gas is clearly
  inhomogeneous within the CGM. We have fitted a power-law to the
  decline in column density with impact parameter, and for $W$(\lya )
  and $\log N$(Si~III), the errors in the fit show that there could be
  no anti-correlation at all. An anti-correlation in $W$(Si~III) with
  $\rho$ appears more plausible. In all cases, the CGM appears clumpy with
  respect to any possible smooth decline with $\rho$, 
  and we characterize the clumpiness to be more than
  $\sim 0.2$ dex for an assumed power-law decline of $W$ and $N$ with
  $\rho$ for both \lya\ and Si~III. 

\item We have modeled the observed column densities and column density
  limits using {\tt CLOUDY}, assuming that the gas is
  photoionized. There are considerable uncertainties in the results,
  either because $N$(H~I) is not well known and the SED of NGC~1097 is
  difficult to account for --- as in the case towards Q0244$-$303 ---
  or because few metals are actually detected (as towards the outer
  sightlines). The {\it simplest} solutions suggest that the gas has
  metallicities that are sub-solar, if the gas is dust-free. 

\item We have attempted to replicate the observed absorption lines using a
  suite of models for the CGM around NGC~1097, including gas that is rotating
  in a huge extended disk, inflowing from the IGM, or outflowing from the
  galaxy. None of the models replicate the observations
  exactly, but one that has a  disk of gas rotating roughly planar to, but
  more slowly, than the inner disk observed from 21~cm data, and which
  also includes gas infalling from the IGM, recreates the data 
  well. The models that provide the poorest predictions of the data
  are those dominated by conical or spherical outflows. The exception
  to this statement arises in modeling the absorption towards
  Q0244$-$303, where the most positive velocity component requires
  additional absorption either from a broad outflow, or from tidal
  debris.  

\item While these models provide a plausible insight into how the CGM
  may be distributed around NGC~1097, it is also possible that some of
  the absorption could arise from ionized gas associated with the
  stellar streams that exist around the galaxy as a result of a
  multiple-bypass minor merger. Gas associated with a cannabalized
  dwarf galaxy may have been stripped during its encounter, but its
  remains may contribute to the observed absorption.

\item In comparison {\it only} to the column densities and equivalent
  widths observed from the CGMs of other galaxies obtained from recent
  COS surveys, the absorption from NGC~1097 seems typical. Given the
  variations in these quantities in the single halo of NGC~1097,
  however, it may be that column density and equivalent width
  measurements are not a particularly sensitive way to disentangle how
  the origin of CGMs depend on the properties of the host galaxies.
  Our data demonstrate the necessity to observe absorption lines from
  the nearest local galaxies where the CGM can be defined a priori
  from multiple techniques, and that multiple probes of single galaxy
  CGMs is now a priority.

\end{enumerate}

\bigskip
\acknowledgments

Big thanks to Gero Rupprecht for help with the calibration of ESO Archive
data, to Katy Oldfield for her expertise with the 3D CAD software
package {\it SolidWorks}, and to Rick Stevenson for permission to use
his image of NGC~1097.
For the work required to identify QSOs behind galaxies, DVB was supported by a
NASA {\it Long Term Space Astrophysics} (LTSA) grant NNG05GE26G, and
by 
NNG05GE31G for the analysis of GALEX Cycle 1 data;
funding for the reduction of the COS spectra obtained from the GO
program was
provided by NASA grant number 12988 from the Space Telescope Science
Institute, which is operated by the Association of Universities for
Research in Astronomy, Inc., under NASA contract NAS5-26555.
The research by DC is partially supported by FP7/IRG PIRG-GA-2009-256434.
Computations presented in this paper have made use of the Hive computer
cluster at the University of Haifa, which is partly funded by ISF grant 2155/15.
 
Extensive use has been made of archival databases in this research:
the {\it SuperCOSMOS Sky Survey} material is based on photographic
data originating from the UK, Palomar and European Southern
Observatory (ESO) Schmidt telescopes and is provided by the Wide-Field
Astronomy Unit, Institute for Astronomy, University of Edinburgh; The
{\it 2dF QSO Redshift Survey} (2QZ) and the {\it 2dF Galaxy Redshift
  Survey} (2dFGRS) was compiled by the 2QZ and sdFGRS survey
teams from observations made with the 2-degree Field on the
Anglo-Australian Telescope (AAT); the {\it NASA/IPAC Extragalactic
  Database} (NED) is operated by the Jet Propulsion Laboratory,
California Institute of Technology, under contract with the NASA;
and the SIMBAD database is
operated at the {\it Centre de Donn\'{e}es astronomiques de
  Strasbourg} (CDS) in Strasbourg, France.

\bibliographystyle{apj}
\bibliography{bib5}

\clearpage


\input{main_tables2}


\clearpage
\appendix

\section{Notes on Individual QSOs and GALEX Grism Observations}
 \label{sect_galex}

As noted in \S\ref{sect_targets}, we observed NGC~1097 
as part of a GALEX Cycle 1 Guest
Investigator program (Program 085) designed to
obtain FUV grism spectra of UV-bright objects in nearby galaxy fields
with the aim of discovering new extragalactic sources. In this
Appendix we summarize these observations, as well as remark on the
original discoveries of QSOs in the field of NGC~1097 which made the
observations discussed in this paper possible.

Many of the 
largest angular-diameter galaxies in our original list of target
fields were to be
observed as part of the GALEX {\it Nearby Galaxy Survey} (NGS)
\citep{paz07}. This made it possible to add grism observations as part of
a GALEX SNAP program without needing to acquire long UV images
beforehand. We observed NGC~1097 for 1786 sec on 26-Oct-2004 as part
of this program, and a total of 117 grism spectra were extracted with
good quality data in both FUV and NUV channels. Of these, 80\% showed
spectra with very irregular continua, decreasing to zero flux
below 2000~\AA , indicative of the spectral energy distribution of
stars. Of the remaining objects with discernable FUV flux, nine
already had published identifications, leaving three
un-identified. These three unfortunately showed no discernable
emission line features and could not be added to our list of potential
targets. Hence for this field, we were not able to discover any new
UV-bright AGN or QSOs near NGC~1097.

Even though our selection of QSOs behind nearby galaxies was
constructed by first selecting UV-bright objects from GALEX data
(\S\ref{sect_targets}), our
eventual choice of targets relied on them having extant
redshifts. While NED catalogs the majority of
redshifts from all published redshift surveys, it turns out that for
NGC~1097, much of the redshift information comes
from surveys conducted three decades ago by \citet{wolstencroft83} and
\citet{arp84}. The apparent over-density of X-bright QSOs in the
field, and a hypothesized alignment of some QSOs with the galaxy's
jets, made this field of particular interest to the latter author
\citep{arp07_n1097, arp_sr43}. Several redshifts were measured for
objects in the field of NGC~1097 as part of that research.

\vtwo\ was originally discovered as a $z=0.528$ X-ray source by
\citet{wolstencroft83}, (their ``Q1097.2''); \citet{arp84} labelled this as
``\#27'' in a follow-up paper. It was later recovered by the 2dF QSO
Redshift Survey \citep{croom04} as 2QZ J024649.8$-$300742, which
confirmed the original detection of Mg~II in emission, as well as
several Balmer emission lines. We detected C~IV emission at the
correct wavelength in our GALEX grism data.

\vfour\ was detected as part of the {\it 2dF Galaxy Redshift Survey}
(2dFGRS) \citep{colless01}, where broad, weak H$\beta$, H$\gamma$ and
Mg~II were clearly detected at $z=0.339$. Equally narrow and weak
\lya\ emission was also detected in our GALEX grism data at the
expected wavelength.

\vsix\  was discovered as an X-ray bright object by \citet{arp84}, their
``\#35''. Their redshift of $z=1.09$ was determined unambiguously from
optical data, and we detected strong and broad \lya\ emission in our
GALEX grism data. 

\vseven\ was discovered by \citet{wisotzki00} as part of the
Hamburg/ESO survey for bright QSOs.   
We obtained no GALEX grism spectra for the QSO, since
it lies just outside of the GALEX field centered on NGC~1097.

\vthree\ was identified as an AGN in the 2dFGRS with $z=0.131$, and we
selected the target based on the cataloged redshift and measured FUV
magnitude.  No spectrum was supplied in the original grism catalog produced
after the GALEX data were taken, but more recent GALEX Data Releases now
provide a grism spectrum for the object. It shows the flux rising
towards the FUV at levels predicted from the FUV magnitude, but is
featureless. After obtaining the HST COS spectrum, it became apparent
that the object is not extragalactic, and is likely to be a local
White Dwarf star, based on the extremely strong damped \lya\
absorption profile at $z=0$ and the weakness of any metal lines that
would normally be expected from a sightline passing through the Milky
Way. A retrospective examination of the optical spectrum from the
2dFGRS database shows that no emission or absorption lines were
detected, making the original redshift measurement erroneous.

One final note: \citet{wolstencroft83} found a relatively bright X-ray
QSO, J024645.60$-$300051.9 (their ``Q1097.3''), with $z=1.00$. We
recovered this QSO independently from our GALEX grism data, with \lya\
emission clearly visible at the expected redshift. However, there is
no FUV flux below a strong discontinuity at 1900~\AA , which could be
due to Lyman limit absorption at the redshift of the QSO itself.  For
this reason, this QSO was never a potential HST COS target.

\section{Notes on NGC~1097B}
\label{sect_n1097b}

Figures~\ref{fig_pretty} and \ref{fig_map} also show the position of
NGC~1097B.  This faint galaxy was reported by \citet{higdon03} to have a
21~cm velocity of 1015$\pm11$~\kms. (The position of the galaxy appears to
be mis-reported in their Table~1 but the position discussed in their text
is only 13$"$ from the optical position of the 2dFGRS source.)  It was also
observed as part of the {\it 2dF Galaxy Redshift Survey} (2dFGRS),
designated S393Z052 with $b_J=18.1$, and assigned a redshift of $z=0.0645$
($\simeq 19,300$~\kms), but there is little evidence from the galaxy
spectrum that this is correct.  Indeed, although the spectrum is of low
signal-to-noise (S/N), there is better evidence for a 4000~\AA\ break and,
perhaps, H$\alpha$ emission, close to the 1015~\kms\ given by
\citet{higdon03}. The galaxy has an absolute magnitude $M(b_J) = -12.5$ if
at the redshift measured by Higdon \& Wallin, reminiscent of the
ultra-compact dwarf galaxies (UCDs) found by \citet{phillipps01} in the
Fornax cluster, but NGC~1097B does not look as compact as those UCDs.  It
is possible that NGC~1097B constitues a third satellite of NGC~1097,
although more accurate redshift information is obviously needed. There is
no evidence for any absorption lines at the velocity of 1015~\kms\ and
there is no reason to suppose that the galaxy contributes directly to the
detected absorption. It would, nevertheless, be of interest to determine if
this is indeed another dwarf galaxy in the halo of NGC~1097.

\section{Notes on photoionization calculations for the absorption
  systems}
\label{sect_photoU}

All of the fits in Tables~\ref{tab_fits_v20}$-$\ref{tab_fits_v70} are
based on allowing the $b$-values of individual species to vary
independently of one another. (The only parameter in common for all
the lines are their fixed velocities.) 
An alternative approach might be to fit all the species 
demanding that the same Doppler parameter be used, an
approach which would be acceptable if all the line widths were
dominated by non-thermal processes. However, at certain
temperatures, the widths of \lya\ lines, and the widths of 
metal lines, can be dominated 
by different processes: it is well known that the width of a line is usually represented by
summing, in quadrature, the width expected from thermal broadening by
gas at a temperature $T$, $b_{\rm{therm}}$, with a width that
describes a turbulent velocity $b_{\rm{turb}}$, such that the total
width in \kms\ is $b^2 = b_{\rm{therm}}^2 + b_{\rm{turb}}^2$, or for an ion of
atomic mass $A$,

\begin{equation}
b^2 = \frac{2 k_B\,T}{A\,m_p} + b_{\rm{turb}}^{2} ,
\label{eqn_doppler}
\end{equation}

--- assuming that the turbulent width can be adequately described by a
Gaussian. (Here, as usual, $k_B$ is the Boltzmann constant and $m_p$ is the mass of
the proton.) Hence if the kinetic temperature of  H~I is, e.g., $3000$~K,
the thermal width of a \lya\ line is $7$~\kms,  
while for metals such as, e.g., silicon, $b_{\rm{therm}} $ is only $\sim 1$~\kms .
So while the thermal widths of the metal lines are largely unresolved at
these temperatures in COS data, the thermal width of H~I can be
a significant fraction of the total line width as $T$
increases.

\begin{figure*}
\centering
\includegraphics[width=16cm]{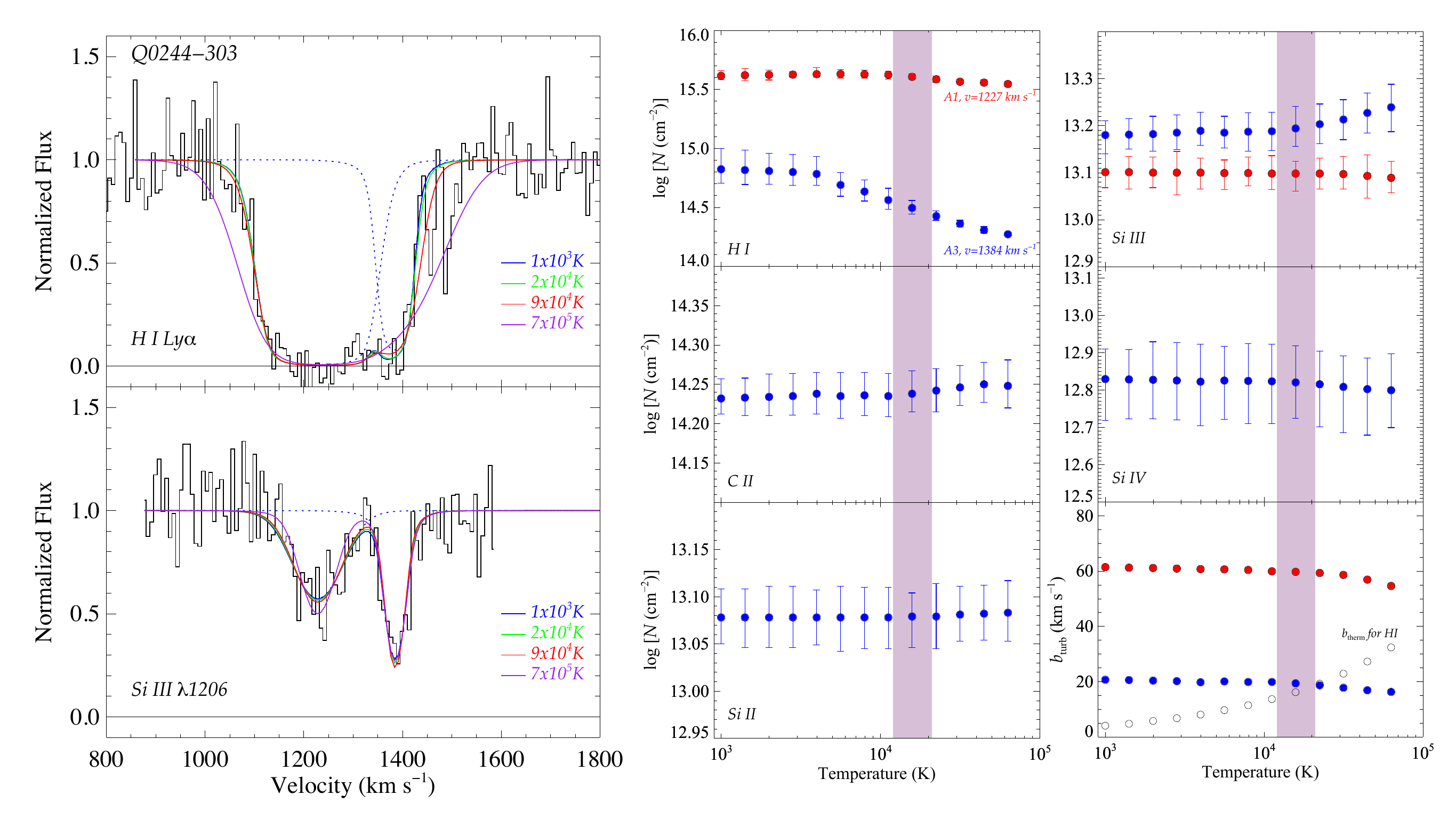}
\caption{ Voigt-profile fits to the lines towards \vtwo , assuming
  that the gas has a range of thermal temperatures. {\it Left:}
  Simultaneous fits
  to both the \lya\ and Si~III~$\lambda 1206$ lines for a two component model
   when the $b$-value is determined by a combination of turbulent
   and thermal broadening, with the thermal component \btherm\ fixed by
   the four indicated temperatures $T$. (The thermal widths are by
   necessity assumed to be the same for both components.) Although
   only two species are shown here, the profiles are a result of
   simultaneous fits to all available lines. The dotted profiles indicate
   the individual components A1 and A3 that comprise the fit for
   \btherm$=1\times 10^3$~K and are shown only to highlight the assumed 2-component
   nature of the line. Fits to the high-velocity weak \lya\ lines (A4
   \& A5 in Fig.~\ref{fig_stack_v20}) are not shown here.
  The profiles demonstrate that the absorbing H~I cannot be at temperatures
  $T \ga 10^5$~K.  {\it Right:} Variations in the column densities when
  \btherm\ --- and hence \bturb --- changes with $T$. Red points
  are for component A1 and blue points are for A3. The metal column
  densities are largely insensitive to different temperatures, but
  $N$(H~I) may change by $\sim 1$~dex for A3, making it difficult to reliably
  constrain the physical properties of the system. The purple-shaded
  regions mark the temperature range where the {\tt CLOUDY}
  predictions discussed in \S\ref{sect_As}
  match the observed values. {\it Bottom-right:} the deduced turbulent
  velocities of the gas are shown for both A1 and A3, while the
  thermal speeds for H~I are shown as black circles.\\
\label{fig_photo_fit}}
\end{figure*}

While $b_{\rm{therm}}$ and $b_{\rm{turb}}$ cannot be determined
independently, we can make a series of fits to the H~I and metal lines
together, 
assuming a set of temperatures which fixes $b_{\rm{therm}}$ for each
line, and which lets $b_{\rm{turb}}$ vary, while always constructing a total
width from Equation~\ref{eqn_doppler} (which is used for fitting the
observed profile). This yields a grid of
column densities and values of  $b_{\rm{turb}}$ for each
$T$ (or \btherm ). Uncertainties on each model can be calculated using the same
Monte-Carlo procedures described in \S\ref{sect_measure}. 
An example of the results from this multiple-$T$ fitting is shown in
Figure~\ref{fig_photo_fit} for both components of A1 and A3 towards \vtwo .
This method enables us to better understand the range of column densities
--- particularly $N$(H~I) in this complicated system --- that are
likely causing the absorption. The only drawback is that we need to
assume that the temperature of the two components (i.e. the \btherm\
values) are the same. This assumption need not be true, but
so long as \btherm\ is not considerably different for the two
components, the results remain largely valid.

Further, 
we can use these results to construct models for
photoionized gas of density $n$ and metallicity $Z$ for each value of
$T$. We do this using the {\tt CLOUDY} software package
\citep[][version C08.01]{ferland13}. The analysis assumes 
all abundances are relative to the total hydrogen density and are
scaled to each other in a solar abundance pattern; the solar values for carbon and
silicon come from \citet{prieto02} and \citet{holweger01}, respectively. 
Unless otherwise stated in the sections that follow, the models are dust-free. 
We adopt a UV background based on the meta-galactic field
  of \citet[][hereafter HM]{haardt96, haardt12}, but discuss the possible contribution from
  NGC~1097 itself when appropriate.
The total hydrogen column
density $N$(H) is obtained by scaling the solution to the measured
$N$(H~I) at that temperature, and the absorber scale-length
$l = N$(H)$/n$ is deduced.

In the following sections we discuss the results of this analysis for
the metal line components toward three of the sightlines.
For \vseven, the two absorption components D1 and D2 
(Table~\ref{tab_fits_v70} and Fig.~\ref{fig_stack_v70}) are only
detected in H~I, and constraints on their physical properties are
of little use. Hence no results are reported for these.

\subsection{Absorption System Towards \vtwo \label{sect_As} }

\begin{figure*}
\includegraphics[width=17cm]{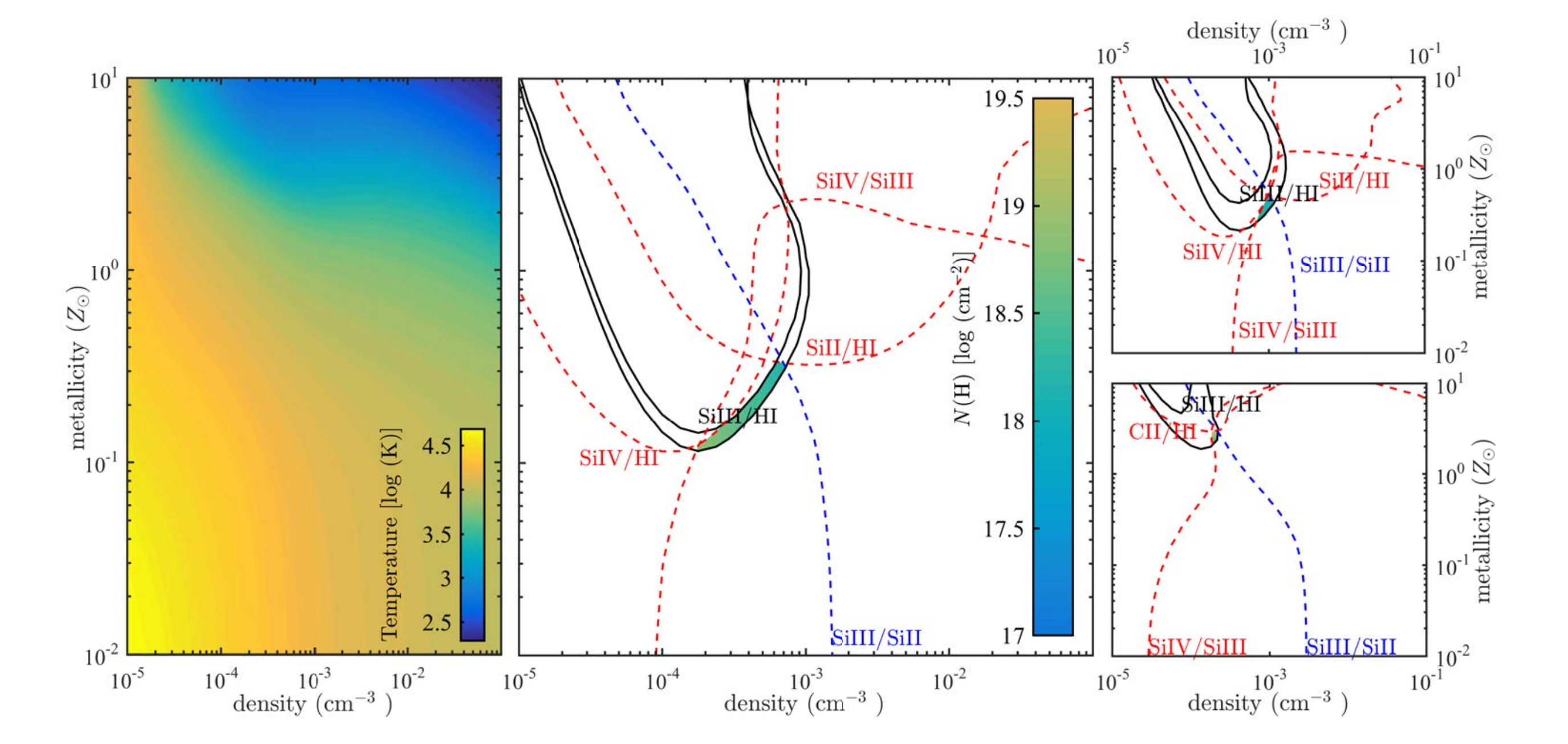}
\caption{ Photoionization models for component A1 towards \vtwo .  {\it
    Left:} variations in temperature for different densities and
  metallicities, $T= T(n,Z)$, for photoionized, dust-free, optically thin
  gas exposed to the metagalactic background radiation field under
  non local thermodynamic equilibrium (non-LTE) conditions.  {\it Middle:}
  the phase-space of $n$ and $Z$ allowed by the different column density
  ratios.  Solid curves mark ratios for which measured column densities
  were available for both ions, plotted for $\pm1\sigma$ variations as a
  result of the errors in the column densities. Dashed curves mark ion
  ratios for which only upper limits are available for one of the ions:
  upper limits to $n$ and $Z$ are marked in red while lower limits are
  shown in blue. The colored shaded region near the center marks the phase space that satisfies
  all the constraints, with the colors representing the total H column density
  (shown by the inset color-bar).  For simplicity, only ion ratios that
  provide useful constraints are plotted. {\it Right:} the upper panel shows
  models for which the contribution of flux from NGC~1097 is included (see
  text) while the lower panel shows the effects of including Milky Way
  dust-to-metals mixtures (the latter yielding no solution). The colors of
  the lines in these panels are the same as those in the middle panel.
  \label{fig_A1}}
\end{figure*}

This system is the most complicated of the four studied. Our analysis is
restricted to the A1 and A3 systems listed in Table~\ref{tab_fits_v20} and
labelled in Figure~\ref{fig_stack_v20}. Component A2 has no unique H~I
column density (and may not be real) while components A4 and A5 show only
weak \lya\ and no metal lines.
Figure~\ref{fig_photo_fit} shows that $N$(H~I) for component A1
remains largely constant for all values of \btherm , $\log
N$(H~I)$\:\simeq 15.6$. 
For component A3, $\log
N$(H~I)$\: = 14.8$ for $T \sim 5\times10^3$~K but decreases at higher
temperatures. The metal-line column densities remain constant
for all values of \btherm . 
We should emphasize, however, that this apparently well-behaved change
in $N$(H~I) with temperature is strongly dependent on a likely
over-simplified 2 component model:
obviously, it is always possible that the entire \lya\ complex
could be made up of many more high-$N$(H~I)/low-$b$ components that 
are blended together, and which we are unable to
separate.

\subsubsection{Photoionization models for Component A1 \label{sect_A1}} 

A single-phase solution (i.e.\ a single cloud with a uniform density) for
this component which satisfies all the observational constraints is
obtained for $n \la 7\times10^{-4}$~cm$^{-3}$ and $Z \la 0.3\: Z_\odot$,
which yields $\log N$(H)$\:\ga\: 18$ (Fig.~\ref{fig_A1}).  Taking the
extreme values of this solution, the resulting length-scale is 0.5 kpc
$\la l \la 10$~kpc, which implies that the gas occupies only a small
fraction of the volume as $l/\rho < 0.2$.

The contribution of NGC\,1097 to the flux ionizing the absorbing gas along
this sightline may not be negligible, given the relatively small impact
parameter of \vtwo . Assessing the galaxy's contribution is difficult as
the UV and X-ray flux that the absorbers see may depend on the galaxy's
inclination relative to where the clouds are, on their distance and
orientation to star-forming regions in the disk, and on any contribution
from the central AGN (all of which are highly uncertain).  The spectral
energy distribution (SED) of the UV background, which is dominated by
galaxies at $z=0$, will likely have, to first order, a similar shape to
that of NGC~1097, so the contribution from the galaxy can be accounted for
by simply scaling-up the background flux level. In this case, a solution is
obtained for the same ratio of photon-to-particle density, and the allowed
density range scales with the flux level (by roughly +0.6 dex given the
luminosity of NGC~1097). 

Alternatively, we can at least qualitatively
assess the effects of assuming that the SED below the He~II edge is similar
to that of the Milky Way \citep{fox05}, by using a flux below the He~II
edge that is twice as high as that given by the HM SED (and which includes
a contribution from both quasars and galaxies); the results are shown in
Figure~\ref{fig_A1}, where a single-phase solution can only be found if the
upper limit to the Si~IV column density is under-estimated by 0.1 dex.  The
resulting model then has $n\sim 10^{-3}$~cm$^{-3}$,
$Z\sim\: 0.5 \:Z_\odot$, $\log N$(H)$\:=18.0$ ($l\sim 0.3$~kpc), which
implies slightly higher gas metallicities and slightly more compact
absorbers.

We also consider the possibility that the elements might be depleted onto
dust grains. Adopting any depletion patterns that are similar to the Milky Way is
difficult, since these are known to differ widely for, e.g. disk and halo
gas \citep[e.g.][]{ARAA_ghrs}.  {\tt CLOUDY} offers a compromise of using
an average of values from both warm and cold gas, and we use these to see how our
calculations might change with dust included. Ignoring any contribution to
the UV flux from NGC~1097, using all the available column density ratios and
limits, there is no single-phase solution. 
Only if
we raise the column density upper limits by a factor of 2, and use $\pm
2\sigma$ intervals around the measured values, do we find a solution that
has $Z \sim 3\: Z_\odot$ and  $n \ga1\times10^{-4}$~cm$^{-3}$. 
We consider it unlikely, however, that the errors and upper limits for the metal
lines in A1 are so badly determined by our analysis,  and we reject these values.

\subsubsection{Photoionization models for Component A3 \label{sect_A3}}

\begin{figure}
\centering
\includegraphics[width=8cm]{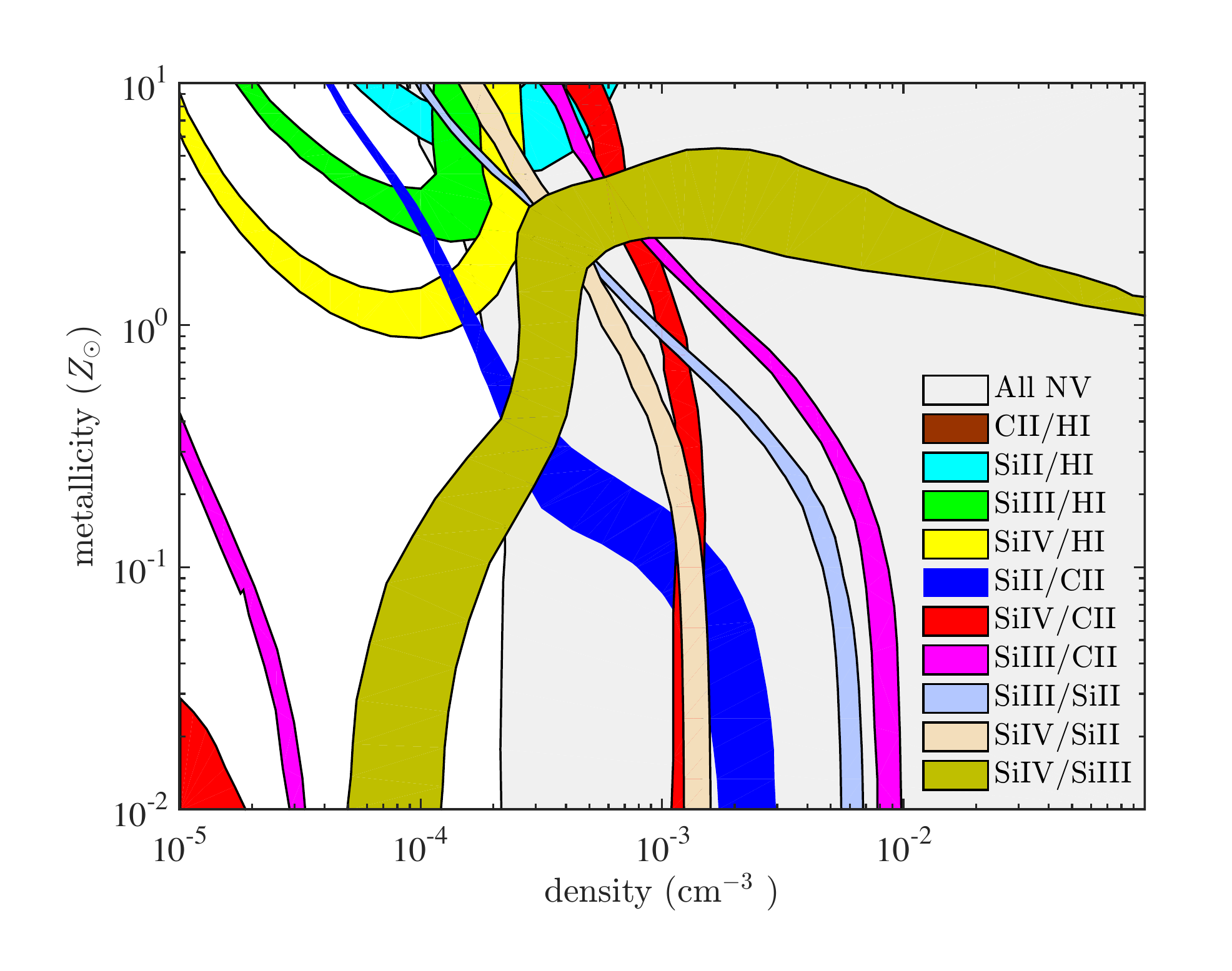}
\caption{Allowed $n,Z$ phase-space regions set by the different ion
  ratios for the A3 system towards \vtwo . The inset legend 
  shows how individual ion ratios are
  colored in the figure. 
  No single-phase solution  solution exists for the
  model because no region exists where all the colored regions
  intersect. 
  The C~II/H~I ratio requires $Z>10\: Z_\odot$ and so is not
  shown at the scales used for this plot (i.e. $10^{-2} \: Z_\odot < Z < 10\:
  Z_\odot$). Possible reasons for the poor agreement between model and
  observations include: {\it a}) that $N$(H~I) is
  poorly determined, {\it
    b}) that the relative metal abundances are different from the solar
  values, and/or {\it c}) that multiple thermal phases are responsible for the
  absorption. If we consider only the silicon lines, a model with
  slightly super-solar abundances and densities of $\ga
  10^{-4}$~cm$^{-3}$ is implied.\\
\label{fig_A3}}
\end{figure}

The analysis of this component yields {\it no} single-phase solution that
can simultaneously account for all the observational constraints. 
The phase spaces for many line ratios are 
shown in Figure~\ref{fig_A3}, and we see that none of the regions
intersect. Considering the $N$(C~II)/$N$(H~I) ratio alone (for example),
the phase space available is completely ruled out because 
the allowed area is completely outside the region plotted
in Figure~\ref{fig_A3} (unless $Z>10\:Z_\odot$). As discussed in
\S\ref{sect_v20fits}, $N$(H~I) is poorly constrained for this component
(although apparently less so than A1), which certainly adds to the
difficulty in finding a suitable solution. Nevertheless, even allowing for a 
 wide range of $N$(H~I), there is no single-phase
solution which accounts for all the {\it metal line ratios} alone
(Fig.~\ref{fig_A3}).

This problem remains even if additional
flux is added from  NGC~1097, or if the absorbing gas is
assumed to be depleted by dust like the Galactic ISM, and it persists
even when pure collisional heating is considered instead of
photoionization. The difficulties can be seen most acutely when
considering low-ionization line ratios [in particular the
$N$(Si~II)/$N$(C~II) ratio, which seems to be more shifted towards
lower densities compared to the other line ratios], 
which suggests that the problem is not simply because we have searched
for a single phase to explain all the line ratios simultaneously.
For the particular case of $N$(Si~II)/$N$(C~II), if
carbon is under-abundant by a factor of $\sim 3$ compared to the solar
value, then the phase-space traced by this ratio is in better
agreement with those of other metal-line column density 
ratios\footnote{Note that our photoionization models are not strictly
  self-consistent in this case because we assume that
the relative abundances of the ions are in solar proportions.
 Nevertheless,
  since such models cannot be reliably constrained by the current data
  we do not consider deviations of relative metal abundances from
  solar composition in this work}.

We can further attempt to constrain the physical conditions of the absorber
by narrowly focussing on only the silicon lines, for which three ionization
levels are observed: the gas density is then restricted by
$N$(Si~IV)/$N$(Si~II) to be $
10^{-4}$~cm$^{-3}
\la n \la 10^{-3}$~cm$^{-3}$
over nearly three decades in metallicity. Constraints for all silicon
column density ratios yield a single-phase solution at $2\times 10^{-4}$
cm$^{-3} < n < 3\times 10^{-4}$~cm$^{-3}$ with
$1.4\: Z_\odot < Z < 3\: Z_\odot$. The total hydrogen column density
depends on the uncertain $N$(H~I), but is in the range $17.3 < \log
N$(H)$
\:< 19.5$,
which gives a scale-length $l\sim 0.2-50$~kpc. 

In conclusion, the physical parameters of the A3 absorbing cloud remain far
from clear. It is possible, for example,  that the gas is not in equilibrium. In that
case, a more detailed treatment of this system would require additional
data.

\subsection{Absorption System Towards \vfour \label{sect_Bs} }

This sightline shows a single system B1 (Table~\ref{tab_fits_v40} and
Fig.~\ref{fig_stack_v40}), and the results of our analysis is shown in
Figure~\ref{fig_otherU} 
(left panel). 
A solution is obtained within a range
of $\pm 0.5$ dex around values centered at
$n \sim 3\times 10^{-4}$~cm$^{-3}$, $Z\sim 0.1 \:Z_\odot$, and
$\log N$(H)$\:\la 19$. This gives $l \sim 10$~kpc ($l/\rho \sim 0.1$) with
deviations of up to $\pm 1$~dex of that value. At $\rho = 84$~kpc, the
contribution of any flux from NGC~1097 is smaller than it is for \vtwo\
(see above), which itself led to only small changes in metallicity and
density for plausible changes in the SED.  Assuming a Galactic-like ISM
gas-to-dust mixture results in a solution with similar densities and total
column densities, but with a metallicity increased to $\sim 0.5 \:Z_\odot$.

\subsection{Absorption System  Towards \vsix \label{sect_Cs}}

This sightline shows two closely blended components C1 and C2
(Table~\ref{tab_fits_v60} and Fig.~\ref{fig_stack_v60}), and a third
more kinematically distinct component C3. Only C2 shows metal line
absorption, so we only discuss our results for that component.

\begin{figure}
\includegraphics[width=8cm]{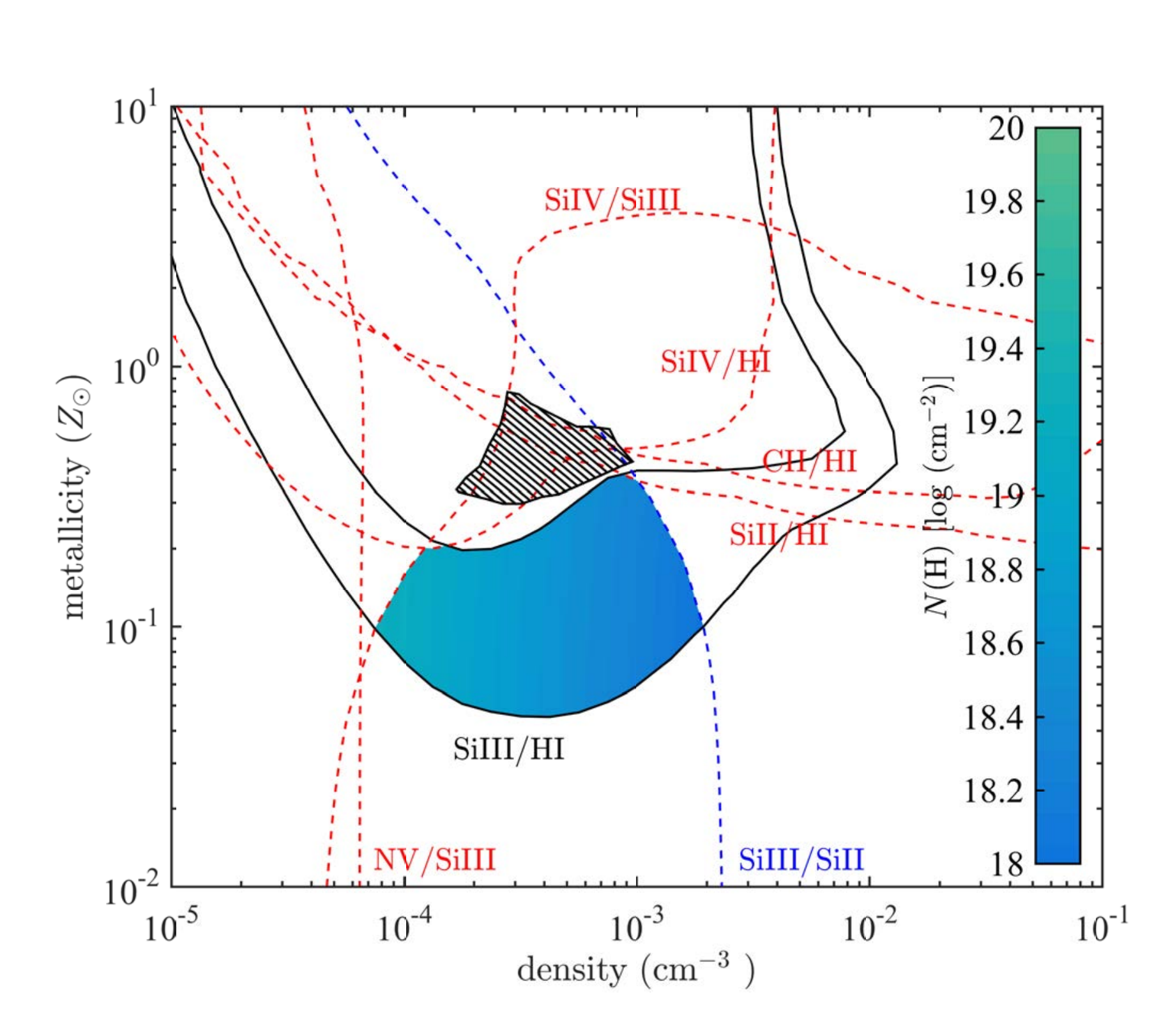}
\includegraphics[width=8cm]{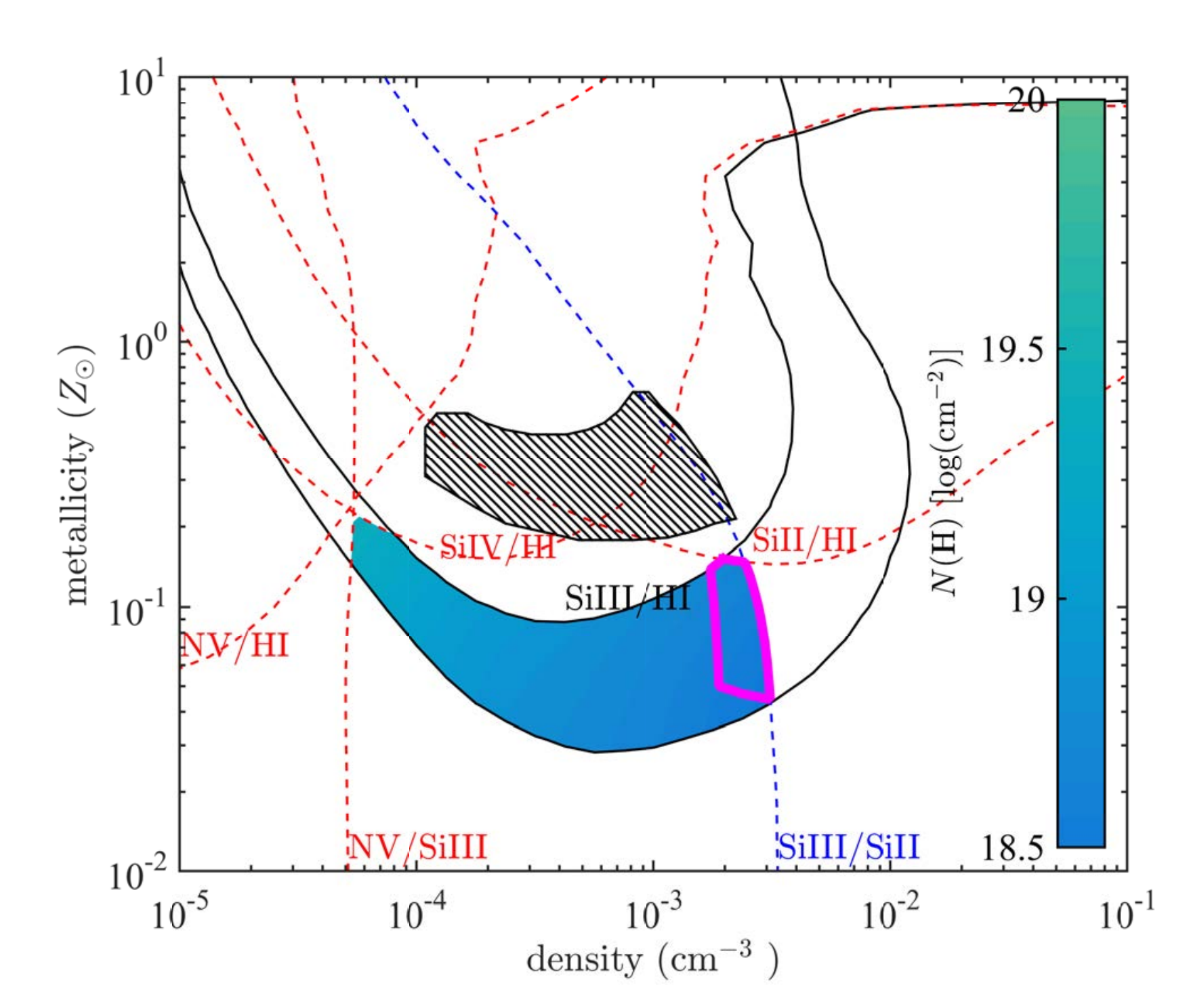}
\caption{Allowed phase-space for the absorption system B1 towards
  \vfour\ 
(left) and C2 towards \vsix\ (right). 
(See  Fig.~\ref{fig_A1} for details.) The hatched surfaces correspond to
  the allowed phase space when dust is included (see text). \\
\label{fig_otherU}}
\end{figure}

The solutions we obtain for this system depends on whether the detection of
the C~II line is real (\S\ref{sect_v60fits}). Without including this line,
we obtain a solution
with $5\times 10^{-5}$~cm$^{-3}$ $ < n < 3\times 10^{-3}$~cm$^{-3}$
and $0.02 \:Z_\odot < Z < 0.2\: Z_\odot$ 
(Fig.~\ref{fig_otherU}, right panel), 
and a total column density in
the range $3 \times 10^{18}$~\pcm $< N$(H)$\: < 3\times 10^{19}$~\pcm,
implying $0.3$~kpc~$< l < 300$~kpc. 
When
including the ISM-like gas-to-dust patterns used in {\tt CLOUDY}, the gas is found to
have a similar density range, but with $0.2\: Z_\odot < Z < 0.6\:
Z_\odot$, and $\log N$(H)$\:\ga 19.0$. As discussed above, there is a wide
variation in the depletion patterns that exist in the Galactic ISM: our
analysis shows merely that for depletions that are at least similar to the
Galaxy, the deduced metallicity still appears to be sub-solar. 

If we assume that the C~II line is real, then a single phase solution is
only found when the confidence levels are increased to 3$\sigma$ for C~II. The
allowed region in shown in magenta in Fig.~\ref{fig_otherU}. The deduced
metallicity changes little, but the density is confined to a much smaller
range.

\end{document}

%% file: main_tables2.tex
\clearpage

\begin{deluxetable*}{c l r r lcclc}
\tabletypesize{\scriptsize}
\tablecolumns{9}
\tablecaption{Equivalent widths and column densities towards v20, \vtwo \label{tab_fits_v20} }
\tablehead{
\colhead{ }          & \colhead{$v$}     & \colhead{$\Delta v$}      & \colhead{$\sigma(v)$}   & \colhead{$W^b$}  & \colhead{$b^b$}    & \colhead{$\sigma(b)_T$} & \colhead{}              & \colhead{}  \\
\colhead{ID$^a$} & \colhead{(\kms)}  & \colhead{(\kms )}          &\colhead{(\kms )}          & \colhead{(\AA)}     & \colhead{(\kms)}    & \colhead{(\kms)}          & \colhead{$\log(N)$} & \colhead{$\sigma(N)_T$}  \\
\colhead{(1)}        & \colhead{(2)}      &  \colhead{(3)}                & \colhead{(4)}               & \colhead{(5)}          & \colhead{(6)}          & \colhead{(7)}               & \colhead{(8)}          & \colhead{(9)}
}
\startdata
\hline
\\
                     &           &    &       &  \multicolumn{5}{c}{H~I \lya } \\
\cline{5-9} \\ 
A1                  & 1227   &  $-44$  & 11      &   \phm{100}$^{-}|$         &    62.7        &    $-10.2,+8.7$    &  \phm{$<$}15.57     & $-0.23,+0.55$       \\ 
A2                  & 1261   &  $-10$  & 5        &   \phm{$<$}1.316         &  \multicolumn{4}{c}{$--$$--$$--$$--$$--$$--$ no unique fit $--$$--$$--$$--$$--$$--$} \\
A3                  & 1384   &  +113   &  2       &   \phm{100}$_{-}|$         &   18.9         &   $\pm 2.0$        &  \phm{$<$}15.07     & $-0.32,+0.24$   \\
A4                  & 1456   &  +185   &  2       &   \phm{$<$}0.035         &    $\leq$9   &   ...                      &  $\geq$13.10   &   ...               \\
A5                  & 1491   &  +220   &  3       &   \phm{$<$}0.045         &    $\leq$9   &   ...                      &  $\geq$13.10   &   ...                 \\
\cline{5-5}\cline{8-9}
total [p]            &            &   &        & \phm{$<$}$1.385 +0.035 -0.058$    &                  &                          &  $\geq$15.69$^c$          &  .. \\ 
total [$N=90$]  &            &    &      & \phm{$<$}$1.365 +0.095 -0.109 $    &                  &                          &                                  &  .. \\ 

\\

                     &            &    &              &  \multicolumn{5}{c}{C~II~$\lambda 1334$}  \\
\cline{5-9} \\
A1                  & 1227   &   $-44$   & 11      &  \phm{$<$}\phn...        & (52)                & ...                &  $<$13.9$^d$        & ...                    \\ 
A2                  & 1261   &   $-10$   & 5       &  \phm{$<$}$0.062$     & 10.9               & $ -4.7$,+5.3  &  \phm{$<$}13.62     & $-0.08$,+0.12   \\
A3                  & 1384   &   +113    &  2       &  \phm{$<$}$0.200$     & 24.4               & $ \pm3.0$     &  \phm{$<$}14.22     & $\pm0.03$       \\
A4                  & 1456   &   +185    &  2       &  $<$0.044                   & (9)                  &   ...               &   $<$13.46               & ...                    \\
A5                  & 1491   &   +220    &  3       &  $<$0.046                   & (9)                  &   ...               &   $<$13.48               & ...                   \\
\cline{5-5}\cline{8-9}
total [p]            &            &   &       &  \phm{$<$}$0.261\pm 0.027$     &                       &                      &  \phm{$<$}14.31     & $\pm0.03$      \\
total [$N=60]$  &            &    &      &  \phm{$<$}$0.266\pm 0.064$     &                       &                      &     &      \\

\\
                   &            &     &       &  \multicolumn{5}{c}{Si~II$^{e}$}  \\
\cline{5-9} \\
A1                 & 1227   &  $-44$   & 11      &   $<$0.028                  &      (52)         &       ...                 &  $<$12.24              &    ...                    \\ 
A2                 & 1261   &  $-10$   &5        &   $<$0.028                  &      (11)         &       ...                 &  $<$12.29              &    ...                    \\
A3                 & 1384   &  +113    & 2       &   \phm{$<$}0.124        &      16.2        &      $\pm2.4$      &  \phm{$<$}13.09    &  $\pm 0.05$     \\
A4                 & 1456   &  +185    & 2       &   $<$0.028                  &      (9)           &      ...                  &  $<$12.30              &   ...                     \\
A5                 & 1491   &  +220    & 3       &   $<$0.028                  &      (9)           &      ...                  &  $<$12.30              &   ...                     \\
\cline{5-5}\cline{8-9}
total [p]          &           &     &      &   \phm{$<$}$0.124\pm0.013$      &                    &                           &  \phm{$<$}13.09    &  $\pm0.05$          \\
total [$N=30$] &           &     &      &   \phm{$<$}$0.118\pm0.020$      &                    &                           &     &         \\

\\ 
                      &           &     &      &  \multicolumn{5}{c}{Si~III~$\lambda 1206$}  \\
\cline{5-9} \\
A1                   & 1227   &   $-44$     & 11      &   \phm{$<$}0.204       & 51.9            &   $-10.3,+11.4$  &  \phm{$<$}13.08       & $\pm0.14$     \\ 
A2                   & 1261   &   $-10$     &  5        &   \phm{$<$}\phn...     & (11)              &  ...                   &   $<$12.2$^f$\phn  &      ...                  \\
A3                   & 1384   &   +113      & 2       &   \phm{$<$}0.183       & 24.9            &  $\pm5.0$        &   \phm{$<$}13.16     &  $\pm 0.07$    \\
A4                   & 1456   &   +185      & 2       &   $<$0.090                 & (9)               &  ...                   &   $<$13.01                &      ...                  \\
A5                   & 1491   &   +220      &   3       &   $<$0.090                 & (9)               &  ...                   &   $<$13.01                &      ...                  \\
\cline{5-5}\cline{8-9}
total [p]            &            &  &        &   \phm{$<$}$0.387 \pm0.084$          &                   &                        &    \phm{$<$}13.42     & $\pm0.03$          \\
total [$N=60$]  &            &   &       &   \phm{$<$}$0.380 +0.134-0.162$     &                   &                        &         &         \\

\\

                     &           &     &       &  \multicolumn{5}{c}{Si~IV~$\lambda 1393^{g}$}\\
\cline{5-9} \\
A1                  & 1227   & $-44$   & 11      &   $<$0.034                  &    (52)           &     ...                   &  $<$12.60               &    ...                   \\ 
A2                  & 1261   & $-10$   & 5        &   $<$0.032                  &     (11)          &     ...                   &  $<$12.62               &   ...                    \\
A3                  & 1384   & +113    &  2       &   \phm{$<$}0.063        &     30.2         &    $-9.2,+19.2$   &  \phm{$<$}12.89     &  $\pm0.13$        \\
A4                  & 1456   & +185    &  2       &   $<$0.032                  &     (9)            &    ...                    &  $<$12.64               &   ...                     \\
A5                  & 1491   & +220    &  3       &   $<$0.032                  &     (9)            &    ...                    &  $<$12.64               &   ...                     \\
\cline{5-5}\cline{8-9}
total [p]           &            &   &       &    \phm{$<$}$0.063\pm0.020$      &                    &                           &   \phm{$<$}12.89         &    $\pm0.13$       \\
total [$N=30$] &            &    &      &    \phm{$<$}$0.058\pm0.024$      &                    &                           &              &          \\

\\ 
                    &           &     &       &  \multicolumn{5}{c}{N~V~$\lambda 1238$} \\
\cline{5-9} \\
A1                & 1227   &  $-44$  & 11      &    $<$0.028                  &     (52)          &     ...                   & $<$13.13                &      ...                  \\ 
A2                & 1261   &  $-10$  & 5        &    $<$0.028                  &     (11)          &    ...                    & $<$13.18                &     ...                   \\
A3                & 1384   &  +113   &  2       &    $<$0.030                  &     (30)          &     ...                   & $<$13.17                &     ...                  \\
A4                & 1456   &  +185   &  2       &    $<$0.030                  &     (9)            &      ...                  & $<$13.23                &     ...                   \\
A5                & 1491   &  +220   &  3       &    $<$0.030                  &     (9)            &     ...                   & $<$13.23                &    ...                    \\
\enddata
\tablenotetext{a}{''total [p]'' in this column refers to the total equivalent width (EW) and total column densities 
calculated from the blended theoretical line profiles used to derive $N$ and $b$ values for individual components;
''total [$N=$]'' refers to the total EW measured over $N$ pixels (see \S\ref{sect_measure}).   }
\tablenotetext{b}{EW limits are $2\sigma(W)_T$ values measured over 14 pixels. Doppler parameters in parentheses are the values used to covert 
these EW limits to column density upper limits.}
\tablenotetext{c}{Given the lack of information on the $N$(H~I) from component A2, this total H~I column density is likely only a lower limit.}
\tablenotetext{d}{This is a conservative estimate based on blending a plausible line profile with $b=52$~\kms\ with the fitted A2 component, 
and is not derived from an EW upper limit. See \S\ref{sect_v20fits_metals}.} 
\tablenotetext{e}{Equivalent width and column density limits are derived from the Si~II~$\lambda 1260$ line, but fitted $b$ and $N$ values are 
measured using all of the $\lambda \lambda 1190, 1193, 1260$ and $1304$
lines.}
\tablenotetext{f}{This is a conservative estimate based on blending plausible line profiles with the fitted A1 component and is not derived from an EW upper limit; 
column densities higher than this would alter the shape of the A1 component. See \S\ref{sect_v20fits_metals}.}
\tablenotetext{g}{Measurements from fitted lines based only on the Si~IV~$\lambda 1393$ line
  because an O~VI~$\lambda 1037$ line at $z=0.35818$ is
  at the expected wavelength of Si~IV~$\lambda 1403$. }
\end{deluxetable*}


\begin{deluxetable*}{c l rc lcclc}
\tabletypesize{\scriptsize}
\tablecolumns{9}
\tablecaption{Equivalent widths and column densities towards v40, \vfour \label{tab_fits_v40} }
\tablehead{
\colhead{ }                 & \colhead{$v$}      & \colhead{$\Delta v$}  & \colhead{$\sigma(v)$}   & \colhead{$W^b$}  & \colhead{$b^b$}    & \colhead{$\sigma(b)_T$} & \colhead{}              & \colhead{}  \\
\colhead{ID$^a$}        & \colhead{(\kms)}  & \colhead{(\kms)}        &\colhead{(\kms )}          & \colhead{(\AA)}     & \colhead{(\kms)}    & \colhead{(\kms)}          & \colhead{$\log(N)$} & \colhead{$\sigma(N)_T$}  \\
\colhead{(1)}              & \colhead{(2)}       & \colhead{(3)}               &  \colhead{(4)}               & \colhead{(5)}          & \colhead{(6)}          & \colhead{(7)}               & \colhead{(8)}          & \colhead{(9)}
}
\startdata
\hline
\\
                     &         &   &       &  \multicolumn{5}{c}{H~I \lya } \\
\cline{5-9} \\ 
B1                   & 1240  & $-31$ & 8    &   \phm{$<$}$0.594\pm0.036$                         &    \multicolumn{2}{c}{$19.1 \leq b \leq 46.2$}   &  \multicolumn{2}{c}{\phm{13}$17.60 \geq \log N \geq 14.80$}       \\ 
\cline{5-5}
total [$N=60$]  &     &       &       &  \phm{$<$}$0.603+0.101 -0.119 $     &                  &                          &                                  &   \\  
\\ 
                     &         &    &                      &  \multicolumn{5}{c}{C~II~$\lambda 1334^c$}  \\
\cline{5-9} \\
B1                  & 1240   & $-31$  & 8       &  $<$0.092                                       & (22)            & ...             &  $<$13.8\phn     & ...  \\
\\
                     &        &     &       &  \multicolumn{5}{c}{Si~II~$\lambda 1260$}  \\
\cline{5-9} \\
B1                  & 1240  & $-31$  & 8     &   $<$0.036                                       &      (22)         &       ...                 &  $<$12.37            &    ...                    \\ 
\\

                       &       &     &        &  \multicolumn{5}{c}{Si~III~$\lambda 1206$}  \\
\cline{5-9} \\
B1                   & 1240 & $-31$  & 8     &    \phm{$<$}0.148$\pm 0.023$          & 22.2          &   $\pm5.0$  &  \phm{$<$}13.04       & $\pm0.08$     \\ 
\cline{5-5}
total [$N=30$]   &      &     &        &   \phm{$<$}$0.140 +0.054-0.060$     &                  &                        &         &         \\
\\

                      &          &  &        &  \multicolumn{5}{c}{Si~IV~$\lambda 1393$}\\
\cline{5-9} \\
B1                  & 1240  & $-31$ & 8      &   $<$0.040                                      &    (22)           &     ...                   &  $<$12.69              &    ...                   \\ 
\\ 
                      &           & &        &  \multicolumn{5}{c}{N~V~$\lambda 1238$} \\
\cline{5-9} \\
B1                  & 1240  & $-31$ & 8      &    $<$0.048                                     &     (22)          &     ...                   & $<$13.41                &      ...                  \\ 
\enddata
\tablenotetext{a}{In this column, ''total [$N=$]'' refers to the total equivalent width (EW) measured over $N$ pixels (see \S\ref{sect_measure}).   }
\tablenotetext{b}{EW limits are $2\sigma(W)_T$ values measured over 14 pixels. Doppler parameters in parentheses are the values used to covert 
these EW limits to column density upper limits.}
\tablenotetext{b}{C~II~$\lambda 1334$ lies at the same wavelength as a line likely to be O~VI~$\lambda 1032$ at $z=0.2986$. The EW listed here is 
a $4\sigma(W)$ limit, and the column density is derived from that limit, assuming $b=22$~\kms. Such a line would be well distinguished from the narrow 
O~VI interloper.}  
\end{deluxetable*}


\begin{deluxetable*}{c l r r lcclc}
\tabletypesize{\scriptsize}
\tablecolumns{9}
\tablecaption{Equivalent widths and column densities towards v60, \vsix \label{tab_fits_v60} }
\tablehead{
\colhead{ }          & \colhead{$v$}     & \colhead{$\Delta v$  } & \colhead{$\sigma(v)$}   & \colhead{$W^b$}  & \colhead{$b^b$}    & \colhead{$\sigma(b)_T$} & \colhead{}              & \colhead{}  \\
\colhead{ID$^a$} & \colhead{(\kms)} & \colhead{(\kms )}        &\colhead{(\kms )}          & \colhead{(\AA)}     & \colhead{(\kms)}    & \colhead{(\kms)}          & \colhead{$\log(N)$} & \colhead{$\sigma(N)_T$}  \\
\colhead{(1)}        & \colhead{(2)}      &  \colhead{(3)}             & \colhead{(4)}               & \colhead{(5)}          & \colhead{(6)}          & \colhead{(7)}               & \colhead{(8)}          & \colhead{(9)}
}
\startdata
\hline
\\
                      &    &       &           &  \multicolumn{5}{c}{H~I \lya } \\
\cline{5-9} \\ 
C1                   & 1269   & $-20$     & 11      &   \phm{$<$}0.269         &   34.7        &  $-2.7,+3.1$    &  \phm{$<$}13.93     & $-0.02,+0.03$       \\ 
C2                   & 1342   & +71       & 3        &   \phm{$<$}0.436         &   16.8        &  $-1.6,+4.7$    &  \phm{$<$}16.89     & $-1.39, +0.50$   \\
C3                   & 1488   & +217      & 14      &   \phm{$<$}0.070         &   28.7        &  $-8.6,+11.7$  &  \phm{$<$}13.16     & $-0.10,+0.13$   \\
\cline{5-5}\cline{8-9}
total [p]            &          &  &          &   \phm{$<$}$0.707 +0.056 -0.047$    &                  &                          &  \phm{$<$}16.89          &  $-1.14,+0.35$ \\ 
total [$N=60$]  &           & &          &    \phm{$<$}$0.693\pm0.055$           &                  &                          &                                  &    \\ 

\\

                      &            & &                      &  \multicolumn{5}{c}{C~II~$\lambda 1334$}  \\
\cline{5-9} \\
C1                  & 1269    & $-20$   & 11       &  \multicolumn{5}{c}{blended}  \\
C2                  & 1342    & +71      & 3         &  \phm{$<$}0.063                       & 14.5            & $ -5.7$,+6.6    &  \phm{$<$}13.59     & $-0.07,+0.09$   \\
C3                  & 1488    & +217    & 14       &  $<$0.044                                  & (29)             & ...                   &    $<$13.37             & ...       \\
\cline{5-5}\cline{8-9}
total [p]           &         &    &           &  \phm{$<$}$0.063\pm0.019$            &               &                      &   \phm{$<$}13.59    &  $-0.07,+0.09$    \\

\\
                     &           & &           &  \multicolumn{5}{c}{Si~II~$\lambda 1260$}  \\
\cline{5-9} \\
C1                  & 1269   &  $-20$     & 11       &   $<$0.036                  &      (16)         &       ...                 &  $<$12.36              &    ...                    \\ 
C2                  & 1342   &  +71       & 3         &   $<$0.034                  &      (17)         &       ...                 &  $<$12.36              &    ...                    \\
C3                  & 1488   &  +217      & 14       &   $<$0.040                  &      (29)         &       ...                 &  $<$12.41              &   ...      \\
\\

                       &     &       &           &  \multicolumn{5}{c}{Si~III~$\lambda 1206$}  \\
\cline{5-9} \\
C1                   & 1269   & $-20$    & 11      &   \phm{$<$}0.040       &  15.5            &   $-12.6,+13.5$   &  \phm{$<$}12.33      & $\pm0.24$     \\ 
C2                   & 1342   & +71      & 3        &   \phm{$<$}0.085       &   9.3             &   $-3.8,+5.1$      &  \phm{$<$}12.92      &    $-0.10,+0.56$                  \\
C3                   & 1488   & +217     & 14      &   $<$0.058                 &  (29)             &   ...                     &   $<$12.49               &   ...   \\
\cline{5-5}\cline{8-9}
total [p]            &            &    &      &   \phm{$<$}$0.125 \pm0.029$          &                            &                                &    \phm{$<$}13.02     & $-0.07,+0.43$          \\
total [$N=35$]  &            &    &      &   \phm{$<$}$0.125 +0.055 -0.060$     &                            &                                &                                 &         \\

\\

                      &           &         &   &  \multicolumn{5}{c}{Si~IV~$\lambda 1393$}\\
\cline{5-9} \\
C1                  & 1269  &   $-20$     & 11      &   $<$0.040                  &     (16)         &     ...                   &  $<$12.68               &    ...                   \\ 
C2                  & 1342   &  +71       & 3        &   $<$0.040                  &     (17)         &     ...                   &  $<$12.70               &   ...                    \\
C3                  & 1488   &  +217      & 14      &   $<$0.044                  &     (29)         &     ...                   &  $<$12.73               &   ...        \\
\\ 
                     &            &         &  &   \multicolumn{5}{c}{N~V} \\
\cline{5-9} \\
C1                 & 1269   &  $-20$      & 11      &    $<$0.030$^c$           &     (16)          &     ...                   & $<$13.47               &      ...                  \\ 
C2                 & 1342   &  +71       & 3        &    $<$0.030$^c$           &     (17)          &    ...                    & $<$13.49                &     ...                   \\
C3                 & 1488   &  +217      & 14      &    $<$0.038                  &     (29)          &     ...                   & $<$13.28                &     ...                  \\
\enddata
\tablenotetext{a}{''total [p]'' in this column refers to the total equivalent width (EW) and total column densities 
calculated from the blended theoretical line profiles used to derive $N$ and $b$ values for individual components;
''total [$N=$]'' refers to the total EW measured over $N$ pixels (see \S\ref{sect_measure}).   }
\tablenotetext{b}{EW limits are $2\sigma(W)_T$ values measured over 14 pixels. Doppler parameters in parentheses are the values used to covert 
these EW limits to column density upper limits.}
\tablenotetext{c}{These are the limits from the N~V~$\lambda 1242$ line, as the $\lambda 1238$ is blended with a higher-redshift line at the velocities of 
C1 \& C2.}
\end{deluxetable*}


\begin{deluxetable*}{c l r c lcclc}
\tabletypesize{\scriptsize}
\tablecolumns{9}
\tablecaption{Equivalent widths and column densities towards v70, \vseven \label{tab_fits_v70} }
\tablehead{
\colhead{ }          & \colhead{$v$}     & \colhead{$\Delta v$}    & \colhead{$\sigma(v)$}   & \colhead{$W^b$}  & \colhead{$b^b$}    & \colhead{$\sigma(b)_T$} & \colhead{}              & \colhead{}  \\
\colhead{ID$^a$} & \colhead{(\kms)}  &\colhead{(\kms )}         &\colhead{(\kms )}          & \colhead{(\AA)}     & \colhead{(\kms)}    & \colhead{(\kms)}          & \colhead{$\log(N)$} & \colhead{$\sigma(N)_T$}  \\
\colhead{(1)}        & \colhead{(2)}      &  \colhead{(3)}              &  \colhead{(4)}               & \colhead{(5)}          & \colhead{(6)}          & \colhead{(7)}               & \colhead{(8)}          & \colhead{(9)}
}
\startdata
\hline
\\
                      &           &       &  &  \multicolumn{5}{c}{H~I \lya } \\
\cline{5-9} \\ 
D1                   & 1219   & $-52$      & 4      &   \phm{$<$}0.087         &   16.0        &  $-6.6,+11.5$     &  \phm{$<$}13.34     & $-0.13,+0.22$       \\ 
D2                   & 1309   & +38        & 2      &   \phm{$<$}0.179         &   19.7        &  $\pm4.3$          &  \phm{$<$}13.82     & $\pm0.04 $   \\
\cline{5-5}\cline{8-9}
total [p]           &            &       & &   \phm{$<$}$0.266 \pm0.050$  &      &                          &  \phm{$<$}13.95      &  $\pm0.04$ \\ 
\\

                      &            & &                     &  \multicolumn{5}{c}{C~II~$\lambda 1334$}  \\
\cline{5-9} \\
D1                  & 1219   & $-52$  & 4       &  $<$0.064                      & (16)             &  ...         &    $<$13.59     & ...   \\
D2                  & 1309   & +38     & 2       &  $<$0.060                      & (20)             & ...          &    $<$13.54     & ...       \\
\\
                      &            &         &  &  \multicolumn{5}{c}{Si~II~$\lambda 1260$}  \\
\cline{5-9} \\
D1                  & 1219   &  $-52$      & 4       &   $<$0.024                  &      (16)         &       ...                 &  $<$12.19              &    ...                    \\ 
D2                  & 1309   &  +38        & 2       &   $<$0.024                  &      (20)         &       ...                 &  $<$12.19              &    ...                    \\
\\
                       &           &       &   &  \multicolumn{5}{c}{Si~III~$\lambda 1206$}  \\
\cline{5-9} \\
D1                   & 1219   &  $-52$      & 4      &   $<$0.044                     &  (16)             &   ...        &   $<$12.39      &    ...                  \\
D2                   & 1309   &  +38       & 2      &   $<$0.046                     &  (20)             &   ...         &   $<$12.40      &   ...   \\
\\

                      &           &         &   &  \multicolumn{5}{c}{Si~IV~$\lambda 1402^c$}\\
\cline{5-9} \\
D1                  & 1219   &  $-52$      & 4      &   $<$0.048                  &     (16)         &     ...                   &  $<$13.10              &    ...                   \\ 
D2                  & 1309   &  +38        & 2      &   $<$0.048                  &     (20)         &     ...                   &  $<$13.08               &   ...                    \\
\\ 
                     &            &        &   &   \multicolumn{5}{c}{N~V~$\lambda 1238$} \\
\cline{5-9} \\
D1                 & 1219   &  $-52$       & 4      &    $<$0.032                  &     (16)          &     ...                   & $<$13.22               &      ...                  \\ 
D2                 & 1309   &  +38         & 2      &    $<$0.032                  &     (20)          &    ...                    & $<$13.22                &     ...                   \\
\enddata
\tablenotetext{a}{''total [p]'' in this column refers to the total equivalent width (EW) and total column densities of \lya\
calculated from the blended theoretical line profiles used to derive $N$ and $b$ values for the 2 individual components;
no measurement from the actual spectrum is given because the components are slightly blended with Ly$\beta$ at $z=0.189$.}
\tablenotetext{b}{EW limits are $2\sigma(W)_T$ values measured over 14 pixels. Doppler parameters in parentheses are the values used to covert 
these EW limits to column density upper limits.}
\tablenotetext{c}{These are the limits from the Si~IV~$\lambda 1402$ line, as the $\lambda 1392$ is blended with a higher-redshift line.}
\end{deluxetable*}